\documentclass[11pt,a4paper]{article}

\usepackage{amssymb}
\usepackage[dvips]{graphicx}
\usepackage{bm}

\begin{document}

\title{The $\beta$-function of ${\cal N}=1$ supersymmetric gauge theories regularized by higher covariant derivatives as an integral of double total derivatives}

\author{
K.V.Stepanyantz\\
{\small{\em Moscow State University,}}\\
{\small{\em Faculty of Physics, Department of Theoretical Physics,}}\\
{\small{\em 119991, Moscow, Russia}}}

\maketitle

\begin{abstract}
For a general ${\cal N}=1$ supersymmetric gauge theory regularized by higher covariant derivatives we prove in all orders that the $\beta$-function defined in terms of the bare couplings is given by integrals of double total derivatives with respect to loop momenta. With the help of the technique used for this proof it is possible to construct a method for obtaining these loop integrals, which essentially simplifies the calculations. As an illustration of this method, we find the expression for the three-loop contribution to the $\beta$-function containing the Yukawa couplings and compare it with the result of the standard calculations made earlier. Also we briefly discuss, how the structure of the loop integrals for the $\beta$-function considered in this paper can be used for the all-loop perturbative derivation of the NSVZ relation in the non-Abelian case.
\end{abstract}

\unitlength=1cm

\section{Introduction}
\hspace*{\parindent}

Ultraviolet divergences in supersymmetric theories are restricted by some non-renormalization theorems. According to one of them, ${\cal N}=4$ supersymmetric Yang--Mills (SYM) theory is finite in all orders \cite{Grisaru:1982zh,Mandelstam:1982cb,Brink:1982pd,Howe:1983sr}. Divergencies in ${\cal N}=2$ theories exist only in the one-loop approximation \cite{Grisaru:1982zh,Howe:1983sr,Buchbinder:1997ib}, so that it is even possible to construct finite ${\cal N}=2$ supersymmetric theories by choosing a gauge group and a matter representation in such a way that the one-loop divergencies cancel \cite{Howe:1983wj}. All these non-renormalization theorems can be derived \cite{Shifman:1999mv,Buchbinder:2014wra} from the equation which relates the $\beta$-function of ${\cal N}=1$ supersymmetric gauge theories with the anomalous dimension of the matter superfields \cite{Novikov:1983uc,Jones:1983ip,Novikov:1985rd,Shifman:1986zi}

\begin{equation}\label{NSVZ_First_Form}
\beta(\alpha,\lambda) = - \frac{\alpha^2\Big(3 C_2 - T(R) + C(R)_i{}^j
\big(\gamma_\phi\big)_j{}^i(\alpha,\lambda)/r\Big)}{2\pi(1- C_2\alpha/2\pi)},
\end{equation}

\noindent
where $\alpha$  is the gauge coupling constant and $\lambda$ denotes the Yukawa couplings. Note that so far we do not specify the definitions of the renormalization group functions (RGFs) and what couplings are considered as their arguments. Eq. (\ref{NSVZ_First_Form}) called the exact NSVZ $\beta$-function can also be considered as a non-renormalization theorem in addition to the well-known statement that the superpotential in ${\cal N}=1$ supersymmetric theories is not renormalized \cite{Grisaru:1979wc}. According to one more non-renormalization theorem derived in \cite{Stepanyantz:2016gtk}, the triple ghost-gauge vertices in ${\cal N}=1$ supersymmetric gauge theories are finite in all orders.\footnote{In the Landau gauge $\xi\to 0$  a similar statement was known earlier for the usual (non-supersymmetric) Yang--Mills theory \cite{Dudal:2002pq} and for ${\cal N}=1$ SYM formulated in terms of the component fields \cite{Capri:2014jqa}. In the former case this statement was explicitly verified by the four-loop calculation in Ref. \cite{Chetyrkin:2004mf}.} With the help of this non-renormalization theorem the exact NSVZ $\beta$-function can be equivalently rewritten in a new form \cite{Stepanyantz:2016gtk},

\begin{equation}\label{NSVZ_Second_Form}
\frac{\beta(\alpha,\lambda)}{\alpha^2} = - \frac{1}{2\pi}\Big(3 C_2 - T(R) - 2C_2 \gamma_c(\alpha,\lambda) - 2C_2 \gamma_V(\alpha,\lambda) + C(R)_i{}^j \big(\gamma_\phi\big)_j{}^i(\alpha,\lambda)/r\Big),
\end{equation}

\noindent
which relates the $\beta$-function to the anomalous dimensions of the quantum gauge superfield ($\gamma_V$), of the Faddeev--Popov ghosts ($\gamma_c$), and of the matter superfields ($\big(\gamma_\phi\big)_i{}^j$).

Some NSVZ-like relations can be written for other theories. For example, in theories with softly broken supersymmetry an analogous equation describes the renormalization of the gaugino mass \cite{Hisano:1997ua,Jack:1997pa,Avdeev:1997vx}. Also it is possible to construct the NSVZ-like equations for the Adler $D$-function in ${\cal N}=1$ SQCD \cite{Shifman:2014cya,Shifman:2015doa} and even for the renormalization of the Fayet--Iliopoulos term in two-dimensional ${\cal N}=(0,2)$ supersymmetric models \cite{Chen:2019eta}.

Various derivations of the exact NSVZ $\beta$-function involve general arguments based on the analysis of the instanton contributions \cite{Shifman:1999mv,Novikov:1983uc}, anomalies \cite{Jones:1983ip,Shifman:1986zi,ArkaniHamed:1997mj}, and non-renormalization of the topological term \cite{Kraus:2002nu}. However, a direct perturbative verification of Eq. (\ref{NSVZ_First_Form}) in all orders appeared to be a highly non-trivial problem. Even to start solving this problem, one should first pay attention to some important subtleties related to the regularization, quantization, and renormalization.

Really, the calculations of quantum corrections made in the $\overline{\mbox{DR}}$-scheme (that is with the help of dimensional reduction \cite{Siegel:1979wq} supplemented by the modified minimal subtractions \cite{Bardeen:1978yd}) in Refs. \cite{Jack:1996vg,Jack:1996cn,Jack:1998uj,Harlander:2006xq,Mihaila:2013wma} demonstrate that the NSVZ relation is not valid for this renormalization prescription. However, the difference can be explained by the scheme dependence of the NSVZ relation, which is described by the general equations derived in \cite{Kutasov:2004xu,Kataev:2014gxa}. Namely, it is possible to tune the renormalization scheme in such a way that the NSVZ equation will take place \cite{Jack:1996vg,Jack:1996cn,Jack:1998uj}.\footnote{A similar result for the Adler $D$-function can be found in Ref. \cite{Aleshin:2019yqj}.} It is important that this possibility is highly not-trivial due to some scheme-independent equations following from the NSVZ relation \cite{Kataev:2014gxa,Kataev:2013csa}. Nevertheless, at present there is no general all-loop prescription giving the NSVZ scheme in the case of using the regularization by dimensional reduction.

The NSVZ renormalization prescription can be naturally formulated in all loops if ${\cal N}=1$ supersymmetric gauge theories are regularized by the higher covariant derivative method \cite{Slavnov:1971aw,Slavnov:1972sq} in the supersymmetric version \cite{Krivoshchekov:1978xg,West:1985jx}. The matter is that using of this regularization reveals the underlying structure of the loop integrals responsible for appearing the NSVZ relation. Namely, in this case the integrals giving the $\beta$-function defined in terms of the bare couplings appear to be integrals of double total derivatives with respect to loop momenta.\footnote{It is important that for theories regularized by dimensional reduction such a factorization does not take place, see Refs. \cite{Aleshin:2015qqc,Aleshin:2016rrr} for the detailed discussion.} This was first noted in calculating quantum corrections for ${\cal N}=1$ supersymmetric electrodynamics (SQED) in Refs. \cite{Soloshenko:2003nc} (the factorization into total derivatives) and \cite{Smilga:2004zr} (the factorization into double total derivatives). Subsequently, this structure of the loop integrals has been confirmed by numerous calculations (see, e.g., Refs. \cite{Pimenov:2009hv,Stepanyantz:2011cpt,Stepanyantz:2011bz,Stepanyantz:2012zz,Stepanyantz:2012us,Kazantsev:2014yna,Shakhmanov:2017soc,Kataev:2017qvk,Kazantsev:2018nbl}). The rigorous all-loop proof for ${\cal N}=1$ SQED has been done in \cite{Stepanyantz:2011jy,Stepanyantz:2014ima}. The same method allowed proving the factorization into integrals of double total derivatives in all orders for the Alder $D$-function in ${\cal N}=1$ SQCD \cite{Shifman:2014cya,Shifman:2015doa} and for the renormalization of the photino mass in softly broken ${\cal N}=1$ SQED \cite{Nartsev:2016nym}. For the non-Abelian supersymmetric gauge theories this will be done in this paper.

The integrals of double total derivatives do not vanish due to the identity

\begin{equation}
\frac{\partial^2}{\partial Q^\mu\,\partial Q_\mu} \frac{1}{Q^2} = -4\pi^2 \delta^4(Q),
\end{equation}

\noindent
where $Q$ is an Euclidean momentum. The $\delta$-function reduces the number of loop integrations by 1, so that in the Abelian case an $L$-loop contribution to the $\beta$-function appears to be related to an $(L-1)$-loop contribution to the anomalous dimension of the matter superfields. The sum of singularities in the Abelian case was calculated in \cite{Stepanyantz:2011jy,Stepanyantz:2014ima}, where it was expressed in terms of the anomalous dimension of the matter superfields. The relation between the $\beta$-function and the anomalous dimension obtained in this way is nothing else than the NSVZ equation for RGFs defined in terms of the bare couplings. Thus, at least in the Abelian case, it naturally appears in the case of using the higher derivative regularization. Note that the RGFs defined in terms of the bare couplings are scheme independent if a regularization is fixed (see, e.g., \cite{Kataev:2013eta}), so that the NSVZ equation for these RGFs is valid for an arbitrary renormalization prescription.\footnote{The NSVZ equation for RGFs defined in terms of the bare couplings is not valid in the case of using dimensional reduction starting from the three-loop approximation \cite{Aleshin:2016rrr}.}

In the non-Abelian case the situation is much more complicated. Eq. (\ref{NSVZ_First_Form}) relates an $L$-loop contribution to the $\beta$-function to the anomalous dimension of the matter superfields in all previous orders. That is why it is more probable that it is Eq. (\ref{NSVZ_Second_Form}) that originally appears in the perturbative calculations. Moreover, unlike Eq. (\ref{NSVZ_First_Form}), Eq. (\ref{NSVZ_Second_Form}) can be visualized in the same way as in the Abelian case (see Refs. \cite{Smilga:2004zr,Kazantsev:2014yna}). Namely, starting from a supergraph without external lines, it is possible to obtain a contribution to the $\beta$-function by attaching two external lines of the background gauge superfield and contributions to the anomalous dimensions by cutting internal lines. Thus obtained contributions are related by Eq. (\ref{NSVZ_Second_Form}).

The similarity between Eq. (\ref{NSVZ_Second_Form}) and the Abelian NSVZ equation \cite{Vainshtein:1986ja,Shifman:1985fi} allows suggesting that the factorization of integrals into double total derivatives also produces the NSVZ equation in the non-Abelian case. This guess was confirmed by numerous calculations in the lowest loops, see, e.g., \cite{Stepanyantz:2011bz,Shakhmanov:2017soc,Kazantsev:2018nbl,Shakhmanov:2017wji}). This implies that all higher order corrections to the $\beta$-function (starting from the two-loop approximation) appear from the $\delta$-singularities. Therefore, to derive the NSVZ relation in the non-Abelian case (for RGFs defined in terms of the bare couplings with the higher covariant derivative regularization), it is necessary only to sum singular contributions and to prove that they give the sum of the anomalous dimensions in the right hand side of Eq. (\ref{NSVZ_Second_Form}). If this is really so, then the NSVZ scheme for RGFs defined in terms of the renormalized couplings is given by the so-called HD+MSL prescription \cite{Stepanyantz:2016gtk} exactly as in the Abelian case \cite{Kataev:2014gxa,Kataev:2013csa,Kataev:2013eta}.\footnote{HD+MSL prescription also gives the NSVZ-like schemes for the Adler $D$-function \cite{Kataev:2017qvk} and for the renormalization of the photino mass in softly broken ${\cal N}=1$ SQED \cite{Nartsev:2016mvn}.} This means that the theory is regularized by higher covariant derivatives supplemented by the minimal subtractions of logarithms, when only powers of $\ln\Lambda/\mu$ are included into the renormalization constants.\footnote{This NSVZ scheme is not unique \cite{Goriachuk:2018cac}. For example, in ${\cal N}=1$ SQED the on-shell scheme is also NSVZ \cite{Kataev:2019olb}.}

The paper is organized as follows: In Sect. \ref{Section_Supersymmetric_Gauge_Theories} we formulate the theory under consideration in ${\cal N}=1$ superspace, regularize it by higher covariant derivatives, and describe the quantization. Also in this section we introduce some auxiliary constructions, which will be needed for the investigation of the loop integrals giving the $\beta$-function. RGFs defined in terms of the bare couplings are introduced in Sect. \ref{Section_Bare_RGFs}. In this section we also present the $\beta$-function and the NSVZ relation for it in the form which is mostly convenient for the analysis. In Sect. \ref{Section_Formal_Derivation} we demonstrate that the $\beta$-function defined in terms of the bare couplings is given by integrals of double total derivatives with respect to loop momenta. Here we also describe the method which allows to construct these integrals in a simple way. This method is applied for calculating the three-loop contribution to the $\beta$-function containing the Yukawa couplings in Sect. \ref{Section_Examples}. In particular, we demonstrate that the result exactly coincides with the one obtained in Ref. \cite{Kazantsev:2018nbl} with the help of the standard supergraph calculation.

\section{${\cal N}=1$ supersymmetric gauge theories: regularization, quantization, and auxiliary parameters}
\hspace*{\parindent}\label{Section_Supersymmetric_Gauge_Theories}

It is convenient to describe ${\cal N}=1$ supersymmetric gauge theories using ${\cal N}~=~1$ superspace with the coordinates $(x^\mu,\theta)$, where $\theta$ is an auxiliary anticommuting Majorana spinor. In this case ${\cal N}=1$ supersymmetry of the theory is manifest. Moreover, it becomes possible to perform the quantization and calculate quantum corrections in a manifestly ${\cal N}=1$ supersymmetric way \cite{Gates:1983nr,West:1990tg,Buchbinder:1998qv}. At the classical level the considered theory in the massless limit is described by the action

\begin{eqnarray}\label{Classical_Action}
&& S_{\mbox{\scriptsize classical}} = \frac{1}{2e^2} \mbox{Re}\,\mbox{tr} \int d^4x\, d^2\theta\, W^a W_a + \frac{1}{4} \int d^4x\, d^4\theta\, \phi^{*i} (e^{2V})_i{}^j \phi_j\qquad\nonumber\\
&&\qquad\qquad\qquad\qquad\qquad\qquad\qquad\qquad\qquad + \Big(\frac{1}{6} \lambda^{ijk} \int d^4x\, d^2\theta\, \phi_i \phi_j \phi_k + \mbox{c.c.}\Big),\qquad
\end{eqnarray}

\noindent
where $V$ is the Hermitian gauge superfield and $\phi_i$ are the chiral matter superfields in a representation $R$ of a gauge group $G$ which is assumed to be simple. In the classical theory (\ref{Classical_Action}) the supersymmetric gauge superfield strength is defined as $W_a \equiv \bar D^2 \left(e^{-2V} D_a e^{2V}\right)/8$. The gauge coupling constant is defined as $\alpha =e^2/4\pi$, and the Yukawa couplings are denoted by $\lambda^{ijk}$. Note that at the classical level we do not distinguish between bare and renormalized couplings. This difference is essential in the quantum theory. Below, considering the quantum theory, we will denote the bare couplings by $\alpha_0 = e_0^2/4\pi$ and $\lambda_0^{ijk}$, while the renormalized couplings will be denoted by $\alpha$ and $\lambda^{ijk}$.

Below $t^A$ and $T^A$ are the generators of the fundamental representation and the representation $R$, respectively. These sets of generators satisfy the conditions

\begin{eqnarray}\label{Normalization_Of_Generators}
&& \mbox{tr}(t^A t^B) = \frac{1}{2} \delta^{AB}; \qquad\qquad\  [t^A, t^B] = i f^{ABC} t^C;\nonumber\\
&& \mbox{tr}(T^A T^B) = T(R)\, \delta^{AB};\qquad [T^A, T^B] = i f^{ABC} T^C.
\end{eqnarray}

\noindent
We will always assume that $\mbox{tr}(T^A) = 0$. Also we will use the notation

\begin{equation}
(T^A T^A)_i{}^j \equiv C(R)_i{}^j;\qquad f^{ACD} f^{BCD} \equiv C_2 \delta^{AB};\qquad r\equiv \mbox{dim}\, G = \delta^{AA},
\end{equation}

\noindent
so that $C(Adj)_A{}^B = C_2 \delta_A^B$. (The generators of the adjoint representation are expressed in terms of the structure constants as $(T_{Adj}^A)_{B}{}^{C} = -i f^{ABC}$.)

Under the condition

\begin{equation}\label{Yukawa_Invariance}
\lambda^{ijm} (T^A)_m{}^k + \lambda^{imk} (T^A)_m{}^j + \lambda^{mjk} (T^A)_m{}^i = 0
\end{equation}

\noindent
the theory (\ref{Classical_Action}) is invariant under the gauge transformations

\begin{equation}
\phi_i \to (e^{A})_i{}^j \phi_j;\qquad e^{2V} \to e^{-A^+} e^{2V} e^{-A}\qquad (\mbox{so that}\quad W_a \to e^{A} W_a e^{-A}),
\end{equation}

\noindent parameterized by a Lie algebra valued chiral superfield $A$.

To quantize the theory (\ref{Classical_Action}), it is also necessary to take into account that the quantum gauge superfield is renormalized in a nonlinear way \cite{Piguet:1981fb,Piguet:1981hh,Tyutin:1983rg} (see also Refs. \cite{Piguet:1981mu,Piguet:1984mv}). The necessity of this nonlinear renormalization has been demonstrated by explicit calculations in Refs. \cite{Juer:1982fb,Juer:1982mp}. Moreover, the two-loop calculation of the Faddeev--Popov ghost anomalous dimension in \cite{Kazantsev:2018kjx} showed that without this nonlinear renormalization the renormalization group equations are not satisfied. Thus, it is really needed for quantum calculations. To take into account the nonlinear renormalization, following Ref. \cite{Piguet:1981hh}, we substitute the gauge superfield $V$ by the function ${\cal F}(V)$ in the action functional. Moreover, it is necessary to replace $e$ and $\lambda$ by the bare couplings $e_0$ and $\lambda_0$, respectively.

For obtaining a manifestly gauge invariant effective action we will use the background field method \cite{DeWitt:1965jb,Abbott:1980hw,Abbott:1981ke} formulated in ${\cal N}=1$ superspace \cite{Grisaru:1982zh,Gates:1983nr}. A distinctive feature of the background field method in the supersymmetric case is the nonlinear background-quantum splitting which in the considered case can be implemented by the substitution

\begin{equation}\label{Background-Quantum_Splitting}
e^{2{\cal F}(V)} \to e^{2{\cal F}(V)} e^{2\bm{V}},
\end{equation}

\noindent
where in the right hand side $V$ and $\bm{V}$ are the quantum and background gauge superfields, respectively.\footnote{The standard form of the background quantum splitting is $e^{2{\cal F}(V)} \to e^{\bm{\Omega}^+} e^{2{\cal F}(V)} e^{\bm{\Omega}}$, the background gauge superfield being defined by the equation $e^{2\bm{V}} = e^{\bm{\Omega}^+} e^{\bm{\Omega}}$. However, after the change of variables $V \to e^{-\bm{\Omega}^+} V e^{\bm{\Omega}^+}$ in the generating functional we arrive to Eq. (\ref{Background-Quantum_Splitting}).} In this case the quantum gauge superfield satisfies the constrain $V^+ = e^{-2\bm{V}} V e^{2\bm{V}}$.

Due to the background-quantum splitting the gauge invariance produces two different types of gauge transformations. Under the background gauge symmetry the superfields of the theory change as

\begin{equation}\label{Background_Gauge_Invariance}
e^{2\bm{V}} \to e^{-A^+} e^{2\bm{V}} e^{-A};\qquad  V\to e^{-A^+} V e^{A^+};\qquad \phi_i \to (e^A)_i{}^j \phi_j.
\end{equation}

\noindent
This invariance remains unbroken at the quantum level and becomes a manifest symmetry of the effective action. Alternatively, the quantum gauge invariance

\begin{equation}\label{Quantum_Gauge_Invariance}
e^{2{\cal F}(V)} \to e^{-A^+} e^{2{\cal F}(V)} e^{2\bm{V}} e^{-A} e^{-2\bm{V}};\qquad
\bm{V} \to \bm{V};\qquad \phi_i \to (e^A)_i{}^j \phi_j
\end{equation}

\noindent is broken by the gauge fixing procedure. It is convenient to introduce the background supersymmetric covariant derivatives $\bm{\nabla}_a$ and $\bm{\bar\nabla}_{\dot a}$ and the gauge supersymmetric covariant derivatives $\nabla_a$ and $\bar\nabla_{\dot a}$ defined by the equations

\begin{equation}\label{Covariant_Derivatives}
\bm{\nabla}_a = \nabla_a \equiv D_a;\qquad \bm{\bar\nabla}_{\dot a} \equiv e^{2\bm{V}} \bar D_{\dot a} e^{-2\bm{V}};\qquad \bar\nabla_{\dot a} \equiv e^{2{\cal F}(V)} e^{2\bm{V}} \bar D_{\dot a} e^{-2\bm{V}} e^{-2{\cal F}(V)}.
\end{equation}

\noindent
Note that for the purposes of this paper it is more convenient to use a different representation for them in comparison with Refs. \cite{Kazantsev:2018kjx,Aleshin:2016yvj}. In the representation (\ref{Covariant_Derivatives}) the covariant derivatives $\nabla_a$ and $\bar\nabla_{\dot a}$ should act on a function $X$ which changes as $X\to e^{-A^+}X$. In this case they transform in the same way under both background and quantum transformations. This is also valid for the background covariant derivatives  $\bm{\nabla_a}$ and $\bm{\bar\nabla}_{\dot a}$, but only in the case of the background gauge transformations.

If we use the background field method and take into account the nonlinear renormalization of the quantum gauge superfield, then the gauge superfield strength is defined as

\begin{equation}\label{W_Definition}
W_a \equiv \frac{1}{8} \bar D^2 \left(e^{-2\bm{V}} e^{-2{\cal F}(V)}\, D_a \left(e^{2{\cal F}(V)}e^{2\bm{V}}\right)\right)
= \frac{1}{8} e^{-2\bm{V}}\bm{\bar\nabla}^2\left( e^{-2{\cal F}(V)} \bm{\nabla}_a e^{2{\cal F}(V)} \right) e^{2\bm{V}} + \bm{W}_a,
\end{equation}

\noindent
where

\begin{equation}\label{W_Bold_Definition}
\bm{W}_a \equiv \frac{1}{8} \bar D^2\left(e^{-2\bm{V}} D_a e^{2\bm{V}}\right).
\end{equation}

Below we will also need some auxiliary parameters. The coordinate-independent complex parameter $g$ describes the continuous deformation of the original theory (corresponding to $g=1$) into the theory in which quantum superfields interact only with the background gauge superfield (corresponding to $g\to 0$). This parameter is introduced by making the substitutions

\begin{equation}
\alpha_0 \to g g^* \alpha_0;\qquad \lambda_0^{ijk}\to g \lambda_0^{ijk};\qquad \lambda^*_{0ijk}\to g^* \lambda^*_{0ijk}.
\end{equation}

\noindent
Then, it is easy to see that an $L$-loop contribution to the two-point Green function of the background gauge superfield is proportional to $(g g^*)^{L-1}$.

Also we introduce the auxiliary chiral superfield\footnote{Note that coordinate-dependent auxiliary parameters were also used in Refs. \cite{Kraus:2002nu,Kraus:2001tg,Kraus:2001kn,Babington:2005vu}.} $\mbox{\sl g}(x,\theta)$. It is added to $g$ in such a way that all quantum corrections containing $g$ will actually depend on the (coordinate-dependent) combination

\begin{equation}\label{Auxiliary_Parameter}
\bm{g} \equiv g+ \mbox{\sl g},
\end{equation}

\noindent
while the background gauge invariance remains unbroken. Various parts of the total action containing the superfield $\mbox{\sl g}$ are written below, see Eqs. (\ref{Regularized_Action}), (\ref{Gauge_Fixing_Term}), and (\ref{Nielsen_Kallosh_Action}).

Now, let us include the parameters $g$ and $\mbox{\sl g}$ into the classical action. For this purpose we write all terms containing the quantum gauge superfield as integrals over $d^4x\, d^4\theta \equiv d^8x$ with the help of Eq. (\ref{W_Definition}). After this we modify the result by introducing the auxiliary parameters in the following way:

\begin{eqnarray}\label{Bare_Action}
&& S_{\mbox{\scriptsize classical}}\ \to\ \frac{1}{2 g g^* e_0^2}\,\mbox{Re}\,\mbox{tr} \int d^6x\, \bm{W}^a \bm{W}_a -\frac{1}{8e_0^2}\,\mbox{Re}\,\mbox{tr} \int d^8x\, \frac{1}{\bm{g} \bm{g}^*} \Big[
\, \frac{1}{8} e^{-2{\cal F}(V)} \bm{\nabla}^a e^{2{\cal F}(V)} \qquad\nonumber\\
&& \times \bm{\bar\nabla}^2\left(e^{-2{\cal F}(V)} \bm{\nabla}_a e^{2{\cal F}(V)} \right) + 2 e^{2\bm{V}} \bm{W}^a e^{-2\bm{V}} e^{-2{\cal F}(V)} \bm{\nabla}_a e^{2{\cal F}(V)} \Big]  + \frac{1}{4} \int d^8x\, \phi^{*i} (e^{2{\cal F}(V)} \nonumber\\
&& \times e^{2\bm{V}})_i{}^j \phi_j + \frac{1}{6} \Big(\lambda_0^{ijk} \int d^6x\, \bm{g}\,\phi_i \phi_j \phi_k + \mbox{c.c.}\Big),\qquad
\end{eqnarray}

\noindent
where the integration measures are

\begin{equation}\label{Integration_Measures_D6}
\int d^6x \equiv \int d^4x\,d^2\theta_x;\qquad \int d^8x \equiv \int d^4x\, d^4\theta_x = -\frac{1}{2}\int d^6x\, \bar D^2.
\end{equation}

\noindent
Note that we do not include the superfield $\mbox{\sl g}$ in the first term of Eq. (\ref{Bare_Action}), which does not contain the  quantum gauge superfield $V$. This allows to avoid breaking of the background gauge invariance (\ref{Background_Gauge_Invariance}). However, the action (\ref{Bare_Action}) is invariant under the quantum gauge transformations (\ref{Quantum_Gauge_Invariance}) only if $\mbox{\sl g}=0$ (but for an arbitrary value of the coordinate independent parameter $g$). Nevertheless, it is not important, because the parameter $\mbox{\sl g}$ is auxiliary and actually we are interested only in the cases when $g=0,1$ and $\mbox{\sl g}=0$.

The most important ingredient needed for deriving the NSVZ $\beta$-function for RGFs defined in terms of the bare couplings is the higher covariant derivative regularization \cite{Slavnov:1971aw,Slavnov:1972sq}. In this paper we will use the version similar to the one considered in Ref. \cite{Aleshin:2016yvj} with some modifications appearing due to the presence of the auxiliary parameters and the function ${\cal F}(V)$. To regularize a theory by higher covariant derivatives, at the first step, it is necessary to add a higher derivative term $S_\Lambda$ to its action. As a result, propagators will contain higher degrees of momenta that, in turn, leads to the finiteness of the regularized theory beyond the one-loop approximation \cite{Faddeev:1980be}. In the case $\mbox{\sl g}=0$ the regularized action $S_{\mbox{\scriptsize reg}} = S+S_\Lambda$ invariant under both background and quantum gauge transformations can be constructed as

\begin{eqnarray}\label{Original_Regularized_Action}
&&\hspace*{-7mm} S_{\mbox{\scriptsize reg}}\Big|_{\mbox{\scriptsize\sl g}=0} = \frac{1}{2g g^* e_0^2}\mbox{Re}\, \mbox{tr} \int d^6x\, W^a \left(e^{-2\bm{V}} e^{-2{\cal F}(V)} \right)_{Adj} R\Big(-\frac{\bar\nabla^2 \nabla^2}{16\Lambda^2}\Big)_{Adj} \left(e^{2{\cal F}(V)}e^{2\bm{V}}\right)_{Adj} W_a \nonumber\\
&&\hspace*{-7mm} + \frac{1}{4} \int d^8x\, \phi^{*i} \Big(F\Big(-\frac{\bar\nabla^2 \nabla^2}{16\Lambda^2}\Big) e^{2{\cal F}(V)}e^{2\bm{V}}\Big)_i{}^j \phi_j
+ \frac{1}{6} \Big(g\lambda_0^{ijk} \int d^6x\, \phi_i \phi_j \phi_k + \mbox{c.c.} \Big),
\end{eqnarray}

\noindent
where the higher derivative regulators $R(x)$ and $F(x)$ are functions rapidly growing at infinity which satisfy the conditions $R(0)=F(0)=1$. In Eq. (\ref{Original_Regularized_Action}) and below the subscript $Adj$ means that

\begin{equation}
\Big(f_0 + f_1 X + f_2 X^2 + \ldots\Big)_{Adj} Y = f_0 Y + f_1 [X, Y] + f_2 [X,[X,Y]] + \ldots
\end{equation}

\noindent
(In particular, this equation implies that $(e^X)_{Adj} Y = e^X Y e^{-X}$.) The superfield $\mbox{\sl g}$ should be included into the regularized action in such a way that the background gauge invariance remains unbroken. This can be done similarly to constructing the action (\ref{Bare_Action}). However, it is more difficult due to the presence of the function $R(x)$. We present this function in the form

\begin{equation}
R(x) \equiv 1 + x r(x), \qquad\mbox{where}\qquad r(x) = \frac{R(x)-1}{x} = \sum\limits_{k=1}^\infty r_k x^{k-1}.
\end{equation}

\noindent
Then the regularized action can be written as

\begin{eqnarray}\label{Regularized_Action}
&&\hspace*{-7mm} S_{\mbox{\scriptsize reg}} = \frac{1}{2 g g^* e_0^2}\,\mbox{Re}\,\mbox{tr} \int d^6x\, \bm{W}^a \bm{W}_a + \frac{1}{e_0^2}\,\mbox{Re}\,\mbox{tr} \int d^8x\, \frac{1}{\bm{g} \bm{g}^*} \Big[ -\frac{1}{4} e^{-2{\cal F}(V)} \bm{\nabla}^a e^{2{\cal F}(V)} e^{2\bm{V}} \bm{W}_a
\nonumber\\
&&\hspace*{-7mm} \times e^{-2\bm{V}} - \frac{1}{64} e^{-2{\cal F}(V)} \bm{\nabla}^a e^{2{\cal F}(V)} \bm{\bar\nabla}^2\left(e^{-2{\cal F}(V)} \bm{\nabla}_a e^{2{\cal F}(V)} \right)
+  W^a \left(e^{-2\bm{V}} e^{-2{\cal F}(V)} \right)_{Adj} \frac{\nabla^2}{16\Lambda^2}\vphantom{\frac{\Lambda^2}{\Lambda^2}}\nonumber\\
&&\hspace*{-7mm} \times\, r\Big(-\frac{\bar\nabla^2 \nabla^2}{16\Lambda^2}\Big)_{Adj} \left(e^{2{\cal F}(V)}e^{2\bm{V}}\right)_{Adj} W_a\Big] + \frac{1}{4} \int d^8x\, \phi^{*i} \Big(F\Big(-\frac{\bar\nabla^2 \nabla^2}{16\Lambda^2}\Big) e^{2{\cal F}(V)}e^{2\bm{V}}\Big)_i{}^j \phi_j\nonumber\\
&&\hspace*{-7mm} + \frac{1}{6} \Big(\lambda_0^{ijk} \int d^6x\, \bm{g}\,\phi_i \phi_j \phi_k + \mbox{c.c.} \Big).
\end{eqnarray}

\noindent
It is important that this action is invariant under the background gauge transformations, but the quantum gauge invariance exists only for $\mbox{\sl g}=0$. In this case the action (\ref{Regularized_Action}) is reduced to Eq. (\ref{Original_Regularized_Action}). Moreover, all terms containing the quantum superfields depend on auxiliary parameters only in the combination $\bm{g} = g+\mbox{\sl g}$. (The first term, which depends on the constant $g$ and does not depend on the superfield $\mbox{\sl g}$, contains only the background gauge superfield.)

To obtain a manifestly gauge invariant effective action, it is necessary to use a gauge fixing term invariant under the background transformations (\ref{Background_Gauge_Invariance}). Taking into account that a higher derivative regulator should be also inserted into this term \cite{Aleshin:2016yvj}, the gauge fixing action can be chosen as

\begin{equation}\label{Gauge_Fixing_Term}
S_{\mbox{\scriptsize gf}} = -\frac{1}{16\xi_0 e_0^2}\, \mbox{tr} \int d^8x\,  \bm{\nabla}^2 V \frac{1}{\bm{g}^*}  K\Big(-\frac{\bm{\bar\nabla}^2 \bm{\nabla}^2}{16\Lambda^2}\Big)_{Adj} \frac{1}{\bm{g}} \bm{\bar\nabla}^2 V.
\end{equation}

\noindent
Certainly, the quantization procedure also requires to introduce the Faddeev--Popov action. The Faddeev--Popov ghosts and the corresponding antighosts in the supersymmetric case are described by the chiral superfields $c^A$ and $\bar c^A$, respectively. The action for them obtained in a standard way takes the form

\begin{eqnarray}\label{Faddeev_Popov_Action}
&& S_{\mbox{\scriptsize FP}} = \frac{1}{2} \int
d^8x\, \frac{\partial {\cal F}^{-1}(\widetilde V)^A}{\partial {\widetilde V}^B}\left.\vphantom{\frac{1}{2}}\right|_{\widetilde V = {\cal F}(V)} \left(e^{2\bm{V}}\bar c e^{-2\bm{V}} +
\bar c^+ \right)^A\nonumber\\
&&\qquad\qquad\qquad\qquad \times \left\{\vphantom{\frac{1}{2}}\smash{\Big(\frac{{\cal F}(V)}{1-e^{2{\cal F}(V)}}\Big)_{Adj} c^+
+ \Big(\frac{{\cal F}(V)}{1-e^{-2{\cal F}(V)}}\Big)_{Adj}
\Big(e^{2\bm{V}} c e^{-2\bm{V}}\Big)}\right\}^B.\qquad
\end{eqnarray}

\noindent
In the case of using the background superfield method it is also necessary to take into account the Nielsen--Kallosh ghost action

\begin{eqnarray}\label{Nielsen_Kallosh_Action}
&& S_{\mbox{\scriptsize NK}} = \frac{1}{2e_0^2}\, \mbox{tr} \int d^8x\,  b^+ \frac{1}{\bm{g}^*} \Big(K\Big(-\frac{\bm{\bar\nabla}^2 \bm{\nabla}^2}{16\Lambda^2}\Big) e^{2\bm{V}}\Big)_{Adj} \frac{1}{\bm{g}} b\nonumber\\
&&\qquad\qquad\qquad\qquad\qquad\qquad \to \frac{1}{2}\,\mbox{tr} \int d^8x\,  b^+ \Big(K\Big(-\frac{\bm{\bar\nabla}^2 \bm{\nabla}^2}{16\Lambda^2}\Big) e^{2\bm{V}}\Big)_{Adj} b.\qquad
\end{eqnarray}

\noindent
Here the Nielsen--Kallosh ghosts $b$ are chiral anticommuting superfields in the adjoint representation, which interact only with the background gauge superfield. The arrow points out that the parameters $\bm{g}$ and $e_0$ can be excluded from the Nielsen--Kallosh action by the change of variables $b\to e_0 \bm{g} b;\ \ b^+ \to e_0 \bm{g}^* b^+$ in the generating functional. (It is easy to see that the corresponding determinant is equal to 1.)

After the gauge fixing procedure the quantum gauge transformations (\ref{Quantum_Gauge_Invariance}) are no longer a symmetry of the total action (that, in particular includes the gauge fixing term and ghosts). The total action is invariant under the BRST transformations \cite{Becchi:1974md,Tyutin:1975qk}. In ${\cal N}=1$ superspace the BRST transformations have been formulated in Ref. \cite{Piguet:1981fb}. For the theory considered in this paper the BRST invariance is a symmetry of the action only in the case $\mbox{\sl g}=0$, but for an arbitrary value of the coordinate independent parameter $g$.

As we mentioned above, the one-loop divergences cannot be regularized by adding the higher derivative term to the action. For this purpose it is necessary to supplement the higher derivative method by the Pauli--Villars regularization which is introduced by inserting the Pauli--Villars determinants into the generating functional \cite{Slavnov:1977zf}. According to Refs. \cite{Aleshin:2016yvj,Kazantsev:2017fdc}, to cancel the one-loop divergences appearing in supersymmetric gauge theories, one should introduce three chiral Pauli--Villars superfields $\varphi_a$ with $a=1,2,3$ in the adjoint representation of the gauge group, and chiral superfields $\Phi_i$ in a certain representation $R_{\mbox{\scriptsize PV}}$ which admits a gauge invariant mass term. The superfields $\varphi_a$ cancel one-loop divergences coming from the loops of the quantum gauge superfield, of the Faddeev--Popov ghosts and of the Nielsen--Kallosh ghosts. The superfields $\Phi_i$ cancel the one-loop divergences coming from the matter loop. This occurs if the generating functional is defined as

\begin{equation}\label{Generating_Functional_Z}
Z = \int D\mu\, \mbox{Det}(PV,M_\varphi)^{-1}\mbox{Det}(PV,M)^c  \exp\Big\{i\Big(S_{\mbox{\scriptsize reg}} + S_{\mbox{\scriptsize gf}} + S_{\mbox{\scriptsize FP}} + S_{\mbox{\scriptsize NK}} + S_{\mbox{\scriptsize sources}}\Big)\Big\},
\end{equation}

\noindent
where $D\mu$ denotes the measure of the functional integration and $c = T(R)/T(R_{\mbox{\scriptsize PV}})$. The sources are included into\footnote{In this paper we present the quantum gauge superfield in the form $V=V^A t^A$ (or $V=V^A T^A$ for the terms with matter superfields).}

\begin{equation}
S_{\mbox{\scriptsize sources}} = \int d^8x\, J^A V^A + \Big(\int d^6x\, \Big(j^i \phi_i + j_c^A c^A + \bar j_c^A \bar c^A\Big) + \mbox{c.c}\Big).
\end{equation}

\noindent
The Pauli--Villars determinants are constructed as

\begin{equation}
\mbox{Det}(PV,M_\varphi)^{-1} \equiv \int D\varphi_1\, D\varphi_2\, D\varphi_3\, \exp(iS_\varphi); \qquad \mbox{Det}(PV,M)^{-1} \equiv \int D\Phi\, \exp(iS_\Phi),
\end{equation}

\noindent
where

\begin{eqnarray}\label{S_varphi}
&&\hspace*{-11mm} S_\varphi = \frac{1}{4} \int d^8x\,\Bigg\{ \varphi_1^{*A} \Big[ \Big(R\Big(-\frac{\bar\nabla^2 \nabla^2}{16\Lambda^2}\Big) e^{2{\cal F}(V)} e^{2\bm{V}}\Big)_{Adj} \varphi_1\Big]_A
+ \varphi_2^{*A} \Big[\Big( e^{2{\cal F}(V)} e^{2\bm{V}}\Big)_{Adj} \varphi_2\Big]_A \nonumber\\
&&\hspace*{-11mm} + \varphi_3^{*A} \Big[\Big(e^{2{\cal F}(V)} e^{2\bm{V}}\Big)_{Adj} \varphi_3\Big]_A
\Bigg\} + \Big(\frac{1}{4} M_\varphi \int d^6x\, \Big((\varphi_1^A)^2 + (\varphi_2^A)^2 + (\varphi_3^A)^2\Big)+\mbox{c.c}\Big);\\
\label{S_Phi}
&&\hspace*{-11mm} S_\Phi = \frac{1}{4} \int d^8x\, \Phi^{*i} \Big(F\Big(-\frac{\bar\nabla^2 \nabla^2}{16\Lambda^2}\Big) e^{2{\cal F}(V)}e^{2\bm{V}}\Big)_i{}^j \Phi_j
+ \Big(\frac{1}{4} M^{ij} \int d^6x\, \Phi_i \Phi_j + \mbox{c.c.}\Big)\vphantom{\Bigg(}
\end{eqnarray}

\noindent
and $M^{jk} M^*_{ki} = M^2 \delta_i^j$. (We assume that the representation $R_{\mbox{\scriptsize PV}}$ is chosen in such a way that this condition can be satisfied. For example, it is possible to use the adjoint representation.) To obtain a regularized theory with a single dimensionful parameter, it is necessary to require that the Pauli--Villars masses $M_\varphi$ and $M$ should be proportional to the parameter $\Lambda$,

\begin{equation}\label{A_Coefficients}
M_\varphi = a_\varphi \Lambda;\qquad M = a\Lambda.
\end{equation}

\noindent
It is important that we consider a regularization for which $a_\varphi$ and $a$ do not depend on couplings.

The effective action is standardly defined as the Legendre transform of the generating functional $W=-i\ln~Z$ for connected Green functions,

\begin{equation}
\Gamma[\bm{V}, V,\phi_i,c,\bar c\,] = W - S_{\mbox{\scriptsize sources}}\Big|_{\mbox{\scriptsize sources}\ \to\ \mbox{\scriptsize fields}},
\end{equation}

\noindent
where the sources should be expressed in terms of (super)fields from the equations

\begin{equation}\label{Sources_Vs_Fields}
\frac{\delta W}{\delta J^A} = V^A;\qquad \frac{\delta W}{\delta j^i} = \phi_i;\qquad \frac{\delta W}{\delta j_c^A} = c^A;\qquad \frac{\delta W}{\delta \bar j_c^A} = \bar c^A.
\end{equation}

\section{Renormalization and RGFs defined in terms of the bare couplings}
\hspace*{\parindent}\label{Section_Bare_RGFs}

In this section we present the $\beta$-function defined in terms of the bare couplings in a form which is the most convenient for proving the factorization of the corresponding loop integrals into integrals of double total derivatives. This factorization is an important step towards constructing the all-loop perturbative derivation of the exact NSVZ $\beta$-function. That is why in this section we also rewrite the NSVZ relation (\ref{NSVZ_Second_Form}) in such a form that can be used as a starting point of this derivation.

To find the $\beta$-function defined in terms of the bare couplings, we consider the two-point Green function of the background gauge superfield. Note that in our conventions the term ``two-point'' in particular means that the auxiliary superfield $\mbox{\sl g}$ is set to 0, but the dependence on the parameter $g$ is kept. It is easy to see that the considered Green function depends on $g$, $\alpha_0$, $\lambda_0$, and $\lambda_0^*$ only via the combinations $g g^*\alpha_0$ and $g g^* \lambda_0^{ijk} \lambda_{0mnp}^*$. (For simplicity, below we will denote the latter one by $g g^*\lambda_0 \lambda_0^*$.) Really, in the case $\mbox{\sl g}=0$ the total action depends on $g g^*\alpha_0$, $g\lambda_0$ and $g^*\lambda_0^*$. However, the numbers of $\lambda_0$ and $\lambda_0^*$ in any supergraph contributing to the considered Green function are equal. Therefore, the Yukawa couplings enter it only in the combination $g g^*\lambda_0\lambda_0^*$. Similar arguments also work for the two-point Green functions of the quantum gauge superfield, of the Faddeev--Popov ghosts, and for the two-point Green function
$\phi^{*i} \phi_j$ of the matter superfields. Below we will use the notation

\begin{equation}
\rho\equiv |g|^2 = g g^*,
\end{equation}

\noindent
so that the above mentioned two-point Green functions actually depend on $\rho\alpha_0$ and $\rho\lambda_0\lambda_0^*$.

Due to the background gauge invariance the two-point Green function of the background gauge superfield is transversal and (in the massless limit) can be written as

\begin{equation}\label{Two_Point_Function_Background}
\Gamma^{(2)}_{\bm{V}} = - \frac{1}{8\pi} \mbox{tr} \int \frac{d^4p}{(2\pi)^4}\, d^4\theta\, \bm{V}(-p,\theta) \partial^2 \Pi_{1/2} \bm{V}(p,\theta)\, d^{-1}(\rho\alpha_0,\, \rho\lambda_0 \lambda_0^*,\, \Lambda/p),
\end{equation}

\noindent
where the supersymmetric transversal projection operator is defined by the equation

\begin{equation}
\Pi_{1/2} \equiv - \frac{D^a \bar D^2 D_a}{8\partial^2} = - \frac{\bar D^{\dot a} D^2 \bar D_{\dot a}}{8\partial^2}.
\end{equation}

With the help of the Slavnov--Taylor identities \cite{Taylor:1971ff,Slavnov:1972fg} (and some other similar equations) it is possible to prove that quantum corrections to the two-point Green function of the quantum gauge superfield are also transversal,

\begin{equation}
\Gamma^{(2)}_V - S_{\mbox{\scriptsize gf}}^{(2)} = -\frac{1}{4 e_0^2 \rho} \int \frac{d^4q}{(2\pi)^4}\, d^4\theta\, V^A(-q,\theta) \partial^2\Pi_{1/2} V^A(q,\theta)\, G_V(\rho\alpha_0,\, \rho\lambda_0 \lambda_0^*,\, \Lambda/q).
\end{equation}

\noindent
Also we will need the two-point Green functions of the Faddeev--Popov ghosts and of the matter superfields,

\begin{eqnarray}\label{Two_Point_Gamma_C}
&&\hspace*{-12mm} \Gamma^{(2)}_c = \frac{1}{4}\int \frac{d^4q}{(2\pi)^4}\, d^4\theta\, \Big(- \bar c^A(-q,\theta) c^{*A}(q,\theta) +  \bar c^{*A}(-q,\theta) c^{A}(q,\theta) \Big) G_c(\rho\alpha_0,\, \rho \lambda_0 \lambda_0^*,\, \Lambda/q);\\
\label{Two_Point_Gamma_Phi}
&&\hspace*{-12mm} \Gamma^{(2)}_\phi = \frac{1}{4}\int \frac{d^4q}{(2\pi)^4}\, d^4\theta\, \phi^{*i}(-q,\theta) \phi_j(q,\theta) \big(G_\phi\big)_i{}^j(\rho\alpha_0,\, \rho\lambda_0 \lambda_0^*,\,\Lambda/q).
\end{eqnarray}

Renormalized couplings $\alpha$, $\lambda$ and the renormalization constants $Z_V$, $Z_c$, $(Z_\phi)_i{}^j$ are defined by requiring finiteness of the functions $d^{-1}$, $Z_V^2 G_V$, $Z_c G_c$, and $\big(Z_\phi\big)_i{}^j (G_\phi)_j{}^k$ expressed in terms of $\alpha$ and $\lambda$ in the limit $\Lambda\to\infty$. Note that due to the non-renormalization of the superpotential \cite{Grisaru:1979wc} the renormalized Yukawa couplings are related to the bare ones by the equation

\begin{equation}\label{Yukawa_Renormalization}
\lambda^{ijk} = \lambda_0^{mnp} \big(\sqrt{Z_\phi}\big)_m{}^i \big(\sqrt{Z_\phi}\big)_n{}^j \big(\sqrt{Z_\phi}\big)_p{}^k.
\end{equation}

\noindent
Similarly, due to the non-renormalization of the triple ghost-gauge vertices \cite{Stepanyantz:2016gtk} the renormalization constants can be chosen in such a way that

\begin{equation}\label{VCC_NonRenormalization}
Z_\alpha^{-1/2} Z_c Z_V = 1, \qquad \mbox{where}\qquad Z_\alpha \equiv \frac{\alpha}{\alpha_0}.
\end{equation}

\noindent
We will always assume that the renormalization constants satisfy Eqs. (\ref{Yukawa_Renormalization}) and (\ref{VCC_NonRenormalization}). (Certainly the renormalization constants are not uniquely defined \cite{Vladimirov:1979my}, and these constrains partially fix an arbitrariness in choosing a subtraction scheme.)

It is important that in the non-Abelian case the quantum gauge superfield is renormalized in a nonlinear way \cite{Piguet:1981fb,Piguet:1981hh,Tyutin:1983rg}. The non-linear renormalization can be realized as a linear renormalization of an infinite set of parameters. For example, in the lowest approximation it is possible to present the function ${\cal F}(V)$ in the form

\begin{equation}
{\cal F}(V) = V + 8 y_0\, G^{ABCD}\, \mbox{tr}(V t^B)\, \mbox{tr}(V t^C)\, \mbox{tr}(V t^D)\, t^A + \ldots,
\end{equation}

\noindent
where $y_0$ is a new bare parameter and

\begin{equation}
G^{ABCD} \equiv \frac{1}{6}\Big(f^{AKL} f^{BLM} f^{CMN} f^{DNK} + \mbox{permutations of $B$, $C$, and $D$} \Big).
\end{equation}

\noindent
Then, the result for the nonlinear renormalization obtained in \cite{Juer:1982fb,Juer:1982mp} can be equivalently written in the form

\begin{equation}\label{Y_Renormalization}
y_0 = y + \frac{\alpha}{90\pi} \Big((2+3\xi) \ln\frac{\Lambda}{\mu} + k_1\Big) + \ldots,
\end{equation}

\noindent
where $\xi$ is the renormalized gauge parameter and $k_1$ is a finite constant which appears due to the arbitrariness in choosing a subtraction scheme. The explicit calculation of Ref. \cite{Kazantsev:2018kjx} demonstrated that the renormalization group equations cannot be satisfied without introducing the parameter $y_0$ (or, possibly, implementing the nonlinear renormalization by some different way). Certainly, in higher orders an infinite set of parameters similar to $y_0$ is needed. All these parameters are similar to the gauge fixing parameter $\xi_0$, because by a proper change of variables in the generating functional it is possible to prove that a nonlinear renormalization is equivalent to a nonlinear change of a gauge \cite{Piguet:1981fb}. That is why below we will include the gauge fixing parameter and the parameters of the nonlinear renormalization inside the function ${\cal F}(V)$ into a single set

\begin{equation}
Y_0 \equiv (\xi_0, y_0,\ldots).
\end{equation}

\noindent
The corresponding renormalized values will be denoted by $Y = (\xi,y,\ldots)$.

We believe that the NSVZ relation is valid for RGFs defined in terms of the bare couplings in the case of using the higher covariant derivative regularization. These RGFs are defined by the equations

\begin{eqnarray}\label{RGFs_Bare}
&& \beta(\rho\alpha_0,\, \rho\lambda_0\lambda_0^*,\,Y_0) \equiv \left.\frac{d(\rho\alpha_0)}{d\ln\Lambda}\right|_{\alpha,\lambda,Y = \mbox{\scriptsize const}};\qquad\nonumber\\
&& \gamma_V(\rho\alpha_0,\, \rho\lambda_0\lambda_0^*,Y_0) \equiv \left. - \frac{d\ln Z_V}{d\ln\Lambda}\right|_{\alpha,\lambda,Y = \mbox{\scriptsize const}};\nonumber\\
&& \gamma_c(\rho\alpha_0,\, \rho\lambda_0\lambda_0^*,\,Y_0) \equiv \left. - \frac{d\ln Z_c}{d\ln\Lambda}\right|_{\alpha,\lambda,Y = \mbox{\scriptsize const}};\nonumber\\
&& (\gamma_\phi)_i{}^j(\rho\alpha_0,\, \rho\lambda_0\lambda_0^*,\,Y_0) \equiv \left. - \frac{d(\ln Z_\phi)_i{}^j}{d\ln\Lambda}\right|_{\alpha,\lambda,Y = \mbox{\scriptsize const}}
\end{eqnarray}

\noindent
and do not depend on a renormalization prescription for a fixed regularization \cite{Kataev:2013eta}. It is easy to see that RGFs defined in terms of the bare couplings can be obtained by differentiating the corresponding Green functions. For example, the $\beta$-function defined in terms of the bare couplings can be constructed by differentiating the quantum corrections in the two-point Green function of the background gauge superfield in the limit of the vanishing external momentum,

\begin{equation}\label{Bare_Beta_Construction}
\left.\frac{d}{d\ln\Lambda} \Big(d^{-1} - \big(g g^*\big)^{-1}\alpha_0^{-1}\Big)\right|_{\alpha,\lambda,Y = \mbox{\scriptsize const};\ p\to 0} = \frac{\beta(\rho\alpha_0,\, \rho\lambda_0 \lambda_0^*, \, Y_0)}{\rho^2 \alpha_0^2}.
\end{equation}

\noindent
Note that the term $1/(g g^*\alpha_0)$ appears in the function $d^{-1}$ in the tree approximation and corresponds to the first term in Eq. (\ref{Regularized_Action}). The limit $p\to 0$ is needed for removing terms proportional to $(p/\Lambda)^k$, where $k$ is a positive integer. The equality follows from the finiteness of the function $d^{-1}$ expressed in terms of the renormalized couplings.

It is well known that for $g=1$ the $\beta$-function can be presented as the series

\begin{equation}
\beta(\alpha_0,\lambda_0\lambda_0^*,Y_0) = \frac{\alpha_0^2}{\pi} \beta_1 + O(\alpha_0^3, \alpha_0^2\lambda_0\lambda_0^*) = \beta_{\mbox{\scriptsize 1-loop}}(\alpha_0) + O(\alpha_0^3,\alpha_0^2\lambda_0\lambda_0^*),
\end{equation}

\noindent
where the ($Y_0$-independent) coefficient

\begin{equation}
\beta_1 = -\frac{1}{2} \Big(3C_2-T(R)\Big)
\end{equation}

\noindent
is obtained by calculating the one-loop contribution to the $\beta$-function. (For the considered regularization the details of this calculation can be found in \cite{Aleshin:2016yvj}.) For $g\ne 1$ it is easy to see that the $L$-loop contribution to the $\beta$-function is proportional to $\big(g g^*\big)^{L+1} = \rho^{L+1}$. Therefore, the dependence of the expression $\beta(\rho\alpha_0,\,\rho\lambda_0\lambda_0^*,\,Y_0)/\rho^2\alpha_0^2$ on $\rho$ is described by a function $f(\rho) = f_0 + f_1 \rho + f_2 \rho^2 + \ldots$ If we consider $g$ and $g^*$  as independent variables, then

\begin{equation}
\frac{\partial^2 f(\rho)}{\partial g\,\partial g^*} = \frac{\partial^2 f(g g^*)}{\partial g\,\partial g^*} = g g^* f''(g g^*) + f'(g g^*) =\frac{d}{d\rho}\Big(\rho \frac{df}{d\rho}\Big),
\end{equation}

\noindent
Consequently,

\begin{equation}\label{Integration_Identity}
\int\limits_{+0}^{1} \frac{d\rho}{\rho}\, \int\limits_{+0}^\rho d\rho\, \frac{\partial^2 f(\rho)}{\partial g\,\partial g^*} = f(1) - f(0),
\end{equation}

\noindent
where $+0$ means that $\rho \ne 0$, but $\rho\to 0$. Taking into account that the limit $\rho\to 0$ corresponds to the theory in which quantum superfields interact only with the background gauge superfield, so that nontrivial quantum corrections exist only in the one-loop approximation, we obtain

\begin{equation}\label{Beta_Corrections_Preliminary}
\int\limits_{+0}^{1} \frac{d\rho}{\rho}\, \int\limits_{+0}^\rho d\rho\, \frac{\partial^2}{\partial g\, \partial g^*} \Big(\frac{\beta(\rho\alpha_0,\,\rho\lambda_0\lambda_0^*,\,Y_0)}{\rho^2\alpha_0^2}\Big)  = \frac{\beta(\alpha_0,\lambda_0\lambda_0^*,Y_0)}{\alpha_0^2} - \frac{\beta_{\mbox{\scriptsize 1-loop}}(\alpha_0)}{\alpha_0^2}.
\end{equation}

\noindent
Therefore, the $\beta$-function defined in terms of the bare couplings (for the original theory which corresponds to $g=1$) can be calculated with the help of the equation

\begin{equation}\label{Beta_Corrections}
\frac{\beta(\alpha_0,\lambda_0\lambda_0^*,Y_0)}{\alpha_0^2} = \frac{\beta_{\mbox{\scriptsize 1-loop}}(\alpha_0)}{\alpha_0^2} + \int\limits_{+0}^{1} \frac{d\rho}{\rho}\, \int\limits_{+0}^\rho d\rho\, \frac{\partial^2}{\partial g\, \partial g^*} \left.\frac{d}{d\ln\Lambda} \Big(d^{-1} - \big(g g^*\big)^{-1}\alpha_0^{-1}\Big)\right|_{\alpha,\lambda,Y = \mbox{\scriptsize const};\ p\to 0}.
\end{equation}

Due to the finiteness of the functions $Z_V^2 G_V$, $Z_c G_c$, and $\big(Z_\phi\big)_i{}^j \big(G_\phi\big)_j{}^k$ the anomalous dimensions of the quantum superfields can also be related to the corresponding Green functions by the equations

\begin{eqnarray}\label{Bare_Gamma_V_Construction}
&& \gamma_V(\rho\alpha_0,\, \rho\lambda_0\lambda_0^*,\,Y_0) = \left.\frac{1}{2}\,\frac{d\ln G_V}{d\ln\Lambda}\right|_{\alpha,\lambda,Y = \mbox{\scriptsize const};\ q\to 0};\\
\label{Bare_Gamma_C_Construction}
&& \gamma_c(\rho\alpha_0,\,\rho\lambda_0\lambda_0^*,\,Y_0) = \left. \frac{d\ln G_c}{d\ln\Lambda}\right|_{\alpha,\lambda,Y = \mbox{\scriptsize const};\ q\to 0};\\
\label{Bare_Gamma_Phi_Construction}
&& \big(\gamma_\phi\big)_i{}^j(\rho\alpha_0,\,\rho\lambda_0\lambda_0^*,\,Y_0) = \left.\frac{d\big(\ln G_\phi\big)_i{}^j}{d\ln\Lambda}\right|_{\alpha,\lambda,Y = \mbox{\scriptsize const};\ q\to 0}.
\end{eqnarray}

\noindent
In the one-loop order these anomalous dimensions contain terms proportional to $\alpha_0$ and $\lambda_0\lambda_0^*$ (the latter ones appear only in $\big(\gamma_\phi\big)_i{}^j$),

\begin{equation}
\gamma(\rho\alpha_0,\rho\lambda_0,Y_0) = O(\alpha_0,\lambda_0\lambda_0^*),
\end{equation}

\noindent
and the terms corresponding to the $L$-loop approximation are proportional to $\big(g g^*\big)^{L} = \rho^L$. Using this fact, from the identity (\ref{Integration_Identity}) we obtain

\begin{equation}
\int\limits_{+0}^{1} \frac{d\rho}{\rho}\,\int\limits_{+0}^{\rho} d\rho\, \frac{\partial^2}{\partial g\, \partial g^*}\, \gamma(\rho\alpha_0,\,\rho\lambda_0\lambda_0^*,\,Y_0) = \gamma(\alpha_0,\lambda_0\lambda_0^*,Y_0).
\end{equation}

\noindent
This implies that for deriving the NSVZ relation (\ref{NSVZ_Second_Form}) it is sufficient to prove that

\begin{eqnarray}\label{Green_Functions_Identity}
&& \frac{\partial^2}{\partial g\, \partial g^*} \left.\frac{d}{d\ln\Lambda} \Big(d^{-1} - \big(g g^*\big)^{-1}\alpha_0^{-1}\Big)\right|_{\alpha,\lambda,Y = \mbox{\scriptsize const};\ p\to 0}\nonumber\\
&&\qquad = \frac{1}{2\pi}\, \frac{\partial^2}{\partial g\,\partial g^*} \left.\frac{d}{d\ln\Lambda} \Big(2C_2 \ln G_c + C_2 \ln G_V - \frac{1}{r} C(R)_i{}^j \big(\ln G_\phi\big)_j{}^i\Big)\right|_{\alpha,\lambda,Y = \mbox{\scriptsize const};\ q\to 0}.\qquad
\end{eqnarray}

\noindent
Eq. (\ref{NSVZ_Second_Form}) is obtained by applying the operator

\begin{equation}
\int\limits_{+0}^{1} \frac{d\rho}{\rho}\,\int\limits_{+0}^{\rho} d\rho\,
\end{equation}

\noindent
to this equation with the help of Eqs. (\ref{Bare_Beta_Construction}) and (\ref{Bare_Gamma_V_Construction}) -- (\ref{Bare_Gamma_Phi_Construction}).

In Eq. (\ref{Green_Functions_Identity}) the derivative with respect to $\ln\Lambda$ is very important, because it removes infrared divergences which could appear in the limit of the vanishing external momentum. Explicit loop calculations (e.g., in Refs. \cite{Shakhmanov:2017soc,Kazantsev:2018nbl}) demonstrate that loop integrals written without $d/d\ln\Lambda$ are not well defined, while after the differentiation all bad terms disappear.

The derivatives with respect to $g$ and $g^*$ are not so important and can be excluded from Eq. (\ref{Green_Functions_Identity}). Certainly, in this case it is necessary to add the constant corresponding to the one-loop contribution,

\begin{eqnarray}
&& \left.\frac{d}{d\ln\Lambda} \Big(d^{-1} - \big(g g^*\big)^{-1}\alpha_0^{-1}\Big)\right|_{\alpha,\lambda,Y = \mbox{\scriptsize const};\ p\to 0} = -\frac{3C_2-T(R)}{2\pi}\nonumber\\
&&\qquad\qquad + \frac{1}{2\pi}\, \left.\frac{d}{d\ln\Lambda} \Big(2C_2 \ln G_c + C_2 \ln G_V - \frac{1}{r} C(R)_i{}^j \big(\ln G_\phi\big)_j{}^i\Big)\right|_{\alpha,\lambda,Y = \mbox{\scriptsize const};\ q\to 0}.\qquad
\end{eqnarray}

\noindent
For $g=1$ this identity was first suggested in Ref. \cite{Stepanyantz:2016gtk}. However, for deriving the NSVZ relation in all loops it is more preferable to use Eq. (\ref{Green_Functions_Identity}).

The left hand side of Eq. (\ref{Green_Functions_Identity}) can be constructed starting from the expression for the two-point Green function of background gauge superfield (\ref{Two_Point_Function_Background}). To extract the function $d^{-1}$, it is convenient to make the formal substitution

\begin{equation}\label{V_Substitution}
\bm{V}^A \to  \theta^4 v^A,\qquad\mbox{where}\qquad \theta^4 \equiv \theta^a \theta_a\, \bar\theta^{\dot a}\bar\theta_{\dot a}.
\end{equation}

\noindent
In this equation $v^A$ are slow varying functions of the space-time coordinates which tend to 0 only at a very large scale $R\to \infty$. For example, it is possible to choose

\begin{equation}\label{Explicit_V}
v^A(X) = v_0^A \exp\Big(-(X^\mu)^2/2 R^2\Big),
\end{equation}

\noindent
where $v_0^A = \mbox{const}$ and $X^\mu = (x^i, ix^0)$ are the Euclidean coordinates. The corresponding Euclidean momenta are denoted by $P^\mu = (p^i, -ip^0)$. In this case

\begin{equation}
v^A(P) \equiv \int d^4X\, v^A(X) \exp(iX^\mu P_\mu) = (2\pi)^2 R^4 v_0^A \exp\Big(-(P^\mu)^2 R^2/2\Big).
\end{equation}

\noindent
From Eq. (\ref{Explicit_V}) we see that $v^A(P)$ is essentially different from 0 only in a small region of the size $1/R \to 0$. This implies that substituting the functions (\ref{Explicit_V}) into Eq. (\ref{Two_Point_Function_Background}) we automatically obtain the limit $P\to 0$ (or, equivalently, $p\to 0$), which is needed for constructing RGFs defined in terms of the bare couplings.

Let us consider quantum corrections encoded in the expression

\begin{equation}\label{Delta_Gamma}
\Delta\Gamma = \Gamma - S_{\mbox{\scriptsize total}},
\end{equation}

\noindent
where $S_{\mbox{\scriptsize total}}$ includes the usual action, the gauge fixing term, and the ghost actions. (Certainly, the terms proportional to $\Lambda^{-k}$, where $k$ is a positive integer, should be omitted). Then we consider a part of $\Delta\Gamma$ corresponding to the two-point Green function of the background gauge superfield. Performing the Wick rotation and making the substitution (\ref{V_Substitution}), after some transformations, in the limit $R\to\infty$ we obtain

\begin{equation}\label{Constructing_Bare_Beta}
\left.\frac{d\Delta\Gamma^{(2)}_{\bm{V}}}{d\ln\Lambda}\right|_{\alpha,\lambda,Y = \mbox{\scriptsize const};\ \bm{V}= \theta^4 v} = \frac{{\cal V}_4}{2\pi} \left.\frac{d}{d\ln\Lambda}\Big(d^{-1} - \big(g g^*\big)^{-1} \alpha_0^{-1}\Big) \right|_{p=0} = \frac{{\cal V}_4}{2\pi}\cdot \frac{\beta(\rho\alpha_0,\,\rho\lambda_0\lambda_0^*,\,Y_0)}{\rho^2 \alpha_0^2},
\end{equation}

\noindent
where we have introduced the notation

\begin{equation}\label{Nu4}
{\cal V}_4 = \int d^4x\, (v^A)^2 \to -i \int d^4X\, (v^A)^2 = -i \int \frac{d^4P}{(2\pi)^4}\, v^A(-P)\, v^A(P).
\end{equation}

\noindent
Evidently, ${\cal V}_4 \sim R^4 \to \infty$. For example, if the functions $v^A$ are chosen in the form (\ref{Explicit_V}), then ${\cal V}_4 = -i \pi^2 (v_0^A)^2 R^4$. Thus, we see that the substitution (\ref{V_Substitution}) allows extracting the $\beta$-function defined in terms of the bare couplings from the considered part of the effective action in the case of using the higher covariant derivative regularization. (In the case of using the  dimensional reduction one should be much more careful, see \cite{Aleshin:2015qqc,Aleshin:2016rrr} for details.)

Differentiating Eq. (\ref{Constructing_Bare_Beta}) with respect to the parameters $g$ and $g^*$ and multiplying the result by the factor $2\pi/{\cal V}_4$, we obtain the left hand side of Eq. (\ref{Green_Functions_Identity}). In turn, the derivatives with respect to the coordinate-independent parameters $g$ and $g^*$ can be expressed in terms of the derivatives with respect to the chiral superfield $\mbox{\sl g}$ and the antichiral superfield $\mbox{\sl g}^*$, respectively. Really, all terms in the action containing quantum superfields depend only on the combinations $\bm{g}$ and $\bm{g}^*$, see Eqs. (\ref{Regularized_Action}), (\ref{Gauge_Fixing_Term}), (\ref{Faddeev_Popov_Action}), and (\ref{Nielsen_Kallosh_Action}). The only term which depends on $g$ and $g^*$ in a different way is the first term in Eq. (\ref{Regularized_Action}), but it does not affect quantum corrections and does not enter $\Delta\Gamma$. Therefore, it is possible to relate the derivatives of $\Delta\Gamma$ with respect to $g$ and $g^*$ to the derivatives with respect to $\mbox{\sl g}$ and $\mbox{\sl g}^*$,

\begin{equation}
\left.\frac{\partial^2\Delta\Gamma}{\partial g\, \partial g^*}\right|_{\mbox{\scriptsize \sl g}=0} = \left.\int d^6z_1\, d^6\bar z_2\, \frac{\delta^2\Delta\Gamma}{\delta\mbox{\sl g}_{z_1} \delta\mbox{\sl g}^*_{z_2}}\right|_{\mbox{\scriptsize \sl g}=0},
\end{equation}

\noindent
where

\begin{equation}
\int d^6\bar x\equiv \int d^4x\, d^2\bar\theta_x.
\end{equation}

Thus, to derive the NSVZ relation, it is sufficient to prove the identity

\begin{eqnarray}\label{Equation_To_Prove_Original}
&& \int d^6z_1\, d^6\bar z_2\, \frac{\delta^2}{\delta\mbox{\sl g}_{z_1} \delta \mbox{\sl g}^*_{z_2}} \left.\frac{d\Gamma^{(2)}_{\bm{V}}}{d\ln\Lambda}\right|_{\parbox{2.5cm}{\scriptsize $\alpha,\lambda,Y = \mbox{\scriptsize const};$\\ $\bm{V}= \theta^4 v;\, \mbox{\scriptsize \sl g}=0$}}
= \frac{\partial^2}{\partial g\,\partial g^*} \left.\frac{d\Delta\Gamma^{(2)}_{\bm{V}}}{d\ln\Lambda}\right|_{\parbox{2.5cm}{\scriptsize $\alpha,\lambda,Y = \mbox{\scriptsize const};$\\ $\bm{V}= \theta^4 v;\, \mbox{\scriptsize \sl g}=0$}}\qquad\nonumber\\
&& = \frac{{\cal V}_4}{4\pi^2} \left.\frac{\partial^2}{\partial g\, \partial g^*}\,\frac{d}{d\ln\Lambda} \Big(2C_2 \ln G_c + C_2 \ln G_V - \frac{1}{r} C(R)_i{}^j \big(\ln G_\phi\big)_j{}^i\Big)\right|_{\alpha,\lambda,Y = \mbox{\scriptsize const};\ q\to 0},\qquad
\end{eqnarray}

\noindent
where $\Gamma^{(2)}_{\bm{V}}$ denotes a part of $\Gamma$ which is quadratic in the background gauge superfield and does not contain the other superfields except for $\mbox{\sl g}$. Note that writing Eq. (\ref{Equation_To_Prove_Original}) we took into account that $S^{(2)}_{\bm{V}}$ is independent of $\mbox{\sl g}$, see Eq. (\ref{Regularized_Action}). It is evident that

\begin{equation}\label{Gamma2_Taylor}
\Gamma^{(2)}_{\bm V} = \frac{1}{2} \int d^8x\, d^8y\, \bm{V}_x^A \bm{V}_y^B \left.\frac{\delta^2\Gamma}{\delta \bm{V}_x^A \delta\bm{V}_y^B}\right|_{\mbox{\scriptsize fields} = 0;\, \mbox{\scriptsize \sl g}\ne 0}.
\end{equation}

\noindent
Note that here we do not set the auxiliary external superfields $\mbox{\sl g}$ and $\mbox{\sl g}^*$ to 0, because Eq. (\ref{Equation_To_Prove_Original}) contains the derivatives with respect to these superfields. In this paper we will consider only ${\cal N}=1$ supersymmetric gauge theories with a simple gauge group. In this case it is easy to see that any invariant tensor $I_{AB}$ should be proportional to $\delta_{AB}$.\footnote{The considered invariant tensor satisfies the equation $[T^A_{Adj}, I] = 0$, so that it commutes with all generators of the adjoint representation. For a simple group the adjoint representation is irreducible. Therefore, $I_{AB}$ should be proportional to $\delta_{AB}$.} Therefore, for simple gauge groups

\begin{equation}\label{Simple_Gauge_Group}
\left.\frac{\delta^2\Gamma}{\delta \bm{V}_x^A \delta\bm{V}_y^B}\right|_{\mbox{\scriptsize fields} = 0;\, \mbox{\scriptsize \sl g}\ne 0} = \frac{1}{r} \delta_{AB} \left.\frac{\delta^2\Gamma}{\delta \bm{V}_x^C \delta\bm{V}_y^C}\right|_{\mbox{\scriptsize fields} = 0;\, \mbox{\scriptsize \sl g}\ne 0}.
\end{equation}

With the help of Eqs. (\ref{Gamma2_Taylor}) and (\ref{Simple_Gauge_Group}) for a simple gauge group it is possible to rewrite Eq. (\ref{Equation_To_Prove_Original}) in the form mostly convenient for proving, namely,

\begin{eqnarray}\label{Equation_To_Prove}
&& \int d^8x\, d^8y\, d^6z_1\, d^6\bar z_2\, (\theta^4)_x (v^B)_x\, (\theta^4)_y (v^B)_y \left. \frac{d}{d\ln\Lambda} \frac{\delta^4\Gamma}{\delta\mbox{\sl g}_{z_1} \delta \mbox{\sl g}^*_{z_2} \delta\bm{V}_x^A \delta \bm{V}_y^A} \right|_{\parbox{2.5cm}{\scriptsize $\alpha,\lambda,Y = \mbox{\scriptsize const};$\\ $\mbox{\scriptsize fields}=0;\,\mbox{\scriptsize \sl g}=0$}} \qquad\nonumber\\
&& = \frac{{\cal V}_4}{2\pi^2} \left. \frac{\partial^2}{\partial g\, \partial g^*}\, \frac{d}{d\ln\Lambda} \Big(2C_2 r\ln G_c + C_2 r\ln G_V - C(R)_i{}^j \big(\ln G_\phi\big)_j{}^i\Big)\right|_{\alpha,\lambda,Y = \mbox{\scriptsize const};\ q\to 0}.\qquad
\end{eqnarray}

\noindent
According to the above discussion, for the theory regularized by higher covariant derivatives this equation is equivalent to the NSVZ relations (\ref{NSVZ_First_Form}) and (\ref{NSVZ_Second_Form}) for RGFs defined in terms of the bare couplings. Below we will prove that the left hand side of Eq. (\ref{Equation_To_Prove}) is given by integrals of double total derivatives.

\section{The $\beta$-function as an integral of double total derivatives}
\label{Section_Formal_Derivation}

\subsection{The Slavnov--Taylor identity for the background gauge invariance}
\hspace*{\parindent}\label{Subsection_STI}

The background gauge invariance is a manifest symmetry of the theory under consideration (even in the presence of the auxiliary superfield $\mbox{\sl g}$). At the quantum level symmetries are encoded in the Slavnov--Taylor identities \cite{Taylor:1971ff,Slavnov:1972fg}. The Slavnov--Taylor identity corresponding to the background gauge transformations constructed in this section is a very important ingredient for the all-loop proof of the factorization into double total derivatives. This identity is derived by standard methods, namely, it is necessary to make the change of variables

\begin{eqnarray}\label{Background_Gauge_Transformations_New}
&& V \to  e^{-A^+} V e^{A^+};\qquad\ \  c \to e^A c e^{-A};\qquad\qquad \bar c \to e^A \bar c e^{-A};\qquad\vphantom{\Big(}\nonumber\\
&& \phi_i \to (e^A)_i{}^j \phi_j;\qquad\qquad \Phi_i \to (e^A)_i{}^j \Phi_j;\qquad\quad \varphi_a \to e^A \varphi_a e^{-A}\qquad\vphantom{\Big(}
\end{eqnarray}

\noindent
in the functional integral (\ref{Generating_Functional_Z}), which does not change the generating functional $Z$. This change of variables coincides with the background gauge transformations of the quantum superfields. Due to the background gauge invariance, the total gauge fixed action

\begin{equation}\label{Total_Action}
S_{\mbox{\scriptsize total}} = S_{\mbox{\scriptsize reg}} + S_{\mbox{\scriptsize gf}} + S_{\mbox{\scriptsize FP}} + S_{\mbox{\scriptsize NK}}
\end{equation}

\noindent
and the Pauli--Villars determinants remain unchanged if the background gauge superfield is also modified as

\begin{equation}\label{Background_Superfield_Transformation}
e^{2\bm{V}} \to e^{-A^+} e^{2\bm{V}} e^{-A}.
\end{equation}

\noindent
However, the source term $S_{\mbox{\scriptsize sources}}$ transforms nontrivially. This implies that in the linear order in $A$ the invariance of the generating functional $W=-i\ln Z$ under the change of variables (\ref{Background_Gauge_Transformations_New}) can be expressed by the equation

\begin{equation}\label{W_Invariance}
\int d^8x\,\delta\bm{V}^B \frac{\delta W}{\delta\bm{V}^B} = \Big\langle \int d^8x\, J^A \delta V^A + \Big[\int d^6x\, \Big(j^i \delta\phi_i + j_c^A \delta c^A + \bar j_c^A \delta\bar c^A \Big) + \mbox{c.c.}\Big] \Big\rangle,
\end{equation}

\noindent
where the variations of various superfields under the infinitesimal background gauge transformations are written as\footnote{The expression for $\delta\bm{V} = \delta \bm{V}^B t^B$ is obtained in the standard way from the identity $0 = \delta[\bm{V}, e^{2\bm{V}}]$.}

\begin{eqnarray}\label{Infinitesimal_Transformations}
&&\hspace*{-7mm} \delta \bm{V} = -\Big(\frac{\bm{V}}{1-e^{-2\bm{V}}}\Big)_{Adj} A + \Big(\frac{\bm{V}}{1-e^{2\bm{V}}}\Big)_{Adj} A^+ = \frac{1}{2} \Big(- A - A^+ - [\bm{V},A] + [\bm{V},A^+]\Big) + O(\bm{V}^2);\nonumber\\
&&\hspace*{-7mm} \delta V = -[A^+,V];\qquad \delta\phi_i = A_i{}^j \phi_j;\qquad \delta c = [A, c];\qquad \delta \bar c = [A, \bar c],\vphantom{\frac{1}{2}}
\end{eqnarray}

\noindent
with $A = A^A t^A$ and $A_i{}^j = A^A (T^A)_i{}^j$. The angular brackets denote

\begin{equation}\label{VEV}
\langle B \rangle \equiv \frac{1}{Z} \int D\mu\, B\, \mbox{Det}(PV,M_\varphi)^{-1} \mbox{Det}(PV, M)^c\,\exp\Big\{i \left(S_{\mbox{\scriptsize total}} + S_{\mbox{\scriptsize sources}} \right)\Big\},
\end{equation}

\noindent
where $B$ is a function(al) depending on the superfields of the theory.

Rewriting Eq. (\ref{W_Invariance}) in terms of (super)fields, we obtain the equation which expresses the manifest background gauge invariance of the effective action,

\begin{equation}\label{Preliminary_Background_STI}
\int d^8x\, \Big(\delta\bm{V}^B \frac{\delta\Gamma}{\delta\bm{V}^B} + \delta V^B \frac{\delta\Gamma}{\delta V^B} \Big) + \left(\int d^6x\, \Big(\delta\phi_i \frac{\delta\Gamma}{\delta\phi_i} + \delta c^B\frac{\delta\Gamma}{\delta c^B} + \delta \bar c^B \frac{\delta\Gamma}{\delta\bar c^B}\Big) +\mbox{c.c.}\right) = 0.
\end{equation}

\noindent
It is important that in this equation (super)fields are not set to 0, so that this equation encodes an infinite set of identities relating Green functions of the theory. That is why we will call it the generating Slavnov--Taylor identity.

Considering $A$ and $A^+$ as independent variables and differentiating Eq. (\ref{Preliminary_Background_STI}) with respect to $A^A$ we obtain

\begin{eqnarray}
&& \frac{\bar D^2}{2} \Bigg\{\Big[\Big(\frac{\bm{V}}{1-e^{-2\bm{V}}}\Big)_{Adj}\Big]_{BA}\, \frac{\delta\Gamma}{\delta \bm{V}^B}\Bigg\} + \phi_j (T^A)_i{}^j \frac{\delta\Gamma}{\delta\phi_i}\nonumber\\
&&\qquad\qquad\qquad\qquad\qquad\qquad\quad + c^C (T^A_{Adj})_{BC} \frac{\delta\Gamma}{\delta c^B} + \bar c^C (T^A_{Adj})_{BC} \frac{\delta\Gamma}{\delta \bar c^B} = 0,\qquad
\end{eqnarray}

\noindent
where the matrix $\left[f(X)_{Adj}\right]_{AB}$ is defined by the equation

\begin{equation}
f(X)_{Adj} (t^A Y^A) \equiv t^A\left[f(X)_{Adj}\right]_{AB} Y^B.
\end{equation}

Expressing the generators of the adjoint representation in terms of the structure constants it is possible to rewrite the generating Slavnov--Taylor identity Eq. (\ref{Preliminary_Background_STI}) corresponding to the background gauge symmetry in the form

\begin{equation}\label{Background_STI}
\bar D^2\, \hat O^A \Gamma = 0,
\end{equation}

\noindent
where the operator $\hat O^A$ is given by the expression

\begin{eqnarray}\label{O_Definition}
&& \hat O^A \equiv \Big[\Big(\frac{2\bm{V}}{1-e^{-2\bm{V}}}\Big)_{Adj}\Big]_{BA}\, \frac{\delta}{\delta \bm{V}^B} - \frac{D^2}{4\partial^2}\phi_j (T^A)_i{}^j \frac{\delta}{\delta\phi_i}\nonumber\\
&&\qquad\qquad\qquad\qquad\qquad\qquad\qquad - i f^{ABC} \frac{D^2}{4\partial^2} c^B \frac{\delta}{\delta c^C} - i f^{ABC} \frac{D^2}{4\partial^2} \bar c^B \frac{\delta}{\delta \bar c^C}.\qquad
\end{eqnarray}

\noindent
To verify Eq. (\ref{Background_STI}), it is necessary to take into account that a derivative with respect to a chiral superfield is also chiral and use the identity

\begin{equation}\label{Phi_Chiral_Identity}
- \frac{\bar D^2 D^2}{16\partial^2}\phi = \phi
\end{equation}

\noindent
valid for an arbitrary chiral superfield $\phi$.

It is important that due to Eq. (\ref{Background_STI}) the effective action satisfies the equation

\begin{equation}\label{Extracting_A_Derivative}
\hat O^A \Gamma = \bar D^{\dot a} \bar{O}_{\dot a}^A \Gamma,
\end{equation}

\noindent
where

\begin{equation}
\bar{O}_{\dot a}^A \equiv \Big(- \frac{\bar D_{\dot a} D^2}{16\partial^2} + \frac{D^2 \bar D_{\dot a}}{8\partial^2}\Big) \hat O^A.
\end{equation}

\noindent
This can be verified with the help of the equality

\begin{equation}
1 = - \frac{D^2\bar D^2}{16\partial^2} - \frac{\bar D^2 D^2}{16\partial^2} - \Pi_{1/2}
\end{equation}

\noindent
and the generating Slavnov--Taylor identity (\ref{Background_STI}).

\subsection{Transforming the left hand side of Eq. (\ref{Equation_To_Prove}) with the help of the supergraph calculation rules}
\hspace{\parindent}

An important observation is that the second derivative of the effective action with respect to the background superfield $\bm{V}$ in Eq. (\ref{Equation_To_Prove}) can be obtained by applying the operator $\big(\hat{O}^A\big)_x \big(\hat{O}^A\big)_y$ to $\Gamma$, where $x$ and $y$ denote the points of the superspace. Really, in the lowest orders in $\bm{V}$ the operator $\hat O^A$ can be written as

\begin{eqnarray}
&& \hat O^A \equiv \frac{\delta}{\delta \bm{V}^A} - if^{ABC} \bm{V}^B \frac{\delta}{\delta \bm{V}^C} + O(\bm{V}^2)\nonumber\\
&&\qquad\qquad\qquad - \frac{D^2}{4\partial^2}\phi_j (T^A)_i{}^j \frac{\delta}{\delta\phi_i} - i f^{ABC} \frac{D^2}{4\partial^2} c^B \frac{\delta}{\delta c^C} - i f^{ABC} \frac{D^2}{4\partial^2} \bar c^B \frac{\delta}{\delta \bar c^C}.\qquad
\end{eqnarray}

\noindent
Therefore, taking into account that $f^{AAC}=0$, after the differentiation we see that

\begin{eqnarray}\label{Two_Point_Function_O}
&& \left.\frac{\delta^2\Gamma}{\delta\bm{V}_x^A \delta\bm{V}_y^A}\right|_{\mbox{\scriptsize fields} = 0;\, \mbox{\scriptsize \sl g}\ne 0} = \left.\vphantom{\frac{1}{2}}\big(\hat O^A\big)_x \big(\hat O^A\big)_y \Gamma\right|_{\mbox{\scriptsize fields} = 0;\, \mbox{\scriptsize \sl g}\ne 0}\nonumber\\
&&\qquad\qquad\qquad\qquad\qquad\quad = \left.\vphantom{\frac{1}{2}} - \big(\bar D^{\dot a}\big)_x \big(\bar D^{\dot b}\big)_y\Big(\big(\bar{O}^A_{\dot a}\big)_x  \big(\bar{O}^A_{\dot b}\big)_y \Gamma\Big)\right|_{\mbox{\scriptsize fields} = 0;\, \mbox{\scriptsize \sl g}\ne 0}.\qquad
\end{eqnarray}

\noindent
Note that here all fields (including the background gauge superfield $\bm{V}$) should be set to 0, but the auxiliary superfield parameter $\mbox{\sl g}$ remains arbitrary. To derive the last equality, it is necessary to use Eq. (\ref{Extracting_A_Derivative}) and the identity

\begin{equation}\label{O_Commutator}
\left.\Big[\big(\hat O^A\big)_x,\, \big(\hat O^A\big)_y\Big]\right|_{\mbox{\scriptsize fields} = 0} = 0,
\end{equation}

\noindent
which can be easily verified. The minus sign in the last expression in Eq. (\ref{Two_Point_Function_O}) appears after anticommuting the Grassmannian odd expressions $\big(\bar D^{\dot b}\big)_y$ and $\big(\bar{O}^A_{\dot a}\big)_x$.

Substituting the expression (\ref{Two_Point_Function_O}) into the left hand side of Eq. (\ref{Equation_To_Prove}) we see that due to the presence of the supersymmetric covariant derivatives $\big(\bar D^{\dot a}\big)_x \big(\bar D^{\dot b}\big)_y$ the overall degree of explicitly written $\theta$-s decreases by 2. (Certainly, $\theta$-s are also present inside the supersymmetric covariant derivatives entering expressions for various supergraphs, but it is the explicitly written $\theta$-s that we are interested in.) Integrating by parts with respect to the above mentioned derivatives it is possible to rewrite the left hand side of Eq. (\ref{Equation_To_Prove}) in the form

\begin{eqnarray}\label{LHS_New}
&& \mbox{LHS of Eq. (\ref{Equation_To_Prove})} = - 4\int d^8x\, d^8y\, d^6z_1\, d^6\bar z_2\, (\theta^2 \bar\theta^{\dot a} v^B)_x \nonumber\\
&&\qquad\qquad\qquad\qquad \times (\theta^2 \bar\theta^{\dot b} v^B)_y \left. \frac{d}{d\ln\Lambda} \frac{\delta^2}{\delta\mbox{\sl g}_{z_1} \delta \mbox{\sl g}^*_{z_2}}\Big(\big(\bar{O}^A_{\dot a}\big)_x  \big(\bar{O}^A_{\dot b}\big)_y \Gamma\Big)\right|_{\mbox{\scriptsize fields} = 0;\, \mbox{\scriptsize \sl g}= 0}.\qquad
\end{eqnarray}

\noindent
This expression can be presented as a sum of certain one particle irreducible (1PI) supergraphs, because the effective action is the generating functional for 1PI Green functions (see, e.g., \cite{Huang:1982ik}). Therefore, it can be calculated using the tools of the perturbation theory, which include standard rules for working with supergraphs. Note that the external lines in the superdiagrams contributing to the expression (\ref{LHS_New}) are attached to the points $x$, $y$, $z_1$, and $z_2$ and correspond to $\big(\theta^2 \bar\theta^{\dot a} v^B\big)_x$, $\big(\theta^2 \bar\theta^{\dot b} v^B\big)_y$, $1$, and $1$, respectively.

Evidently, any two points of an 1PI graph can be connected by a chain of vertices and propagators. This allows to shift $v^B$ in an arbitrary point of the supergraph, because additional terms produced by such shifts are suppressed by powers of $1/R$. Really, propagators contain derivatives with respect to the superspace coordinates acting on $\delta^8_{xy}$. Certainly, $v^B$ commutes with $\partial/\partial\theta^a$ and $\partial/\partial\bar\theta^{\dot a}$ due to the independence of $\theta$. As for the derivatives with respect to the space-time coordinates $x^\mu$, the shifting of $v^B$ from the superspace point 1 to the point 2 is made according to the procedure

\begin{equation}\label{Shifting_Of_V}
\big(v^B\big)_1 \big(\partial_{\mu}\big)_1\delta^8_{12} = \big(\partial_{\mu}\big)_1 \left(\big(v^B\big)_1 \delta^8_{12}\right)  - \left(\partial_{\mu} v^B\right)_1 \delta^8_{12} = \big(v^B\big)_2 \big(\partial_{\mu}\big)_{1} \delta^8_{12} + O(1/R),
\end{equation}

\noindent
where we took into account that the space-time derivatives of $v^B$ are proportional to powers of $1/R$, see, e.g., Eq. (\ref{Explicit_V}). (To be exact, the dimensionless parameter in this case is $1/(\Lambda R)$.) Certainly, the terms proportional to $1/R$ can be omitted in the limit $R\to\infty$, which is actually equivalent to the limit $p\to 0$ in equations like Eq. (\ref{Bare_Beta_Construction}). Below we will always ignore them.

With the help of equations like (\ref{Shifting_Of_V}) we can shift $v^B$ to an arbitrary point of the supergraph. Let us shift both $v^B$ in Eq. (\ref{LHS_New}) to the point $z_1$,

\begin{equation}\label{V_Shift}
(v^B)_x (v^B)_y \to (v^B)^2_{z_1}
\end{equation}

\noindent
Note that in this case the usual coordinates $x^\mu$ on which $v^B$ depends should be replaced by the chiral coordinates $y^\mu = x^\mu + i\bar\theta^{\dot a} (\gamma^\mu)_{\dot a}{}^b \theta_b$ to obtain a manifestly supersymmetric expression. Certainly, this is possible, because the difference is proportional to powers $1/R$ and vanishes in the limit $R\to \infty$.

Also it is possible to prove that $\bar\theta^{\dot a}$ and $\bar\theta^{\dot b}$ in Eq. (\ref{LHS_New}) can be shifted in an arbitrary point. Really, let us consider a supergraph contributing to the expression (\ref{LHS_New}). It is calculated according to the well-known algorithm (see, e.g., \cite{West:1990tg}), the result being given by an integral over the full superspace.\footnote{Note that even the vertices corresponding to the points $z_1$ and $z_2$ can be presented as integrals over the full superspace, although the integrands in this case are nonlocal.} The integral over the full superspace includes integration over $d^4\theta$ and does not vanish only if the integrand contains $\theta^4 = \theta^2 \bar\theta^2$. Note that new $\theta$-s cannot be produced in calculating the supergraphs, in spite of their presence inside the supersymmetric covariant derivatives. Therefore, any supergraph with $\theta$-s on external lines does not vanish only if it contains at least two right components $\theta_a$ and two left components $\bar\theta_{\dot a}$. The expression (\ref{LHS_New}) is quadratic in $\bar\theta$, which can be shifted along a pass consisting of vertices and propagators using equations like

\begin{equation}\label{Shift_Identities}
\big(\bar\theta^{\dot a}\big)_1 \frac{D_1^2 \bar D_1^2}{4\partial^2} \delta^8_{12} = \frac{D_1^2 \bar D_1^2}{4\partial^2} \Big(\big(\bar\theta^{\dot a}\big)_1\delta^8_{12}\Big) + O(1)\  \to\  \frac{D_1^2 \bar D_1^2}{4\partial^2} \Big(\big(\bar\theta^{\dot a}\big)_2\delta^8_{12}\Big) = \big(\bar\theta^{\dot a}\big)_2 \frac{D_1^2 \bar D_1^2}{4\partial^2} \delta^8_{12}.
\end{equation}

\noindent
Here $O(1)$ denotes terms which do not contain $\bar\theta$. They appear when the covariant derivatives are commuted with $\bar\theta$-s with the help of the identity $\{\bar\theta^{\dot a}, \bar D_{\dot b}\} = \delta^{\dot a}_{\dot b}$. The arrow in Eq. (\ref{Shift_Identities}) points that we omit them, because these terms do not contribute to Eq. (\ref{LHS_New}). Really, the original expression is quadratic in $\bar\theta$, so that the contributions of $O(1)$ terms are no more than linear in $\bar\theta$-s. This implies that they are removed by the final integration over $d^4\theta$.

Thus, we see that $\bar\theta$-s in supergraphs contributing to Eq. (\ref{LHS_New}) can be shifted in an arbitrary way using equations like (\ref{Shift_Identities}). This allows shifting $\bar\theta^{\dot a}$ and $\bar\theta^{\dot b}$ from the points $x$ and $y$ to the point $z_2$,

\begin{equation}\label{Bar_Theta_Shift}
\big(\bar\theta^{\dot a}\big)_x \big(\bar\theta^{\dot b}\big)_y \to \big(\bar\theta^{\dot a}\,\bar\theta^{\dot b}\big)_{z_2}.
\end{equation}

\noindent
After this, we use the identity

\begin{equation}
\big(\bar\theta^{\dot a}\big)_{z_2}\, \big(\bar\theta^{\dot b}\big)_{z_2}\cdot \bar\psi_{\dot a}\, \bar\xi_{\dot b} = - \big(\bar\theta^{\dot a}\big)_{z_2}\, \big(\bar\theta_{\dot b}\big)_{z_2}\cdot  \bar\psi_{\dot a}\, \bar\xi^{\dot b} = - \frac{1}{2} \big(\bar\theta^2\big)_{z_2}\cdot \bar\psi_{\dot b}\, \bar\xi^{\dot b} = \frac{1}{2} \big(\bar\theta^2\big)_{z_2}\cdot \bar\psi^{\dot a}\, \bar\xi_{\dot a}.
\end{equation}

\noindent
(Here we essentially use that both $\bar\theta$-s are placed into a single point $z_2$.) As a result, we obtain that after the shifts (\ref{V_Shift}) and (\ref{Bar_Theta_Shift}) the considered expression is written as

\begin{eqnarray}\label{LHS_Before_Identity}
&& \mbox{LHS of Eq. (\ref{Equation_To_Prove})} = -2\int d^8x\, d^8y\, d^6z_1\, d^6\bar z_2\, \big(\bar\theta^2\big)_{z_2} \big(v^B\big)^2_{z_1}\, \nonumber\\
&&\qquad\qquad\qquad\qquad \times \big(\theta^2\big)_x\, \big(\theta^2\big)_y \left. \frac{d}{d\ln\Lambda} \frac{\delta^2}{\delta\mbox{\sl g}_{z_1} \delta\mbox{\sl g}^*_{z_2}}\Big(\big(\bar{O}^{\dot a,A}\big)_{x}  \big(\bar{O}^A_{\dot a}\big)_{y} \Gamma\Big)\right|_{\mbox{\scriptsize fields} = 0;\, \mbox{\scriptsize \sl g}= 0}.\qquad
\end{eqnarray}

\noindent
Note that due to the antichirality of $\bar\theta^2$ this expression remains manifestly supersymmetric.

\begin{figure}[h]
\begin{picture}(0,5)
\put(6,0){\includegraphics[scale=0.25]{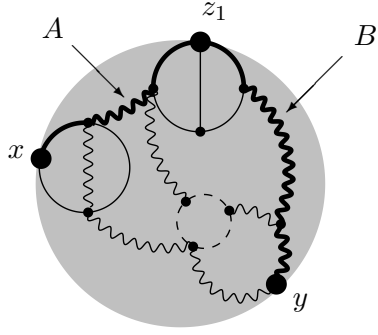}}
\put(5.85,2.4){$x$} \put(9.6,0.45){$y$} \put(8.4,4.3){$z_1$}
\put(6.3,4.0){$A$} \put(10.4,3.9){$B$} \put(6.7,3.9){$\vector(1,-1){0.68}$} \put(10.25,3.75){$\vector(-1,-1){0.73}$}
\end{picture}
\caption{The points $x$, $z_1$, and $y$ of a supergraph can be connected by a pass which consists of the gauge, matter, and ghost propagators. $A$ corresponds to its part connecting the points $x$ and $z_1$, and $B$ corresponds to the part connecting the points $z_1$ and $y$.}\label{Figure_Passes}
\end{figure}

The right components $\theta$ cannot be shifted in an arbitrary way, because the considered expression is quartic in $\theta_{a}$ (here we count only the degree of the right components). However, in this case it is possible to use a special identity derived in Ref. \cite{Stepanyantz:2014ima}. Let us consider an 1PI supergraph contributing to the expression (\ref{LHS_Before_Identity}) and construct two passes connecting the point $x$ with $z_1$ and the point $z_1$ with $y$, see Fig. \ref{Figure_Passes}. The corresponding sequences of vertices and propagators we will denote by $A$ and $B$, respectively. Actually, $A$ and $B$ are products of the expressions in which various derivatives (namely, $\partial_\mu$, $D_a$, $\bar D_{\dot a}$, and $1/\partial^2$) act on superspace $\delta$-functions. Then according to Ref. \cite{Stepanyantz:2014ima}

\begin{equation}\label{Auxiliary_Theta_Identity}
\theta^2 AB \theta^2 + 2(-1)^{P_A+P_B}\theta^a A \theta^2 B \theta_a - \theta^2 A \theta^2 B - A\theta^2 B \theta^2 =O(\theta),
\end{equation}

\noindent
where $(-1)^{P_X}$ is the Grassmannian parity of an expression $X$, and $O(\theta)$ denotes terms which are no more than linear in $\theta$. For completeness, we also present the proof of this identity in Appendix \ref{Appendix_Identity}. (The point $x$ is on the left of each term, the point $y$ is on the right, and the point $z_1$ is between $A$ and $B$.)

Evidently, the $O(\theta)$ terms in Eq. (\ref{Auxiliary_Theta_Identity}) do not contribute to Eq. (\ref{LHS_Before_Identity}), because the integral over $d^4\theta$ which remains after the calculation of the supergraph removes them. Therefore, with the help of Eq. (\ref{Auxiliary_Theta_Identity}) the left hand side of Eq. (\ref{Equation_To_Prove}) can be rewritten in the form

\begin{eqnarray}\label{LHS_After_Identity}
&&\hspace*{-5mm} \mbox{LHS of Eq. (\ref{Equation_To_Prove})} = - 2\int d^8x\, d^8y\, d^6z_1\, d^6\bar z_2\, \big(\theta^2\big)_{z_1} \big(v^B\big)^2_{z_1}\, \big(\bar\theta^2\big)_{z_2}  \nonumber\\
&&\hspace*{-5mm} \qquad\qquad \times\Big((\theta^2)_x + (\theta^2)_y - 2(\theta^b)_x (\theta_b)_y \Big) \left.  \frac{d}{d\ln\Lambda} \frac{\delta^2}{\delta\mbox{\sl g}_{z_1} \delta \mbox{\sl g}^*_{z_2}}\Big(\big(\bar{O}^{\dot a,A}\big)_{x}  \big(\bar{O}^A_{\dot a}\big)_{y} \Gamma\Big)\right|_{\mbox{\scriptsize fields} = 0;\, \mbox{\scriptsize \sl g}= 0},\qquad
\end{eqnarray}

\noindent
where we take into account that all propagators are Grassmannian even. This expression can be equivalently expressed in terms of the operator $\hat{O}^A$ as

\begin{eqnarray}\label{Double_O}
&&\hspace*{-7mm}  -2\int d^8x\, d^8y\, d^6z_1\, d^6\bar z_2\, \big(\theta^2\big)_{z_1} \big(v^B\big)^2_{z_1}\, \big(\bar\theta^2\big)_{z_2} \Big((\theta^2 \bar\theta^{\dot a})_x (\bar\theta_{\dot a})_y  + (\bar\theta^{\dot a})_x (\theta^2 \bar\theta_{\dot a})_y \nonumber\\
&&\hspace*{-7mm}\qquad\qquad\qquad\qquad\qquad + 2(\theta^b \bar\theta^{\dot a})_x (\theta_b \bar\theta_{\dot a})_y \Big) \left. \frac{d}{d\ln\Lambda} \frac{\delta^2}{\delta\mbox{\sl g}_{z_1} \delta\mbox{\sl g}^*_{z_2}}\Big(\big(\hat{O}^{A}\big)_{x}  \big(\hat{O}^A\big)_{y} \Gamma\Big)\right|_{\mbox{\scriptsize fields} = 0;\, \mbox{\scriptsize \sl g}= 0}.\qquad
\end{eqnarray}

\noindent
To see this, it is necessary to use the identity

\begin{equation}
\big(\hat{O}^A\big)_x \big(\hat{O}^A\big)_y \Gamma = - (\bar D^{\dot c})_x (\bar D^{\dot d})_y \big(\bar{O}_{\dot c}^A\big)_x \big(\bar{O}_{\dot d}^A\big)_y \Gamma,
\end{equation}

\noindent
which follows from Eqs. (\ref{Extracting_A_Derivative}) and (\ref{O_Commutator}), and integrate by parts with respect to the derivatives $(\bar D^{\dot c})_x$ and $(\bar D^{\dot d})_y$. With the help of Eq. (\ref{Two_Point_Function_O}) the expression (\ref{Double_O}) can be presented in the form

\begin{eqnarray}\label{LHS_After_Final_Transformation}
&&\hspace*{-7mm} \mbox{LHS of Eq. (\ref{Equation_To_Prove})} = - 2\int d^8x\, d^8y\, d^6z_1\, d^6\bar z_2\, \big(\theta^2\big)_{z_1} \big(v^B\big)^2_{z_1}\, \big(\bar\theta^2\big)_{z_2}\, \Big((\theta^2 \bar\theta^{\dot a})_x (\bar\theta_{\dot a})_y  \nonumber\\
&&\hspace*{-7mm} + (\bar\theta^{\dot a})_x (\theta^2 \bar\theta_{\dot a})_y - \big(\bar\theta^{\dot a} (\gamma^\mu)_{\dot a}{}^b \theta_b\big)_x \big(\bar\theta^{\dot c} (\gamma_\mu)_{\dot c}{}^d \theta_d\big)_y \Big) \left. \frac{d}{d\ln\Lambda} \frac{\delta^4 \Gamma}{\delta\mbox{\sl g}_{z_1} \delta{\mbox{\sl g}}{}^*_{z_2} \delta\bm{V}_x^A \delta\bm{V}_y^A}\right|_{\mbox{\scriptsize fields} = 0;\, \mbox{\scriptsize \sl g}= 0},\qquad
\end{eqnarray}

\noindent
where we also took the identity

\begin{equation}
(\gamma^\mu)_{\dot a}{}^b (\gamma_\mu)_c{}^{\dot d} = 2 \delta_{\dot a}^{\dot d} \delta_c^b
\end{equation}

\noindent
into account. Eq. (\ref{LHS_After_Final_Transformation}) is a convenient starting point for presenting the left hand side of Eq. (\ref{Equation_To_Prove}) in the form of an integral of double total derivatives. This will be made in the next section.

\subsection{Formal calculation}
\hspace{\parindent}\label{Subsection_Formal_Calculation}

Numerous explicit calculations of the $\beta$-function reveal that it is given by integrals of double total derivatives in the momentum space for both the Abelian \cite{Smilga:2004zr,Kazantsev:2014yna} and non-Abelian \cite{Stepanyantz:2011bz,Stepanyantz:2012zz,Stepanyantz:2012us,Shakhmanov:2017soc,Kazantsev:2018nbl} ${\cal N}=1$ supersymmetric theories regularized by higher covariant derivatives. In the Abelian case this factorization into integrals of double total derivatives has been proved in all orders in Refs. \cite{Stepanyantz:2011jy,Stepanyantz:2014ima}. For generalizing this result to the non-Abelian case we consider the left hand side of Eq. (\ref{Equation_To_Prove}) related to $\beta/\alpha_0^2$ by the equation

\begin{equation}\label{Beta_Vs_Gamma}
\mbox{LHS of Eq. (\ref{Equation_To_Prove})} = \frac{r {\cal V}_4}{\pi}\, \frac{\partial^2}{\partial g\,\partial g^*}\Big(\frac{\beta(\rho\alpha_0,\rho\lambda_0\lambda_0^*,Y_0)}{\rho^2\alpha_0^2}\Big)
\end{equation}

\noindent
(where $\rho = g g^*$) and present it in the form (\ref{LHS_After_Final_Transformation}). Below we will demonstrate that it is given by integrals of double total derivatives in the momentum space in all orders.

An important observation is that the expression (\ref{LHS_After_Final_Transformation}) {\it formally} vanishes as a consequence of the Slavnov--Taylor identity (\ref{Preliminary_Background_STI}). In fact, it is not true because of singular contributions, which will be discussed in Sect. \ref{Subsection_Singularities}. However, first, we describe the formal calculation.

As a starting point we consider the Slavnov--Taylor identity (\ref{Preliminary_Background_STI}) in which we set the superfields $V$, $\phi_i$, $c^A$, and $\bar c^A$ to 0. However, the auxiliary superfields remain arbitrary. This gives the equation

\begin{equation}\label{STI_For_V}
\left.\int d^8x\, \delta \bm{V}^A_x \frac{\delta\Gamma}{\delta\bm{V}_x^A}\right|_{\mbox{\scriptsize quantum fields}=0} = 0.
\end{equation}

\noindent
Its left hand side is a functional of the background gauge superfield $\bm{V}$ and the auxiliary external superfields $\mbox{\sl g}$ and $\mbox{\sl g}^*$. Next, we differentiate Eq. (\ref{STI_For_V}) with respect to $\bm{V}^B_y$ and, after this, set the background gauge superfield to 0. Then using Eq. (\ref{Infinitesimal_Transformations}) we obtain

\begin{equation}\label{STI_For_2Point_Function}
\left.\int d^8x\, \left(A^B_x + (A^{B}_x)^*\right) \frac{\delta^2\Gamma}{\delta \bm{V}_y^A \delta\bm{V}_x^A} \right|_{\mbox{\scriptsize quantum fields}=0,\,\bm{V}=0} = 0,
\end{equation}

\noindent
where we also took into account that (even for $\mbox{\sl g}\ne 0$)

\begin{eqnarray}\label{1Point_Function}
&& \left.\frac{\delta\Gamma}{\delta\bm{V}^A_y}\right|_{\mbox{\scriptsize quantum fields}=0,\,\bm{V}=0} = 0;\\
\label{2Point_Function}
&& \left.\frac{\delta^2\Gamma}{\delta \bm{V}_y^B \delta\bm{V}_x^A}\right|_{\mbox{\scriptsize quantum fields}=0,\,\bm{V}=0} = \frac{1}{r} \delta_{AB} \left.\frac{\delta^2\Gamma}{\delta \bm{V}_y^C \delta\bm{V}_x^C}\right|_{\mbox{\scriptsize quantum fields}=0,\,\bm{V}=0}.\qquad
\end{eqnarray}

\noindent
These equations follow from the group theory considerations. Really, if we take into account that the auxiliary superfield $\mbox{\sl g}$ is gauge invariant, then the expressions in Eqs. (\ref{1Point_Function}) and (\ref{2Point_Function}) are proportional to tensors invariant under the gauge group $G$. However, there is no invariant tensors with a single index $A$, and the expression in the left hand side of Eq. (\ref{1Point_Function}) vanishes. (Let us recall that in the case under consideration all generators are traceless.) In this paper we assume that the gauge group is simple, so that the only invariant tensor with two indices $A$ and $B$ is $\delta_{AB}$. This immediately gives Eq. (\ref{2Point_Function}).

Let us choose the parameter $A$ in Eqs. (\ref{Background_Gauge_Transformations_New}) and (\ref{Background_Superfield_Transformation}) in the form

\begin{equation}\label{First_A}
A = \varepsilon^{a B} \theta_a t^B;\qquad A^+ = \bar\varepsilon^{\dot a B} \bar\theta_{\dot a} t^B,
\end{equation}

\noindent
where $\varepsilon^{a B}$ is a coordinate independent anticommuting parameter. This implies that $A^B = \varepsilon^{a B} \theta_a$. Substituting these parameters into Eq. (\ref{STI_For_2Point_Function}) and differentiating with respect to $\bar\varepsilon^{\dot a B}$, we obtain the equation

\begin{equation}
\left.\int d^8x\, \big(\bar\theta_{\dot a}\big)_x \frac{\delta^2\Gamma}{\delta \bm{V}_y^A \delta\bm{V}_x^A} \right|_{\mbox{\scriptsize quantum fields}=0,\,\bm{V}=0} = 0,
\end{equation}

\noindent
the left hand side of which being a functional of the auxiliary superfield $\mbox{\sl g}$. Therefore, it is possible to differentiate with respect to $\mbox{\sl g}$ and $\mbox{\sl g}^*$, so that the part of Eq. (\ref{LHS_After_Final_Transformation}) obtained from the second term in the round brackets vanishes. The part obtained from the first term vanishes due to the same reason. This implies that

\begin{eqnarray}\label{LHS_Third_Term}
&& \mbox{LHS of Eq. (\ref{Equation_To_Prove})} = 2\int d^8x\, d^8y\, d^6z_1\, d^6\bar z_2\, \big(\theta^2\big)_{z_1} \big(v^B\big)^2_{z_1}\, \big(\bar\theta^2\big)_{z_2}\,\nonumber\\
&&\qquad\qquad\qquad\qquad \times \big(\bar\theta^{\dot a} (\gamma^\mu)_{\dot a}{}^b \theta_b\big)_x \big(\bar\theta^{\dot c} (\gamma_\mu)_{\dot c}{}^d \theta_d\big)_y \left. \frac{d}{d\ln\Lambda} \frac{\delta^4\Gamma}{\delta\mbox{\sl g}_{z_1} \delta{\mbox{\sl g}}{}^*_{z_2} \delta\bm{V}_x^A \delta\bm{V}_y^A}\right|_{\mbox{\scriptsize fields} = 0;\, \mbox{\scriptsize \sl g}= 0}.\qquad\quad
\end{eqnarray}

The similar arguments can be used for this expression (which corresponds to the third term in the round brackets in Eq. (\ref{LHS_After_Final_Transformation})). In this case it is necessary to choose the superfield $A$ as

\begin{equation}\label{Second_A}
A = i a_\mu^B t^B y^\mu;\qquad A^+ = -i a_\mu^B t^B (y^\mu)^*,
\end{equation}

\noindent
where $a_\mu^B$ are real coordinate-independent parameters. Therefore, $A^B = i a_\mu^B y^\mu$, where the chiral coordinates $y^\mu$ and the antichiral coordinates $(y^\mu)^*$ are defined as

\begin{equation}\label{Chiral_Coordinates}
y^\mu \equiv x^\mu + i\bar\theta^{\dot a} (\gamma^\mu)_{\dot a}{}^b \theta_b;\qquad (y^\mu)^* = x^\mu - i\bar\theta^{\dot a} (\gamma^\mu)_{\dot a}{}^b \theta_b,
\end{equation}

\noindent
respectively. In this case from Eq. (\ref{STI_For_2Point_Function}) for arbitrary $\bm{g}$ we {\it formally}\footnote{This identity is not actually valid, because the parameter $A$ too rapidly grows at infinity.} obtain the identity

\begin{equation}\label{LHS_Third_Term_Formally_Vanishes}
\left.\int d^8x\, \big(\bar\theta^{\dot a} (\gamma^\mu)_{\dot a}{}^b \theta_b \big)_x \frac{\delta^2\Gamma}{\delta \bm{V}_y^A \delta\bm{V}_x^A} \right|_{\mbox{\scriptsize quantum fields}=0,\,\bm{V}=0} \to \mbox{(formally)} \to 0.
\end{equation}

\noindent
Consequently, the expression (\ref{LHS_Third_Term}) seems to vanish. This implies (see Eqs. (\ref{Beta_Corrections_Preliminary}) and (\ref{Beta_Vs_Gamma})) that all higher order corrections to the $\beta$-function vanish and the $\beta$-function is completely defined by the one-loop approximation. Certainly, it is not true. The matter is that the above calculation was made formally and something very important was missed.

The origin of the incorrect result can be found analyzing the explicit calculations made with the higher covariant derivative regularization \cite{Pimenov:2009hv,Stepanyantz:2011cpt,Stepanyantz:2011bz,Kazantsev:2014yna,Shakhmanov:2017soc,Kataev:2017qvk,Kazantsev:2018nbl}. They demonstrate that all integrals giving the $\beta$-function are integrals of double total derivatives in the momentum space, and that all loop corrections come from $\delta$-singularities. Below in Sect. \ref{Subsection_Double_Total_Derivatives} we will see that the integrals of (double) total derivatives appear due to the presence of $x^\mu$ in Eq. (\ref{Second_A}). These total derivatives produce singular contributions which were ignored in the formal calculation. Note that Eq. (\ref{First_A}) does not contain $x^\mu$, so that the momentum total derivatives do not appear in the first two terms of Eq. (\ref{LHS_After_Final_Transformation}). This implies that the higher ($L\ge 2$) loop corrections to the $\beta$-function are completely determined by the third term inside the round brackets in Eq. (\ref{LHS_After_Final_Transformation}). It is this term that produces the double total derivatives in the momentum space. To derive this fact in Sect. \ref{Subsection_Double_Total_Derivatives}, here we relate this term with the second variation of the functional integral giving the effective action under the change of variables corresponding to the background gauge transformations.

Let us set all quantum superfields to 0. Then the effective action will depend only on the external superfields $\bm{V}$ and $\mbox{\sl g}$. Taking into account that (at least, in the perturbation theory) the vanishing of the quantum (super)fields corresponds to the vanishing of the sources, we obtain

\begin{equation}\label{Effective_Action_For_Vanishing_Fields}
\Gamma\Big|_{\mbox{\scriptsize quantum fields}=0} = -i\ln Z\Big|_{\mbox{\scriptsize sources} = 0},
\end{equation}

\noindent
where $Z$ is given by the functional integral (\ref{Generating_Functional_Z}).

Similarly to the derivation of the Slavnov--Taylor identity in Sect. \ref{Subsection_STI}, we perform the change of variables (\ref{Background_Gauge_Transformations_New}) in this functional integral, but the parameter $A$ will be chosen in the form (\ref{Second_A}). Let us denote the variation of the effective action under the background gauge transformations of the {\it quantum} superfields by $\bar\delta_a$. (This variation does not include the transformation of the background gauge superfield $\bm{V}$.) Taking into account that the generating functional (\ref{Effective_Action_For_Vanishing_Fields}) remains the same after the considered change of variables, while the total action is invariant under the background gauge transformation, we obtain the equation similar to Eq. (\ref{Preliminary_Background_STI}),

\begin{equation}\label{Gamma_Variation}
0 = \bar\delta_a \Gamma\Big|_{\mbox{\scriptsize quantum fields}=0} = - \int d^8y\, \delta_a\bm{V}_y^A \left. \frac{\delta\Gamma}{\delta\bm{V}^A_y}\right|_{\mbox{\scriptsize quantum fields}=0},
\end{equation}

\noindent
which is certainly a mere consequence of the Slavnov--Taylor identity. (Note that the background superfield $\bm{V}$ and the external superfield $\mbox{\sl g}$ are not so far set to 0.) Differentiating Eq. (\ref{Gamma_Variation}) with respect to $a_\mu^B$ gives

\begin{eqnarray}\label{Gamma_Variation_After_Differentiation}
&&\hspace*{-4mm} 0 = \frac{\partial}{\partial a_\mu^B} \bar\delta_a \Gamma\Big|_{\mbox{\scriptsize quantum fields}=0} = i \int d^8y\, \left\{y^\mu \Big[\Big(\frac{\bm{V}}{1-e^{-2\bm{V}}}\Big)_{Adj}\Big]_{AB} \right.
\nonumber\\
&&\hspace*{-4mm}\qquad\qquad\qquad\qquad\qquad\qquad\qquad \left.
+ (y^\mu)^* \Big[\Big(\frac{\bm{V}}{1-e^{2\bm{V}}}\Big)_{Adj}\Big]_{AB} \right\}_y \left. \frac{\delta\Gamma}{\delta\bm{V}^A_y}\right|_{\mbox{\scriptsize quantum fields}=0}.\qquad\quad
\end{eqnarray}

\noindent
The derivative of the effective action with respect to $\bm{V}^A$ entering this equation can be presented as the functional integral

\begin{equation}\label{Gamma_Derivative}
\frac{\delta\Gamma}{\delta \bm{V}^A}\Big|_{\mbox{\scriptsize quantum fields}=0} = \Big\langle \frac{\delta S_{\mbox{\scriptsize total}}}{\delta \bm{V}^A} + \frac{\delta S_\varphi}{\delta \bm{V}^A} - c\, \Big\langle \frac{\delta S_\Phi}{\delta \bm{V}^A} \Big\rangle_\Phi  \Big\rangle\Big|_{\mbox{\scriptsize quantum fields}=0},
\end{equation}

\noindent
where the angular brackets are defined by Eq. (\ref{VEV}) and we also introduced the notation

\begin{equation}
\langle B \rangle_\Phi \equiv \mbox{Det}(PV,M) \int D\Phi\, B\, \exp(i S_\Phi).
\end{equation}

\noindent
In this functional integral it is possible to perform again the change of variables (\ref{Background_Gauge_Transformations_New}) with the parameter $A = i b_\mu^B t^B y^\mu$. After this change of variables we set the background gauge superfield $\bm{V}$ to 0. As a result, we obtain the identity

\begin{eqnarray}\label{First_Derivative}
&& 0 = \left.\frac{\partial^2}{\partial b^{\mu B}\, \partial a_\mu^B}\, \bar\delta_b \bar\delta_a \Gamma \right|_{\mbox{\scriptsize fields}=0} = \frac{i}{2} \int d^8y\, \left(y^\mu -
(y^\mu)^*\right)_y \left. \frac{\partial}{\partial b^{\mu B}} \bar\delta_b\Big(\frac{\delta\Gamma}{\delta\bm{V}^B_y}\Big)\right|_{\mbox{\scriptsize fields}=0}\qquad\nonumber\\
&& = - \int d^8y\, \big(\bar\theta^{\dot c} (\gamma^\mu)_{\dot c}{}^d \theta_d\big)_y\, \frac{\partial}{\partial b^{\mu B}} \bar\delta_b \left.\Big\langle \frac{\delta S_{\mbox{\scriptsize total}}}{\delta \bm{V}^B_y} + \frac{\delta S_\varphi}{\delta \bm{V}^B_y} - c\, \Big\langle \frac{\delta S_\Phi}{\delta \bm{V}^B_y} \Big\rangle_\Phi  \Big\rangle\right|_{\mbox{\scriptsize fields}=0}.
\end{eqnarray}

\noindent
As usual, the subscript ``fields = 0'' means that the superfields $V$, $\phi_i$, $c$, $\bar c$, and $\bm{V}$ are set to 0, while the chiral superfield $\mbox{\sl g}$ can take arbitrary values. The symbol $\bar\delta_b$ denotes the variation under the transformations (\ref{Background_Gauge_Transformations_New}) of the quantum superfields parameterized by $A = i b_\mu^A t^A y^\mu$, the background gauge superfield $\bm{V}$ being fixed.

Let us transform the right hand side of this expression taking into account that the total action (\ref{Total_Action}) and the Pauli--Villars actions $S_\varphi$ and $S_\Phi$ (given by Eqs. (\ref{S_varphi}) and (\ref{S_Phi}), respectively) are invariant under the background gauge transformations. Due to the background gauge invariance

\begin{equation}\label{Invariance_Actions}
\Big(\bar\delta_b + \int d^8x\, \delta_b \bm{V}^A_x \frac{\delta}{\delta\bm{V}^A_x}\Big) S_{\mbox{\scriptsize total},\,\varphi,\,\Phi} = 0,
\end{equation}

\noindent
where $\delta_b \bm{V}$ is given by Eq. (\ref{Infinitesimal_Transformations}). From Eq. (\ref{Invariance_Actions}) it is possible to obtain the identities

\begin{eqnarray}\label{Invariance_Derivatives}
&& \frac{\partial}{\partial b_\mu^{B}} \Big(\bar\delta_b + \int d^8x\, \delta_b \bm{V}^A_x \frac{\delta}{\delta\bm{V}^A_x}\Big) \left.\frac{\delta S_{\mbox{\scriptsize total},\,\varphi,\,\Phi}}{\delta\bm{V}^B_y}\right|_{\bm{V}=0} = 0.
\end{eqnarray}

\noindent
They can be derived by commuting the derivative with respect to $\bm{V}^B_y$ to the left, if we take into account that it commutes with $\bar\delta_b$ and use the equation

\begin{eqnarray}
&& \left.\Big[ \frac{\partial}{\partial b_\mu^B}\, \delta_b\bm{V}^A_x \frac{\delta}{\delta \bm{V}^A_x},\, \frac{\delta}{\delta \bm{V}^B_y} \Big]\right|_{\bm{V}=0}
\nonumber\\
&& = \left.\Big[ -i\Big\{ y^\mu \Big(\frac{\bm{V}_x}{1-e^{-2\bm{V}_x}}\Big)_{Adj} + (y^\mu)^* \Big(\frac{\bm{V}_x}{1-e^{2\bm{V}_x}}\Big)_{Adj} \Big\}_{AB} \frac{\delta}{\delta \bm{V}^A_x},\, \frac{\delta}{\delta \bm{V}^B_y} \Big]\right|_{\bm{V}=0} =0\qquad
\end{eqnarray}

\noindent
which is valid because $f^{AAC}=0$.

The operator $\bar\delta_b$ in Eq. (\ref{First_Derivative}) acts on the expression inside the angular brackets and on the actions $S_{\mbox{\scriptsize total}}$, $S_\varphi$, and $S_\Phi$ in the exponents. Eqs. (\ref{Invariance_Actions}) and (\ref{Invariance_Derivatives}) allow expressing the result in terms of the derivatives with respect to the background gauge superfield. From the other side, the derivative of the angular brackets with respect to $\bm{V}$ also acts on the expression inside these brackets and on the actions in the exponents. This implies that

\begin{eqnarray}\label{B_Variation}
&& \hspace*{-9mm} \left.\frac{\partial}{\partial b^{\mu B}} \bar\delta_b\Big(\frac{\partial}{\partial a_\mu^B} \bar\delta_a\Gamma \Big) \right|_{\mbox{\scriptsize fields}=0} = \frac{\partial}{\partial b^{\mu B}}\int d^8x\,  \delta_b\bm{V}^A_x \frac{\delta}{\delta\bm{V}_x^A}\int d^8y\, \big(\bar\theta^{\dot c} (\gamma^\mu)_{\dot c}{}^d \theta_d\big)_y\, \Big\langle \frac{\delta S_{\mbox{\scriptsize total}}}{\delta \bm{V}^B_y} + \frac{\delta S_\varphi}{\delta \bm{V}^B_y}\nonumber\\
&& \hspace*{-9mm} \left. - c\, \Big\langle \frac{\delta S_\Phi}{\delta \bm{V}^B_y}\Big\rangle_\Phi  \Big\rangle\right|_{\mbox{\scriptsize fields}=0}
= \left. \frac{\partial}{\partial b^{\mu B}} \int d^8x\, d^8y\, \delta_b\bm{V}^A_x \frac{\delta}{\delta \bm{V}^A_x} \left(\big(\bar\theta^{\dot c} (\gamma^\mu)_{\dot c}{}^d \theta_d\big)_y\, \frac{\delta\Gamma}{\delta \bm{V}^B_y} \right)\right|_{\mbox{\scriptsize fields}=0}.\qquad
\end{eqnarray}

\noindent
The expression $\delta_b\bm{V}$ entering this equation is given by Eq. (\ref{Infinitesimal_Transformations}). Differentiating it with respect to $b_\mu^B$ and setting the background gauge superfield to 0, we obtain

\begin{equation}\label{DeltaB_V}
\frac{\partial}{\partial b_\mu^B} \delta_b\bm{V}^A_x \Big|_{\bm{V}=0} = -\frac{i}{2}\left(y^\mu - (y^\mu)^*\right)_x \delta^{AB} = \big(\bar\theta^{\dot a} (\gamma^\mu)_{\dot a}{}^b \theta_b\big)_x \delta^{AB}.
\end{equation}

\noindent
Therefore, taking into account Eq. (\ref{Gamma_Variation}), we see that the formal calculation gives

\begin{eqnarray}\label{Second_Variation}
&& \int d^8x\, d^8y\, \big(\bar\theta^{\dot a} (\gamma_\mu)_{\dot a}{}^b \theta_b\big)_x\, \big(\bar\theta^{\dot c} (\gamma^\mu)_{\dot c}{}^d \theta_d\big)_y\, \left.\frac{\delta^2\Gamma}{\delta\bm{V}^A_x\,\delta\bm{V}^A_y}\right|_{\mbox{\scriptsize fields}=0} \to \mbox{(formally)}\qquad\nonumber\\
&&\qquad\qquad\qquad\qquad\qquad\qquad\qquad\qquad\qquad\qquad\qquad \to \left.\frac{\partial^2}{\partial b^{\mu B}\, \partial a_\mu^B}\, \bar\delta_b \bar\delta_a \Gamma \right|_{\mbox{\scriptsize fields}=0} = 0.\qquad
\end{eqnarray}

\noindent
(Note that in this expression we do not set the external superfield $\mbox{\sl g}$ to 0.) However, in what follows we will see that the first equality is not true, because doing the formal calculation we ignore singular contributions. These singular contributions will be discussed below.

If we apply the operator

\begin{equation}
2\, \frac{d}{d\ln\Lambda} \int d^6z_1\, d^6\bar z_2\, \big(\theta^2\big)_{z_1} \big(v^B\big)_{z_1}^2\, \big(\bar\theta^2\big)_{z_2}\, \frac{\delta^2}{\delta\mbox{\sl g}_{z_1} \delta{\mbox{\sl g}}{}^*_{z_2}}
\end{equation}

\noindent
to the left hand side of Eq. (\ref{Second_Variation}) and, after this, set the auxiliary external superfield $\mbox{\sl g}$ to 0, then we obtain the expression (\ref{LHS_Third_Term}),

\begin{eqnarray}\label{Double_Total_Derivative}
&& \mbox{LHS of Eq. (\ref{Equation_To_Prove})} \to \mbox{(formally)} \to 2\, \frac{d}{d\ln\Lambda} \int d^6z_1\, d^6\bar z_2\,  \nonumber\\
&&\qquad\qquad\qquad\qquad \times \big(\theta^2\big)_{z_1} \big(v^B\big)_{z_1}^2\, \big(\bar\theta^2\big)_{z_2}\, \frac{\delta^2}{\delta\mbox{\sl g}_{z_1} \delta{\mbox{\sl g}}{}^*_{z_2}} \left.\frac{\partial^2}{\partial b^{\mu B}\, \partial a_\mu^B}\, \bar\delta_b \bar\delta_a \Gamma \right|_{\mbox{\scriptsize fields}=0;\ \mbox{\scriptsize {\sl g}} = 0} = 0.\qquad
\end{eqnarray}

\noindent
According to this equation all higher order corrections to the $\beta$-function vanish. Certainly, it is not true. As we have already mentioned above, such a result appears, because singular contributions were missed in the formal calculation described above.

Although from Eq. (\ref{Double_Total_Derivative}) we obtain the same (incorrect) formal result as from Eq. (\ref{LHS_Third_Term_Formally_Vanishes}), Eq. (\ref{Double_Total_Derivative}) will be very useful below, because it allows explaining the factorization of the loop integrals giving the $\beta$-function into integrals of double total derivatives.

\subsection{Integrals of double total derivatives}
\hspace{\parindent}\label{Subsection_Double_Total_Derivatives}

Although the calculation described in the previous section is formal, it allows explaining why the $\beta$-function (defined in terms of the bare couplings with the higher derivative regularization) is given by integral of double total derivatives in the momentum space. This can be done starting from Eq. (\ref{Double_Total_Derivative}). Its left hand side is related to the $\beta$-function by Eq. (\ref{Beta_Vs_Gamma}). In this section we present the right hand side of Eq. (\ref{Double_Total_Derivative}) as a sum of integrals of double total derivatives and formulate a prescription for constructing these integrals.

Let $\varphi_I$ denotes the whole set of superfields of the theory, where the index $I$ corresponds to quantum numbers with respect to the gauge group, and $j^I$ are the corresponding sources. In the momentum representation the propagators can be presented in the form

\begin{equation}
\left. - \frac{1}{Z_0} \frac{\delta^2 Z_0}{\delta (j^I)_1 \delta (j^J)_2}\right|_{j=0} \equiv P_{IJ}(1,2) \equiv \int\frac{d^4k}{(2\pi)^4} \exp\Big(-ik_\alpha \left(x^\alpha_1-x^\alpha_2\right)\Big)\, P_{IJ}(k,\theta_1-\theta_2),
\end{equation}

\noindent
where $Z_0$ is the generating functional for the free theory.

Let us make the change of the integration variables (\ref{Background_Gauge_Transformations_New}) with the parameter $A$ given by Eq. (\ref{Second_A}) in the generating functional $Z$ with the sources and the background gauge superfield set to 0. Although under this change of variables the generating functional remains invariant, the propagators and vertices transform nontrivially. Really, if $S_2$ and $S_{\mbox{\scriptsize int}}$ are the quadratic part of the action and the interaction, respectively, then

\begin{eqnarray}
&& Z = Z' = \int D\varphi'\exp \Big(i \left(S_2[\varphi'] + S_{\mbox{\scriptsize int}}[\varphi']\right)\hspace*{-1mm}\Big)\nonumber\\
&&\qquad\qquad\left. = \exp\Big(iS_{\mbox{\scriptsize int}}\left[\varphi'(\varphi \to -i\delta/\delta j)\right]\hspace*{-1mm}\Big) \int D\varphi \exp\Big(iS_2\left[\varphi'(\varphi)\right] + i\varphi \cdot j\Big)\right|_{j=0}.\qquad
\end{eqnarray}

\noindent
(The corresponding Jacobian does not depend on the superfields of the theory and can be omitted.) The new vertices obtained from $S_{\mbox{\scriptsize int}}[\varphi'(\varphi)]$ are evidently different from the old ones coming from $S_{\mbox{\scriptsize int}}[\varphi]$. The new propagators

\begin{equation}
\left. P_{IJ}'(1,2) = - \frac{1}{Z_0'} \frac{\delta^2 Z_0'}{\delta (j^I)_1 \delta (j^J)_2}\right|_{j=0},\quad \mbox{where}\quad Z_0' \equiv  \int D\varphi \exp\Big(iS_2\left[\varphi'(\varphi)\right] + i\varphi \cdot j\Big),
\end{equation}

\noindent
are also different from the old ones.

Now, let us try to understand how the evident equality $Z=Z'$ appears at the level of superdiagrams. For this purpose we write the transformation (\ref{Background_Gauge_Transformations_New}) with the parameter (\ref{Second_A}) and concentrate on the terms linear in $x^\mu$,

\begin{equation}\label{Change_Of_Variable}
\varphi_I \to \varphi_I' = \varphi_I + i a_\mu^A x^\mu (T^A)_I{}^J \varphi_J + \ldots,
\end{equation}

\noindent
where $(T^A)_I{}^J$ are the generators of the gauge group in a relevant representation, and the terms which do not explicitly depend on $x^\mu$ are denoted by dots.\footnote{Note that if the sources are not set to 0, then $Z' \equiv \int D\varphi\, \exp(iS[\varphi']+i\varphi\cdot j) = Z[j']$, where $j'{}^I = j^I - ia_\mu^A x^\mu (T^A)_J{}^I j^J + \ldots$ In this case the arguments of the effective action change as $\varphi_I = \delta W/\delta j^I \to \varphi'_{I} = \varphi_I +ia_\mu^A x^\mu (T^A)_I{}^J \varphi_J +\dots$ This implies that the considered change of the integration variables actually generates the transformation $\bar\delta_a$.} Then the propagator changes as

\begin{eqnarray}\label{Proparator_Variation_Original}
&& \bar\delta_a P_{IJ}(1,2) = -i a_\mu^A \int\frac{d^4k}{(2\pi)^4} \exp\Big(-ik_\alpha \big(x^\alpha_1 - x^\alpha_2\big)\Big)\nonumber\\
&&\qquad\qquad\qquad \times \Big(x^\mu_1\, (T^A)_I{}^K P_{KJ}(k,\theta_1-\theta_2) + x^\mu_2\, (T^A)_J{}^K P_{IK}(k,\theta_1-\theta_2)\Big)+\ldots \qquad
\end{eqnarray}

\noindent
Next, we note that both the quadratic part of the action and all vertices are invariant under the {\it global} gauge transformations $\delta\varphi_I = i\alpha^A (T^A)_I{}^J \varphi_J$, where $\alpha^A \ne \alpha^A(x,\theta)$ are the real parameters. This implies that the propagators should be proportional to tensors invariant under the gauge group transformations,

\begin{equation}
(T^A)_I{}^K P_{KJ} + (T^A)_J{}^K P_{IK} = 0.
\end{equation}

\noindent
Using this equation it is possible to demonstrate that in the momentum representation the change of the propagator (\ref{Proparator_Variation_Original}) is related to its derivative with respect to the momentum,

\begin{equation}\label{Propagator_Variation}
\bar\delta_a P_{IJ}(k,\theta_1-\theta_2) = - a_\mu^A (T^A)_I{}^K\, \frac{\partial}{\partial k_\mu} P_{KJ}(k,\theta_1-\theta_2) + \ldots.
\end{equation}

Next, let us proceed to the interaction vertices. An $n$-point vertex can be formally written in the form

\begin{eqnarray}\label{N_Point_Vertex}
&& \int d^8x\, \hat V^{I_1 I_2 \ldots I_n}(x_1, x_2,\ldots, x_n;\theta_1, \theta_2, \ldots, \theta_n)\nonumber\\
&&\qquad\qquad\qquad\qquad \left.\vphantom{\frac{1}{2}} \times \Big(\varphi_{I_1}(x_1,\theta_1)\, \varphi_{I_2}(x_2,\theta_2) \ldots \varphi_{I_n}(x_n,\theta_n)\Big)\right|_{\parbox{2.5cm}{\scriptsize $x_1=x_2=\ldots = x;$\\
$\theta_1=\theta_2 = \ldots=\theta$}},\qquad
\end{eqnarray}

\noindent
where the operator $\hat V^{I_1 I_2 \ldots I_n}$ contains various derivatives $D_a$, $\bar D_{\dot a}$, and $\partial_\mu$. Certainly, it can also have Lorentz indices which have been omitted in the above expression. The invariance of the vertex under the above mentioned global gauge transformations leads to the identity

\begin{equation}\label{Vertex_Invariance}
\hat V^{K I_2 \ldots I_n} (T^A)_{K}{}^{I_1} + \hat V^{I_1 K \ldots I_n} (T^A)_{K}{}^{I_2} + \ldots + \hat V^{I_1 I_2 \ldots K} (T^A)_{K}{}^{I_n} = 0.
\end{equation}

\noindent
To rewrite the vertex (\ref{N_Point_Vertex}) in the momentum representation, we present all superfields entering it as

\begin{equation}
\varphi_I(x,\theta) = \int \frac{d^4k}{(2\pi)^4} \exp\left(-ik_\alpha x^\alpha\right) \varphi_I(k,\theta).
\end{equation}

\noindent
Then after some transformations the considered vertex takes the form

\begin{eqnarray}\label{N_Point_Vertex_Momentum}
&& \int d^4\theta\, \int \frac{d^4k_1}{(2\pi)^4}\, \frac{d^4k_2}{(2\pi)^4} \ldots \frac{d^4k_n}{(2\pi)^4}\, \hat V^{I_1 I_2 \ldots I_n}(k_1,k_2,\ldots,k_n;\theta_1,\theta_2,\ldots,\theta_n)\qquad\nonumber\\
&&\qquad\qquad\qquad\qquad\qquad\qquad\quad \times \left.
\Big(\varphi_{I_1}(k_1,\theta_1)\, \varphi_{I_2}(k_2,\theta_2)\ldots \varphi_{I_n}(k_n,\theta_n)\Big)\right|_{\theta_1=\theta_2\ldots = \theta_n =\theta},\qquad\quad
\end{eqnarray}

\noindent
where the operator

\begin{eqnarray}\label{Vertex_With_Delta_Function}
&& \hat V^{I_1 I_2 \ldots I_n}(k_1,k_2,\ldots,k_n;\theta_1,\theta_2,\ldots,\theta_n)\nonumber\\
&&\left. \equiv  \int d^4x\, \hat V^{I_1 I_2 \ldots I_n}(x_1,x_2,\ldots,x_n;\theta_1,\theta_2,\ldots,\theta_n) \exp\Big(-i\sum\limits_{i=1}^n(k_i)_\alpha (x_i)^\alpha\Big)\right|_{x_1=x_2\ldots = x_n =x}\qquad
\nonumber\\
&& = (2\pi)^4 \delta^4(k_1+k_2+\ldots+k_n)\, \hat W^{I_1 I_2 \ldots I_n}(k_1,k_2,\ldots,k_n;\theta_1,\theta_2,\ldots,\theta_n)\qquad\
\end{eqnarray}

\noindent
contains derivatives with respect to $\theta$-s and the $\delta$-function responsible for the four-momentum conservation,

\begin{equation}\label{Momentum_Conservation}
k_1^\mu + k_2^\mu +\ldots + k_n^\mu = 0.
\end{equation}

Under the change of the integration variables (\ref{Change_Of_Variable}) in the generating functional (\ref{Generating_Functional_Z}) the vertex transforms as

\begin{eqnarray}
&& \bar\delta_a \hat V^{I_1 I_2 \ldots I_n}(k_1,k_2,\ldots,k_n;\theta_1,\theta_2,\ldots,\theta_n)\vphantom{\frac{1}{2}} \nonumber\\
&&\qquad = - a^{\mu A} \left((T^A)_K{}^{I_1} \frac{\partial}{\partial k_1^\mu} \hat V^{K I_2 \ldots I_n}(k_1,k_2,\ldots,k_n;\theta_1,\theta_2,\ldots,\theta_n) \right.\qquad\nonumber\\
&&\qquad + (T^A)_K{}^{I_2} \frac{\partial}{\partial k_2^\mu} \hat V^{I_1 K \ldots I_n}(k_1,k_2,\ldots,k_n;\theta_1,\theta_2,\ldots,\theta_n) + \ldots \nonumber\\
&&\qquad\left. + (T^A)_K{}^{I_n} \frac{\partial}{\partial k_n^\mu} \hat V^{I_1 I_2 \ldots K}(k_1,k_2,\ldots,k_n;\theta_1,\theta_2,\ldots,\theta_n)\right)+\ldots,
\end{eqnarray}

\noindent
where the last dots correspond to the terms which were not written explicitly in Eq. (\ref{Change_Of_Variable}). Using Eq. (\ref{Vertex_Invariance}) it is possible to rewrite this expression in the form

\begin{eqnarray}
&& a^{\mu A} \left((T^A)_K{}^{I_2} \Big(\frac{\partial}{\partial k_1^\mu} - \frac{\partial}{\partial k_2^\mu}\Big)\hat V^{I_1 K \ldots I_n}(k_1,k_2,\ldots,k_n;\theta_1,\theta_2,\ldots,\theta_n) + \ldots \right. \qquad\nonumber\\
&&\left. + (T^A)_K{}^{I_n} \Big(\frac{\partial}{\partial k_1^\mu} - \frac{\partial}{\partial k_n^\mu}\Big) \hat V^{I_1 I_2 \ldots K}(k_1,k_2,\ldots,k_n;\theta_1,\theta_2,\ldots,\theta_n) \right)+\ldots
\end{eqnarray}

\noindent
Then with the help of Eq. (\ref{Vertex_With_Delta_Function}) we obtain

\begin{eqnarray}\label{Vertex_Variation}
&& \bar\delta_a\hat V^{I_1 I_2 \ldots I_n}(k_1,k_2,\ldots,k_n;\theta_1,\theta_2,\ldots,\theta_n)  = - (2\pi)^4 \delta^4(k_1+k_2+\ldots+k_n)\, \vphantom{\frac{1}{2}}\nonumber\\
&& \times a^{\mu A} \left((T^A)_K{}^{I_2} \frac{\partial}{\partial k_2^\mu} \hat W^{I_1 K \ldots I_n}(-k_2-\ldots-k_n,k_2,\ldots,k_n;\theta_1,\theta_2,\ldots,\theta_n) + \ldots \right. \qquad\nonumber\\
&&\left. + (T^A)_K{}^{I_n} \frac{\partial}{\partial k_n^\mu} \hat W^{I_1 I_2 \ldots K}(-k_2-\ldots-k_n,k_2,\ldots,k_n;\theta_1,\theta_2,\ldots,\theta_n) \right).
\end{eqnarray}

\noindent
Next, it is necessary to note a resemblance between Eq. (\ref{Momentum_Conservation}) and Eq. (\ref{Vertex_Invariance}). In Eq. (\ref{Vertex_Invariance}) each generator actually corresponds to a propagator coming from the considered vertex exactly as momenta in Eq. (\ref{Momentum_Conservation}). This implies that such equations appear in pairs. Say, if the considered vertex is placed inside a certain graph in which the momentum $k^\mu_2$ can be expressed in terms of $k^\mu_3,\ldots, k^\mu_n$, then

\begin{eqnarray}
&& k^\mu_2\ \to\ c_3 k^\mu_3 + \ldots + c_n k^\mu_n;\vphantom{\Big(}\\
&& (T^A)_K{}^{I_2} \hat V^{I_1 K I_3 \ldots I_n}\ \to\ c_3 (T^A)_K{}^{I_3} \hat V^{I_1 I_2 K \ldots I_n} + \ldots + c_n (T^A)_K{}^{I_n} \hat V^{I_1 I_2 I_3\ldots K},\qquad
\end{eqnarray}

\noindent
where $c_3,\ldots c_n$ are some numerical coefficients. In this case $\bar\delta_a \hat V^{I_1 I_2\ldots I_n}$ will be proportional to

\begin{eqnarray}\label{Vertex_Derivatives}
&& (T^A)_K{}^{I_3} \Big(\frac{\partial}{\partial k^\mu_3} + c_3 \frac{\partial}{\partial k^\mu_2}\Big) W^{I_1 I_2 K\ldots I_n} + \ldots + (T^A)_K{}^{I_n} \Big(\frac{\partial}{\partial k^\mu_n} + c_n \frac{\partial}{\partial k^\mu_2}\Big) W^{I_1 I_2 I_3 \ldots K} \nonumber\\
&& = (T^A)_K{}^{I_3} \frac{\partial}{\partial k^\mu_3} W^{I_1 I_2 K\ldots I_n}\Big(k_2^\nu \to c_3 k^\nu_3 + \ldots + c_n k^\nu_n\Big) + \ldots\nonumber\\
&&\qquad\qquad\qquad\qquad\qquad\qquad\qquad + (T^A)_K{}^{I_n} \frac{\partial}{\partial k^\mu_n} W^{I_1 I_2 I_3 \ldots K}\Big(k_2^\nu \to c_3 k^\nu_3 + \ldots + c_n k^\nu_n\Big).\qquad
\end{eqnarray}

\noindent
Thus, the variations $\bar\delta_a$ of vertices inside a supergraph contain only derivatives with respect to independent momenta.

\begin{figure}[h]
\begin{picture}(0,5.8)
\put(8.5,3.2){\includegraphics[scale=0.21]{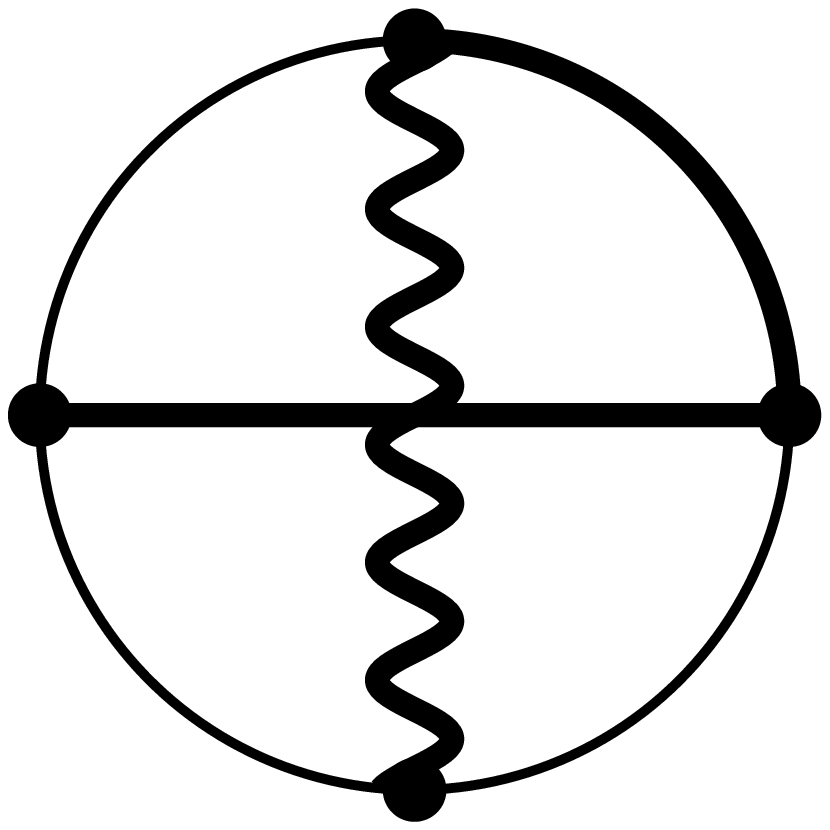}}
\put(6.4,0.3){\includegraphics[scale=0.21]{graph_example.eps}}
\put(9.6,0.3){\includegraphics[scale=0.21]{graph_example.eps}}
\put(12.8,0.3){\includegraphics[scale=0.21]{graph_example.eps}}
\put(3.8,3.95){${\displaystyle \int \frac{d^4k}{(2\pi)^4}\, \frac{d^4l}{(2\pi)^4}\, \frac{d^4q}{(2\pi)^4}}\, \bar\delta_a\, \Bigg\{\hspace*{2.5cm}\Bigg\}$}
\put(10.53,3.13){\vector(-3,1){1}}\put(10.6,3.0){$k_\mu$}
\put(11.1,5.0){\vector(-3,-1){1}}\put(11.2,4.95){$q_\mu$}
\put(8.5,4.7){\vector(1,-1){0.5}}\put(8.2,4.9){$l_\mu$}
\put(0.5,1.05){${\displaystyle = - a_\mu^A \int \frac{d^4k}{(2\pi)^4}\, \frac{d^4l}{(2\pi)^4}\, \frac{d^4q}{(2\pi)^4}\, \Bigg\{\frac{\partial}{\partial l^\mu}\hspace*{2.1cm} + \frac{\partial}{\partial k^\mu}
\hspace*{2.1cm} + \frac{\partial}{\partial q^\mu}\hspace*{1.9cm}\Bigg\}}$}
\put(11.65,0.33){\vector(-3,1){1}}\put(11.8,0.15){$T_{Adj}^A$}
\put(15.35,2.1){\vector(-3,-1){1}}\put(15.45,2.0){$T^A$}
\put(6.3,1.8){\vector(1,-1){0.5}}\put(6.0,1.9){$T^A$}
\end{picture}
\caption{This figure illustrates how the total derivatives in the momentum space appear as a result of the variable change (\ref{Change_Of_Variable}). Propagators with independent momenta $k^\mu$, $l^\mu$, and $q^\mu$ are depicted by the bold lines. Note that the integrations over the loop momenta are written explicitly and (in this figure) are not included into the supergraph.}\label{Figure_Graph_Variation}
\end{figure}

It is well known that due to the momentum conservation in each vertex (encoded in equations like Eq. (\ref{Momentum_Conservation})) in an $L$ loop graph without external lines only $L$ momenta are independent. (In our case this is also true, because the momenta of all external lines vanish.) Therefore, we can mark $L$ propagators whose momenta are considered as independent parameters, see Fig.~\ref{Figure_Graph_Variation} (which corresponds to the case $L=3$). Then, using the resemblance between Eq. (\ref{Momentum_Conservation}) and Eq. (\ref{Vertex_Invariance}), it is possible to construct $L$ independent structures in which the generators correspond to certain propagators, e.g., to the propagators whose momenta we consider as independent parameters. Any graph in which $T^A$ stands on a certain propagator can be expressed in terms of these structures.

Let us consider a closed loop, consisting of vertices and propagators, which includes one of the independent momenta, say, $k^\mu$. Then according to Eqs. (\ref{Propagator_Variation}), (\ref{Vertex_Variation}) and (\ref{Vertex_Derivatives}), from the terms containing the derivative $\partial/\partial k^\mu$ we obtain the contribution to the first variation of the considered supergraph given by an integral of a total derivative

\begin{equation}
- a^A_\mu T^A \int \frac{d^4k}{(2\pi)^4} \frac{\partial}{\partial k_\mu},
\end{equation}

\noindent
where the generator $T^A$ should be inserted on the propagator with the momentum $k^\mu$. This is graphically illustrated in Fig. \ref{Figure_Graph_Variation}.

The second variation is calculated similarly.

Thus, we have a prescription, how to find integrals of double total derivatives which contribute to the $\beta$-function. The starting point is the expression

\begin{equation}\label{Starting_Expression}
\left. \frac{d}{d\ln\Lambda} \int d^6z_1\, d^6\bar z_2\, \big(\theta^2\big)_{z_1} \big(v^B\big)^2_{z_1}\, \big(\bar\theta^2\big)_{z_2}\, \frac{\delta^2 \Gamma}{\delta \mbox{\sl g}_{z_1} \delta \mbox{\sl g}^*_{z_2}}\right|_{\mbox{\scriptsize fields}=0;\, \mbox{\scriptsize{\sl g}}=0}.
\end{equation}

\noindent
First, we consider a certain $L$ loop supergraph contributing to it and (in an arbitrary way) mark $L$ propagators with the (Euclidean) momenta $Q^\mu_i$ considered as independent. Let $a_i$ be the indices corresponding to their begginings. Next, it is necessary to calculate the supergraph using the standard rules. The result includes a coefficient which contains couplings and some group factors. This coefficient should be replaced by a certain differential operator which is obtained by calculating the ``second variation'' of the expression $\prod_i \delta_{a_i}^{b_i}$, where $\delta_{a_i}^{b_i}$ comes from the marked propagators, formally setting

\begin{equation}
\delta(\delta_{a_i}^{b_i}) \to (T^A)_{a_i}{}^{b_i} \frac{\partial}{\partial Q_i^\mu}.
\end{equation}

\noindent
In other words, we make the replacement

\begin{equation}
\prod\limits_{i=1}^L \delta_{a_i}^{b_i}\ \to\ \sum\limits_{k,l=1}^L \prod\limits_{i\ne k,l} \delta_{a_i}^{b_i}\, (T^A)_{a_k}{}^{b_k} (T^A)_{a_l}{}^{b_l} \frac{\partial}{\partial Q_k^\mu} \frac{\partial}{\partial Q_{l\, \mu}}.
\end{equation}

\noindent
Next, one should multiply the result by the factor

\begin{equation}
-\frac{2\pi}{r{\cal V}_4},
\end{equation}

\noindent
where the sign ``$-$'' appears, because

\begin{equation}\label{Sign}
\frac{\partial}{\partial q_k^\mu} \frac{\partial}{\partial q_{l\, \mu}} = - \frac{\partial}{\partial Q_k^\mu} \frac{\partial}{\partial Q_{l\, \mu}}.
\end{equation}

\noindent
Finally, it is necessary to rewrite the result in terms of $\rho=g g^*$ and perform the integration

\begin{equation}
\int\limits_{+0}^1 \frac{d\rho}{\rho} \int\limits_{+0}^1 d\rho.
\end{equation}

The expression obtained according to the algorithm described above coincides with a contribution to $\beta/\alpha_0^2$ coming from the sum of all superdiagrams which are obtained from the original vacuum supergraphs by attaching two external lines of the background gauge superfield in all possible ways.

Below in Sect. \ref{Section_Examples} we will verify this algorithm for some particular examples.

\subsection{The role of singularities}
\hspace{\parindent}\label{Subsection_Singularities}

From the discussion of the previous section we can conclude that in the case of using the higher derivative regularization the integrals giving the $\beta$-function are integrals of double total derivatives. This agrees with the results of explicit calculations which also reveal that all higher order corrections to the $\beta$-function originate from singularities of the momentum integrals. Actually it is the contributions of the singularities that have been missed in the formal calculation of Sect. \ref{Subsection_Formal_Calculation}. Let us demonstrate, how they appear, by considering the integral

\begin{equation}\label{Toy_Integral}
I \equiv \int \frac{d^4Q}{(2\pi)^4} \frac{\partial}{\partial Q^\mu} \frac{\partial}{\partial Q_\mu} \Big[\frac{f(Q^2)}{Q^2}\Big] = -2 \int \frac{d^4Q}{(2\pi)^4} \frac{\partial}{\partial Q^\mu} \Big[\frac{Q^\mu}{Q^4}\, \Big(f(Q^2) - Q^2 f'(Q^2)\Big)\Big]
\end{equation}

\noindent
as a simple example. In Eq. (\ref{Toy_Integral}) $Q_\mu$ denotes the Euclidean momentum, and $f(Q^2)$ is a nonsingular function which rapidly tends to 0 in the limit $Q^2\to\infty$.

If we calculate the integral (\ref{Toy_Integral}) formally, then it vanishes, because it is an integral of a total derivative. Actually, using the divergence theorem, we reduce the integral under consideration to the integral over the infinitely large sphere $S^3_\infty$ in the momentum space. Evidently, the result is equal to 0, because the function $f$ vanishes on this sphere,

\begin{equation}\label{I_Formal_Calculation}
I \to \mbox{(formally)} \to - \frac{1}{8\pi^4} \oint\limits_{S^3_\infty} dS_\mu\, \frac{Q^\mu}{Q^4}\, \Big(f(Q^2) - Q^2 f'(Q^2)\Big) = 0,
\end{equation}

\noindent
where $dS_\mu$ is the integration measure on $S^3_\infty$. Actually, in Sect. \ref{Subsection_Formal_Calculation} we made a similar calculation. However, the result obtained in Eq. (\ref{I_Formal_Calculation}) is evidently incorrect due to a singularity of the integrand at $Q^\mu=0$.

To correct the above calculation, it is necessary to surround the singularity by a sphere $S^3_\varepsilon$ of an infinitely small radius $\varepsilon$ (with the inward-pointing normal) and take into account the integral over this sphere,

\begin{eqnarray}\label{I_Calculation}
&& I = - \frac{1}{8\pi^4} \oint\limits_{S^3_\infty} dS_\mu\, \frac{Q^\mu}{Q^4}\, \Big(f(Q^2) - Q^2 f'(Q^2)\Big) - \frac{1}{8\pi^4} \oint\limits_{S^3_\varepsilon} dS_\mu\, \frac{Q^\mu}{Q^4}\, \Big(f(Q^2) - Q^2 f'(Q^2)\Big)\qquad
\nonumber\\
&& = \frac{1}{8\pi^4} \oint\limits_{S^3_\varepsilon} dS\, \frac{1}{Q^3}\, \Big(f(Q^2) - Q^2 f'(Q^2)\Big) = \frac{1}{4\pi^2} f(0).
\end{eqnarray}

Let us visualize this result by reobtaining it in a different way. First, we note that defining the integral $I$ we actually do not distinguish between the expression (\ref{Toy_Integral}) and the integral

\begin{equation}
I = -2 \int \frac{d^4Q}{(2\pi)^4}\, \frac{Q^\mu}{Q^4}\, \frac{\partial}{\partial Q^\mu} \Big(f(Q^2) - Q^2 f'(Q^2)\Big).
\end{equation}

\noindent
However, it is possible to introduce the operator $\bm{\partial}/\bm{\partial Q^\mu}$ which is similar to $\partial/\partial Q^\mu$, but, by definition, the integral of it is always reduced to the integral over the sphere $S^3_\infty$ only. Moreover, we assume that this operator is commuted with $Q^\mu/Q^4$ in the integrand with the help of the identity

\begin{equation}\label{Delta_Identity}
\Big[\frac{\bm{\partial}}{\bm{\partial Q^\mu}},\, \frac{Q^\mu}{Q^4}\Big] = \frac{\bm{\partial}}{\bm{\partial Q^\mu}}\Big(\frac{Q^\mu}{Q^4}\Big) = 2\pi^2 \delta^4(Q).
\end{equation}

\noindent
In terms of the operator $\bm{\partial}/\bm{\partial Q^\mu}$ the considered integral is defined as

\begin{equation}
I \equiv -2 \int \frac{d^4Q}{(2\pi)^4}\, \frac{Q^\mu}{Q^4}\, \frac{\bm{\partial}}{\bm{\partial Q^\mu}} \Big(f(Q^2) - Q^2 f'(Q^2)\Big).
\end{equation}

\noindent
Then, if we integrate by parts taking into account vanishing of the integral of a total derivative and Eq. (\ref{Delta_Identity}), we obtain

\begin{eqnarray}
&& I = -2 \int \frac{d^4Q}{(2\pi)^4}\,\left\{\frac{\bm{\partial}}{\bm{\partial Q^\mu}}\Big[\frac{Q^\mu}{Q^4}\,  \Big(f(Q^2) - Q^2 f'(Q^2)\Big)\Big] - \frac{\bm{\partial}}{\bm{\partial Q^\mu}}\Big(\frac{Q^\mu}{Q^4}\Big) \Big(f(Q^2) - Q^2 f'(Q^2)\Big) \right\}\qquad\nonumber\\
&& = 0 + 4\pi^2 \int \frac{d^4Q}{(2\pi)^4}\, \delta^4(Q) \Big(f(Q^2) - Q^2 f'(Q^2)\Big) = \frac{1}{4\pi^2} f(0).
\end{eqnarray}

\noindent
From this equation we see that the integral $I$ is determined by a contribution of the $\delta$-singularity.

Note that in the coordinate representation

\begin{equation}\label{Bold_Derivatives}
\int \frac{d^4Q}{(2\pi)^4} \frac{\bm{\partial}^2 a}{\bm{\partial Q^\mu\, \partial Q_\mu}} = -i\, \mbox{Tr}\, \big[x_\mu,\big[x^\mu, a\big]\big] = 0,
\end{equation}

\noindent
where $a$ is a certain function, while

\begin{equation}\label{Usual_Detivatives}
\int \frac{d^4Q}{(2\pi)^4} \frac{\partial^2 a}{\partial Q^\mu\, \partial Q_\mu} = -i\, \mbox{Tr}\, \big[x_\mu,\big[x^\mu, a\big]\big] -\mbox{singularities} = -\mbox{singularities}.
\end{equation}

\noindent
Such a structure of loop integrals appears in the Abelian case (see, e.g., \cite{Stepanyantz:2011jy}). In the non-Abelian case the structure analogous to (\ref{Bold_Derivatives}) is the right hand side of Eq. (\ref{Double_Total_Derivative}), while its left hand side is an analog of the expression (\ref{Usual_Detivatives}). Therefore, it becomes clear that making the calculations formally in the previous section we ignored the $\delta$-singularities. Thus, to make the calculation properly, it is necessary to take into account singular contributions, which generate all terms containing the anomalous dimensions in the NSVZ equation (\ref{NSVZ_Second_Form}) for RGFs defined in terms of the bare couplings. We hope to describe how to sum these singularities in a future publications.

\section{Verification in the lowest orders}
\hspace*{\parindent}\label{Section_Examples}

To confirm the correctness of the general arguments presented above, it is desirable to verify them by explicit calculations in the lowest orders. In Sect. \ref{Subsection_Double_Total_Derivatives} we have formulated the prescription, how to construct integrals of double total derivatives which appear in calculating the $\beta$-function in the case of using the higher covariant derivative regularization. For obtaining these integrals one usually calculates a set of superdiagrams which are obtained from a given graph by attaching two external lines of the background gauge superfield in all possible ways. For example, in Ref. \cite{Shakhmanov:2017soc} this has been done for the three-loop contributions quartic in the Yukawa couplings. All three-loop terms containing the Yukawa couplings have been subsequently found in Ref. \cite{Kazantsev:2018nbl}. (Both these calculations were made in the Feynman gauge $\xi=1$ for the higher derivative regulator $K=R$.) Unfortunately, at present no other three-loop contributions to the $\beta$-function are known in the case of using the higher covariant derivative regularization. Nevertheless, the results of Refs. \cite{Shakhmanov:2017soc,Kazantsev:2018nbl} allow verifying the general argumentation of the present paper by comparing the algorithm described in Sect. \ref{Subsection_Double_Total_Derivatives} with the result of the standard calculation.

\begin{figure}[h]
\begin{picture}(0,9.2)
\put(0.5,5.3){\includegraphics[scale=0.2]{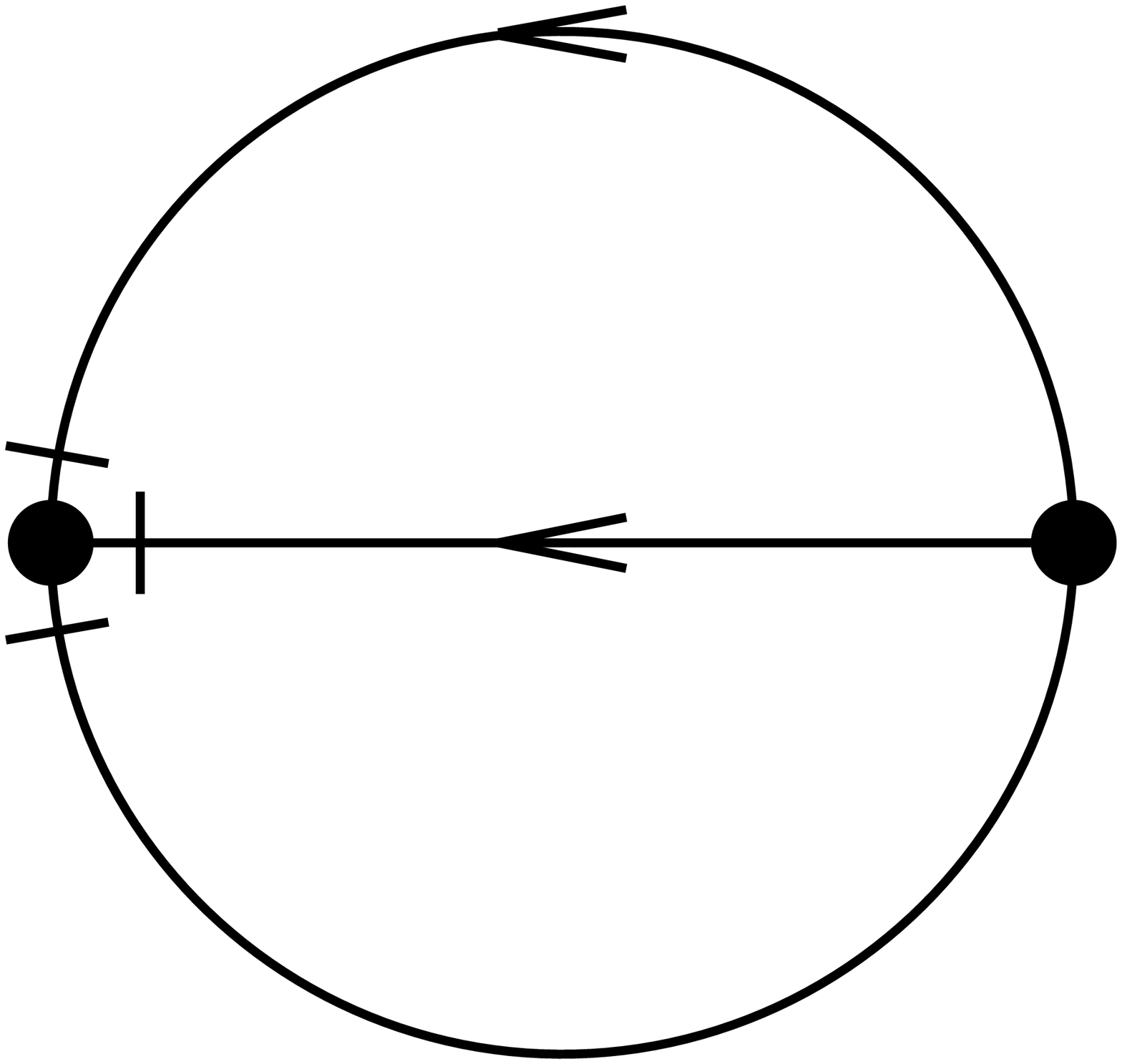}}
\put(0.3,8.7){$(1)$} \put(2.2,8.2){$K^\mu$} \put(2.2,6.5){$Q^\mu$} \put(4.1,7.3){$m$} \put(3.5,6.7){$n$}
\put(6.2,5.2){\includegraphics[scale=0.2]{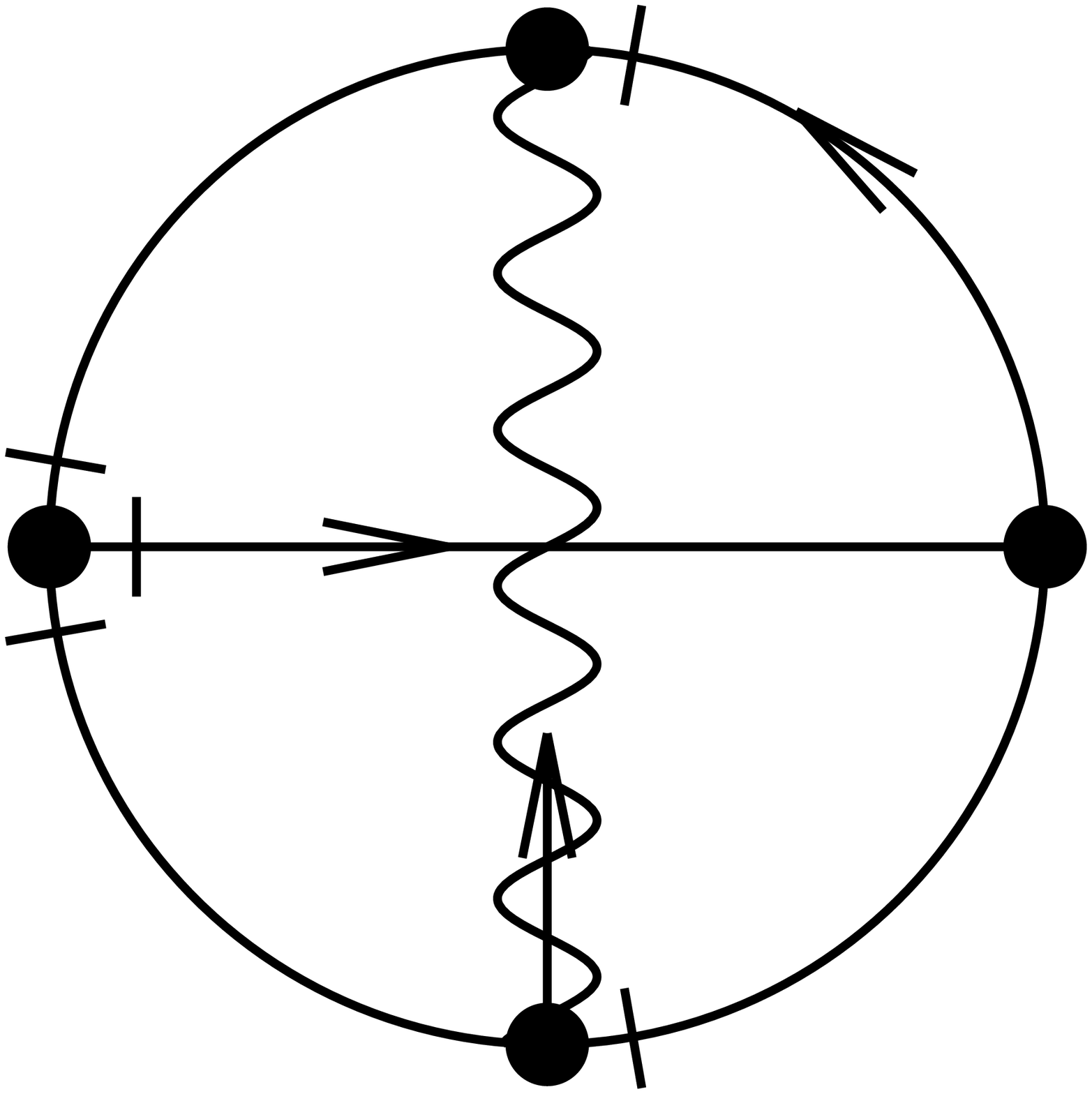}}
\put(6.0,8.7){$(2)$} \put(9.4,8.4){$Q^\mu$} \put(8.4,6.1){$K^\mu$} \put(7.1,6.5){$L^\mu$}
\put(6.7,7.3){$i$} \put(9.9,7.3){$m$}
\put(12,5.2){\includegraphics[scale=0.201]{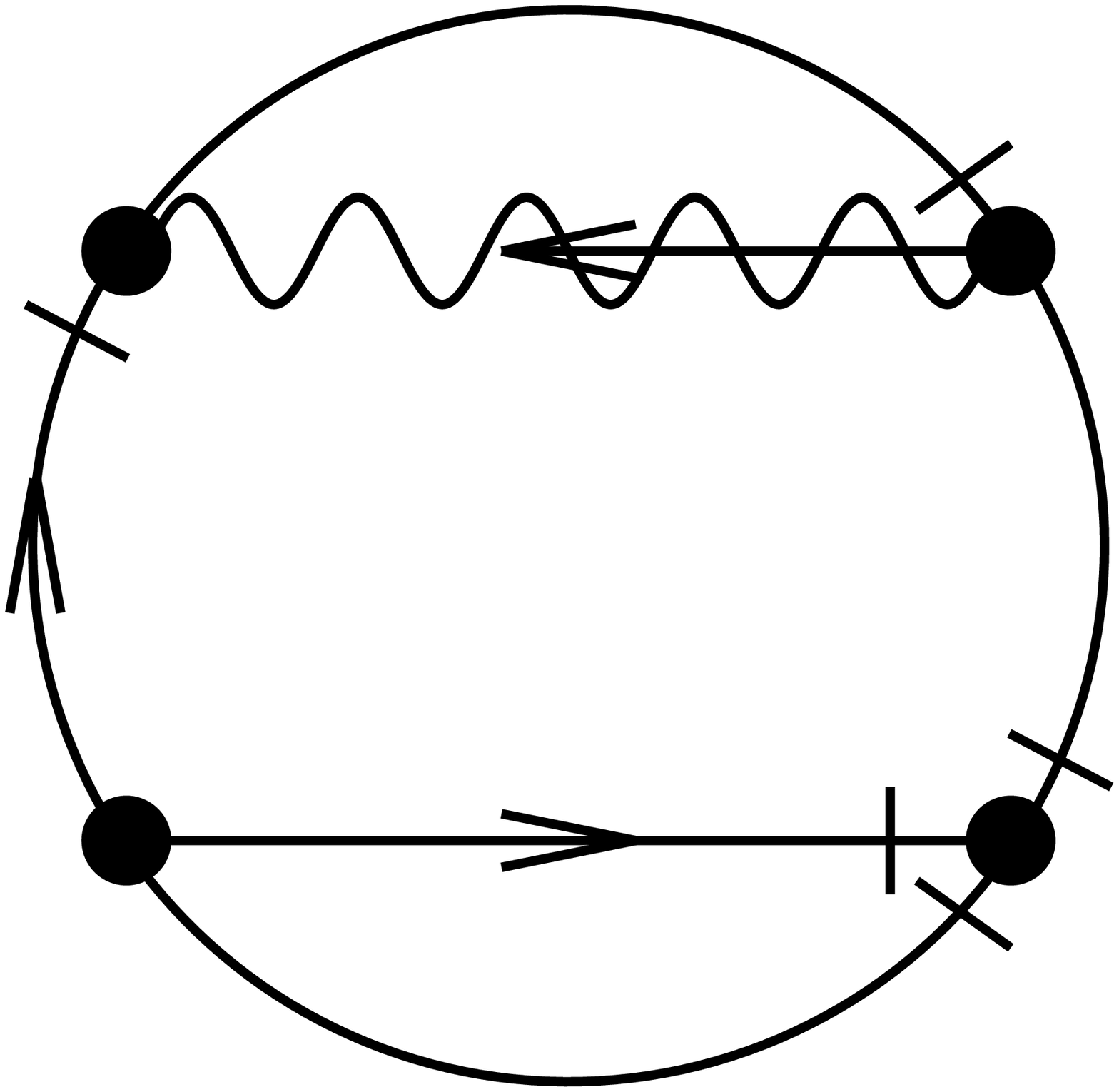}}
\put(11.8,8.7){$(3)$}  \put(13.5,7.3){$K^\mu$}  \put(13.6,6.2){$L^\mu$} \put(12.3,6.8){$Q^\mu$}
\put(15.3,8.0){$B$}  \put(12.0,6.1){$l$} \put(12.6,6.1){$j$}
\put(3.5,0){\includegraphics[scale=0.35]{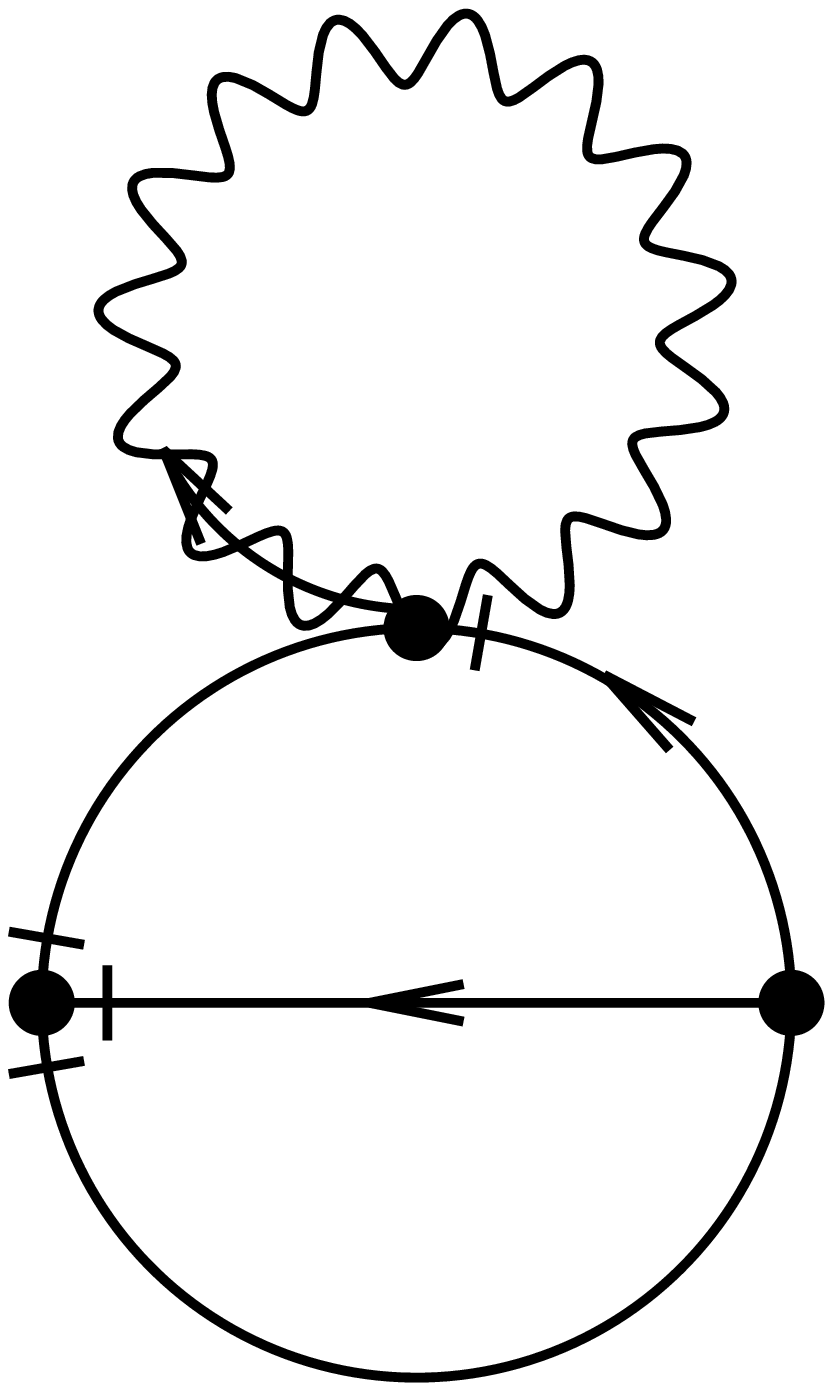}}
\put(2.8,3.6){$(4)$} \put(3.4,2.8){$K^\mu$} \put(6.0,2.4){$Q^\mu$} \put(4.8,0.88){$L^\mu$}
\put(4.8,2.18){$B$} \put(6.4,1.6){$l$} \put(5.8,1.0){$j$}
\put(9,0.2){\includegraphics[scale=0.2]{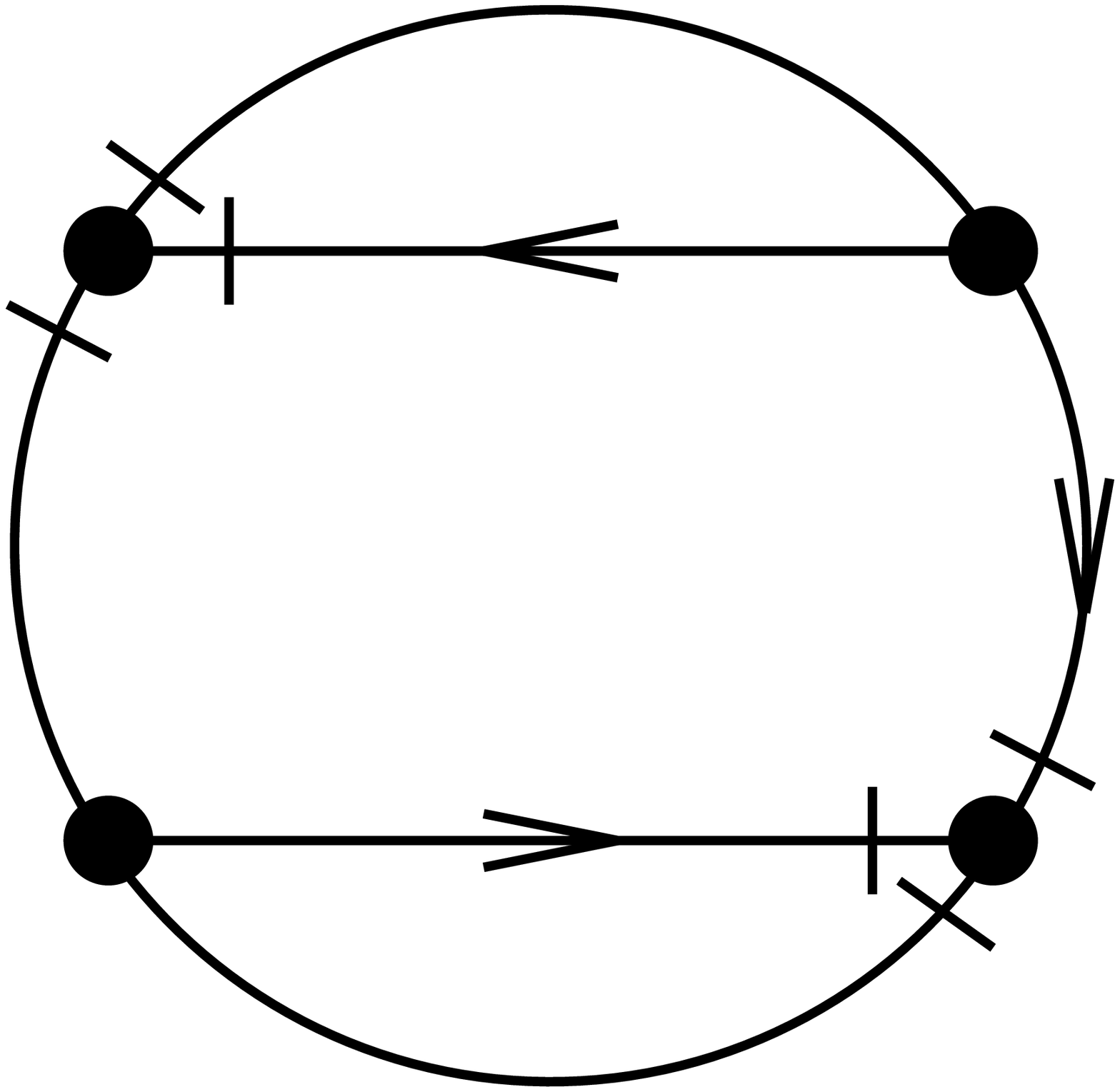}}
\put(8.8,3.6){$(5)$}  \put(11.7,1.8){$K^\mu$} \put(10.5,2.4){$Q^\mu$} \put(10.5,1.2){$L^\mu$}
\put(11.7,2.5){$i$} \put(12.4,2.5){$l$} \put(9.5,1.18){$m$}
\put(7.9,4.8){$B$}
\end{picture}
\caption{Graphs generating terms containing the Yukawa couplings in the three-loop $\beta$-function. We point out independent momenta and indices corresponding to beginnings of the respective propagators using the same notations as in the calculation described in the text.}\label{Figure_Yukawa}
\end{figure}

A part of the three-loop $\beta$-function which contains the Yukawa couplings originates from the supergraphs presented in Fig. \ref{Figure_Yukawa}. Within the standard technique used in Refs. \cite{Shakhmanov:2017soc,Kazantsev:2018nbl} they generate large sets of superdiagrams with two external lines corresponding to the background gauge superfield which have to be calculated. However, now it is possible to derive the result for their sums by a different (and much simpler) way. Namely, we should calculate the (specially modified) superdiagrams without external lines and, after this, follow the algorithm described in Sect. \ref{Subsection_Double_Total_Derivatives}. Here we describe this calculation for the graph $(1)$ in details and present the similar results for the remaining graphs $(2)$ --- $(5)$.

As a starting point we find the contribution of the graph $(1)$ to the expression (\ref{Starting_Expression}). Due to the derivatives with respect to the superfields $\mbox{\sl g}$ and $\mbox{\sl g}^*$ and subsequent integrations, two vertices in this graph take the form

\begin{equation}
\frac{1}{6} \lambda_0^{ijk} \int d^6z_1\, \theta^2\,\big(v^A\big)^2 \phi_i \phi_j \phi_k\qquad\mbox{and}\qquad \frac{1}{6} \lambda^*_{0pmn} \int d^6\bar z_2\, \bar\theta^2\, \phi^{*p} \phi^{*m} \phi^{*n}.
\end{equation}

\noindent
Then, after some standard calculations, for the contribution of the supergraph (1) (in the Euclidean space after the Wick rotation) we obtain

\begin{equation}\label{Graph1_Integral}
\mbox{graph}(1) = \frac{2}{3}\,{\cal V}_4\, \frac{d}{d\ln\Lambda}\int \frac{d^4Q}{(2\pi)^4} \frac{d^4K}{(2\pi)^4}\, \lambda_0^{ijk} \lambda_{0ijk}^*\, \frac{1}{Q^2 F_Q K^2 F_K (Q+K)^2 F_{Q+K}}.
\end{equation}

\noindent
Note that although here the superfield $\mbox{\sl g}$ is set to 0, the coordinate independent parameter $g$ can in general be present in the Yukawa vertices and gauge propagators. However, the graph (1) appears to be independent on $g$ and, therefore, on $\rho = g g^*$.

According to the prescription described in Sect. \ref{Subsection_Double_Total_Derivatives} for obtaining the contribution to the $\beta$-function, at the first step, it is necessary to replace the factor $\lambda_0^{ijk} \lambda_{0ijk}^*$ (which in the original graph comes from the expression $\lambda_0^{ijk} \lambda_{0pmn}^*\, \delta_i^p\, \delta_j^m\, \delta_k^n$) by a certain differential operator acting on the integrand in Eq. (\ref{Graph1_Integral}). To construct this operator, we consider the propagators with the independent momenta $K^\mu$ and $Q^\mu$. Let they are proportional to $\delta_j^m$ and $\delta_k^n$, respectively. Then, we construct the second ``variation'' formally replacing

\begin{equation}
\delta\big(\delta_j^m\big)\ \to\ (T^A)_j{}^m \frac{\partial}{\partial K^\mu};\qquad \delta\big(\delta_k^n\big)\ \to\ (T^A)_k{}^n \frac{\partial}{\partial Q^\mu}.
\end{equation}

\noindent
This operation changes the Yukawa coupling dependent factor  in Eq. (\ref{Graph1_Integral}) as

\begin{eqnarray}
&&\hspace*{-8mm} \lambda_0^{ijk} \lambda_{0ijk}^*\ \to\ \lambda_0^{ijk} \lambda_{0imk}^* (T^A)_j{}^m \frac{\partial}{\partial K^\mu} + \lambda_0^{ijk} \lambda_{0ijn}^* (T^A)_k{}^n \frac{\partial}{\partial Q^\mu}\ \to\ \lambda_0^{ijk} \lambda_{0imk}^* C(R)_j{}^m \frac{\partial}{\partial K^\mu}\frac{\partial}{\partial K_\mu} \nonumber\\
&&\hspace*{-8mm} + 2 \lambda_0^{ijk} \lambda_{0imn}^* (T^A)_j{}^m (T^A)_k{}^n \frac{\partial}{\partial K^\mu}\frac{\partial}{\partial Q_\mu} + \lambda_0^{ijk} \lambda_{0ijn}^* C(R)_k{}^n \frac{\partial}{\partial Q^\mu} \frac{\partial}{\partial Q_\mu}.
\end{eqnarray}

\noindent
Replacing the factor $\lambda_0^{ijk} \lambda_{0ijk}^*$ in Eq. (\ref{Graph1_Integral}) by this operator and taking into account that the Euclidean momenta $K^\mu$ and $Q^\mu$ enter the integrand of Eq. (\ref{Graph1_Integral}) symmetrically, we obtain the expression

\begin{eqnarray}
&&\hspace*{-8mm} \frac{4}{3} {\cal V}_4\,  \frac{d}{d\ln\Lambda} \int \frac{d^4Q}{(2\pi)^4} \frac{d^4K}{(2\pi)^4}\, \lambda_0^{ijk} \lambda^*_{0imn} (T^A)_j{}^m (T^A)_k{}^n\, \frac{\partial}{\partial Q^\mu}\frac{\partial}{\partial K_\mu} \Big(\frac{1}{Q^2 F_Q K^2 F_K (Q+K)^2 F_{Q+K}}\Big)\nonumber\\
&&\hspace*{-8mm} + \frac{4}{3} {\cal V}_4\,  \frac{d}{d\ln\Lambda} \int \frac{d^4Q}{(2\pi)^4} \frac{d^4K}{(2\pi)^4}\, \lambda_0^{ijk} \lambda^*_{0ijl} C(R)_k{}^l\, \frac{\partial}{\partial Q^\mu}\frac{\partial}{\partial Q_\mu} \Big(\frac{1}{Q^2 F_Q K^2 F_K (Q+K)^2 F_{Q+K}}\Big).
\end{eqnarray}

\noindent
To simplify it, we use two identities. The first one,

\begin{equation}
\lambda_0^{ijk} \lambda^*_{0imn} (T^A)_j{}^m (T^A)_k{}^n = -\frac{1}{2} \lambda_0^{ijk} \lambda^*_{0ijl} C(R)_k{}^l,
\end{equation}

\noindent
follows from Eq. (\ref{Yukawa_Invariance}), while the second one,

\begin{eqnarray}
&& \frac{d}{d\ln\Lambda}\int \frac{d^4Q}{(2\pi)^4} \frac{d^4K}{(2\pi)^4}\, \frac{\partial}{\partial Q^\mu}\frac{\partial}{\partial K_\mu} \Big(\frac{1}{Q^2 F_Q K^2 F_K (Q+K)^2 F_{Q+K}}\Big)\nonumber\\
&&\qquad\qquad\quad = \frac{1}{2} \frac{d}{d\ln\Lambda}\int \frac{d^4Q}{(2\pi)^4} \frac{d^4K}{(2\pi)^4}\, \frac{\partial}{\partial Q^\mu}\frac{\partial}{\partial Q_\mu} \Big(\frac{1}{Q^2 F_Q K^2 F_K (Q+K)^2 F_{Q+K}}\Big),\qquad
\end{eqnarray}

\noindent
can be verified by direct differentiating after some changes of integration variables in the resulting integrals. Then the expression under consideration takes the form

\begin{equation}
{\cal V}_4\,\frac{d}{d\ln\Lambda} \int \frac{d^4Q}{(2\pi)^4} \frac{d^4K}{(2\pi)^4}\, \lambda_0^{ijk} \lambda^*_{0ijl} C(R)_k{}^l\, \frac{\partial}{\partial Q^\mu}\frac{\partial}{\partial Q_\mu} \Big(\frac{1}{Q^2 F_Q K^2 F_K (Q+K)^2 F_{Q+K}}\Big).
\end{equation}

\noindent
To find the contribution to the function $\beta(\alpha_0,\lambda_0 \lambda_0^*,Y_0)/\alpha_0^2$, it is necessary to multiply this expression by $-2\pi/r {\cal V}_4$ and apply the operator

\begin{equation}\label{Integration_Operator}
\int\limits_{+0}^1 \frac{d\rho}{\rho} \int\limits_{+0}^\rho d\rho
\end{equation}

\noindent
to the result. For the graph (1) this integration gives the factor 1, because the expression for this graph does not depend on $\rho$. Therefore,

\begin{equation}\label{Graph1_Result}
\Delta_1\Big(\frac{\beta}{\alpha_0^2}\Big) = - \frac{2\pi}{r} \frac{d}{d\ln\Lambda} \int \frac{d^4Q}{(2\pi)^4} \frac{d^4K}{(2\pi)^4}\, \lambda_0^{ijk} \lambda^*_{0ijl} C(R)_k{}^l \frac{\partial}{\partial Q_\mu} \frac{\partial}{\partial Q^\mu} \Big(\frac{1}{Q^2 F_Q K^2 F_K (Q+K)^2 F_{Q+K}}\Big).
\end{equation}

\noindent
This result exactly coincides with the one derived in Ref. \cite{Shakhmanov:2017soc} by direct summation of the superdiagrams contributing to the two-point Green function of the background gauge superfield. Certainly, the calculation described here is much simpler, because we had to calculate the only superdiagram without external lines. The agreement of the results confirms the correctness of the general arguments presented in this paper. However, it is desirable to verify also the three-loop results corresponding to the graphs $(2)$ --- $(5)$ in Fig. \ref{Figure_Yukawa}. As in Refs. \cite{Shakhmanov:2017soc,Kazantsev:2018nbl} we will use the Feynman gauge, so that in what follows the parameter $\xi_0$ is set to $1$ and the higher derivative regulator $K$ is chosen equal to $R$.

\begin{figure}[h]
\begin{picture}(0,4)
\put(0.2,1.7){${\displaystyle \frac{1}{(3-1)^2}\, \left\{\begin{array}{c} \vphantom{1}\\ $\vphantom{1}$\\ $\vphantom{1}$\\ $\vphantom{1}$\\ $\vphantom{1}$\\ $\vphantom{1}$ \\ $\vphantom{1}$
\end{array} \hspace*{9.3cm}\right\} }$}
\put(2.8,2){\includegraphics[scale=0.09]{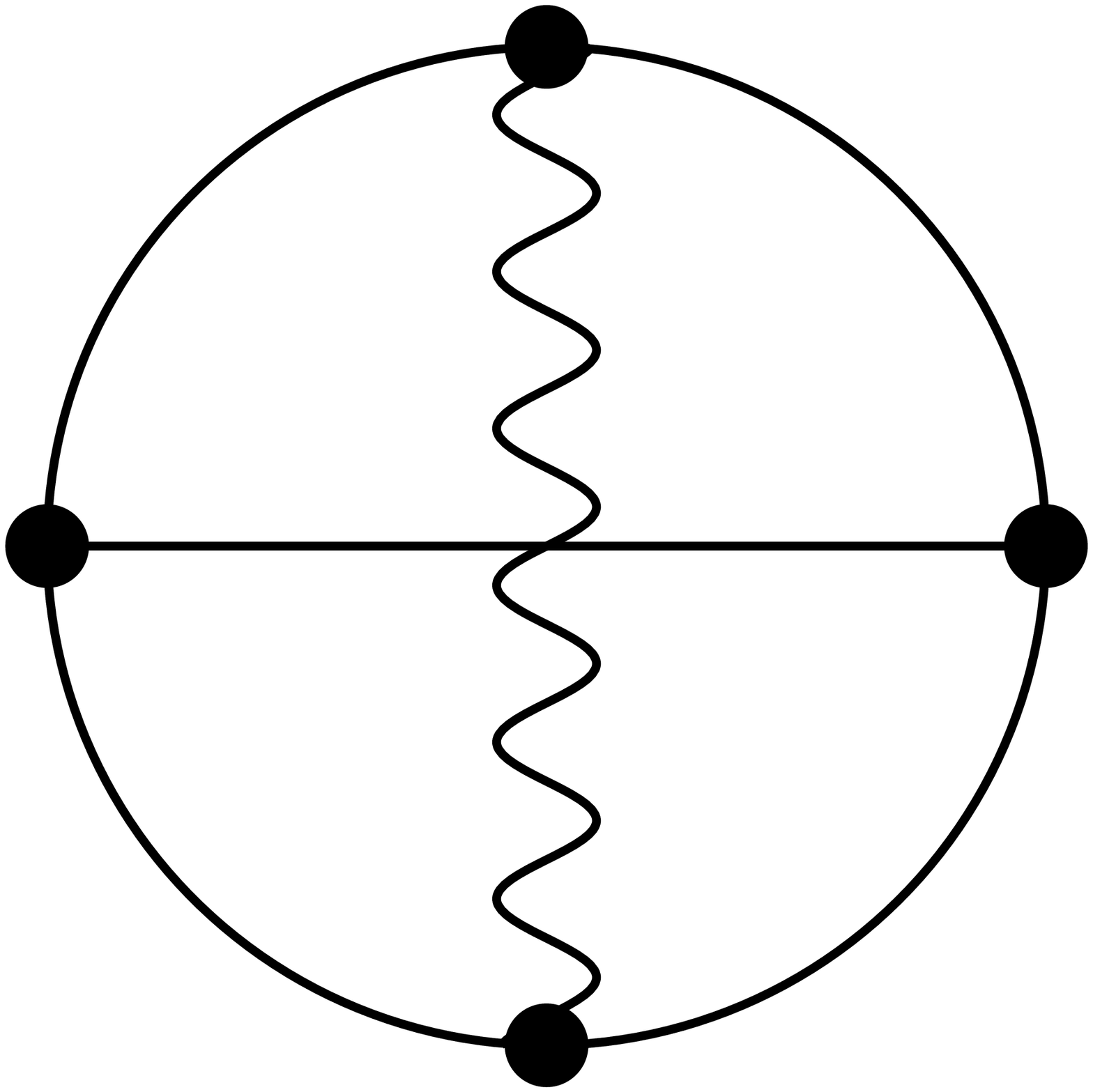}}
\put(2.4,2.8){$\theta^2$} \put(4.65,2.8){$\bar\theta^2$}
\put(6.0,2){\includegraphics[scale=0.09]{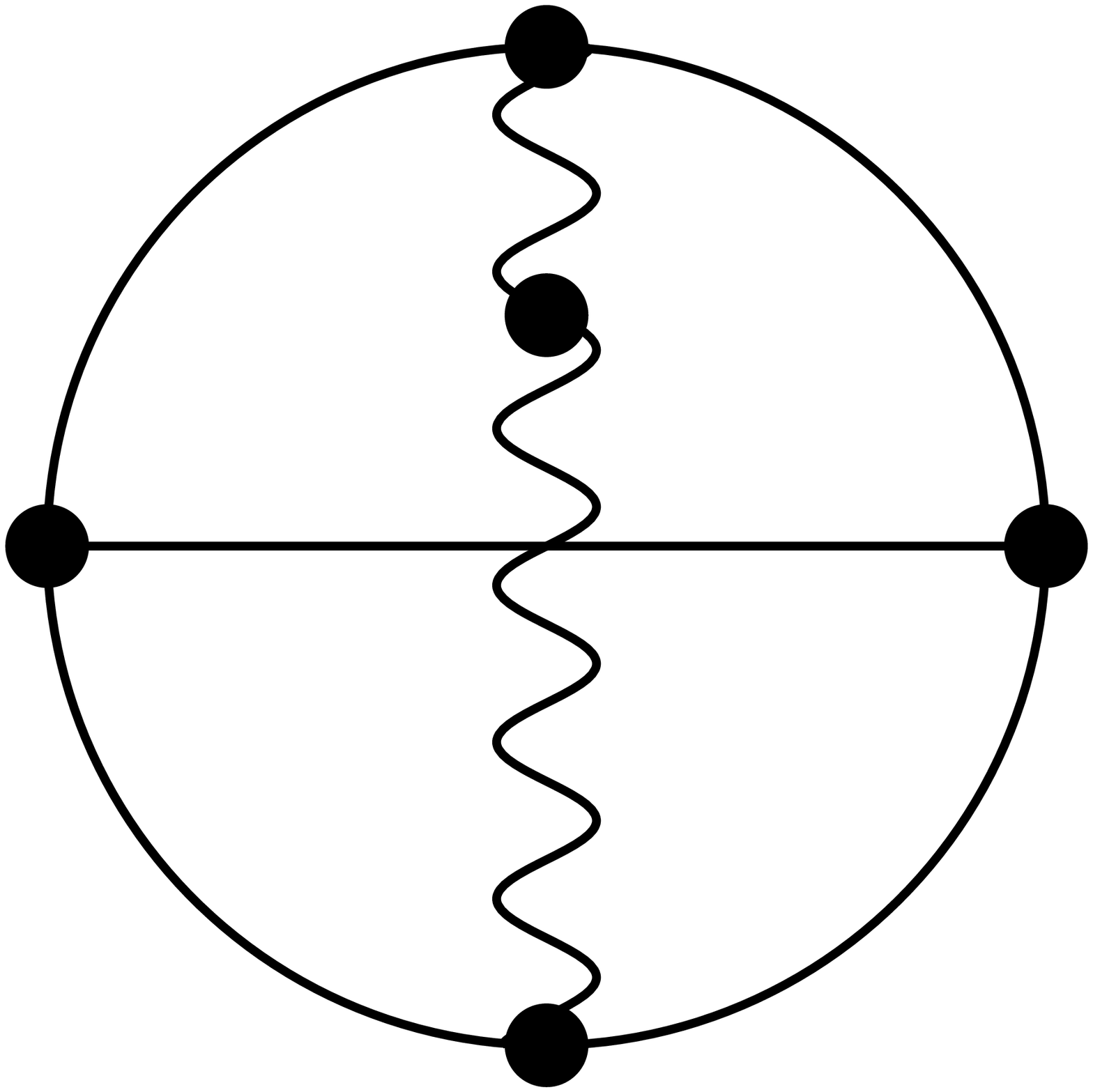}}
\put(5.6,2.8){$\theta^2$} \put(7.05,3.1){$\bar\theta^2$}
\put(9.2,2){\includegraphics[scale=0.09]{graph2d2.eps}}
\put(9.6,3.1){$\theta^2$} \put(11.0,2.8){$\bar\theta^2$}
\put(4.4,0){\includegraphics[scale=0.09]{graph2d2.eps}}
\put(4.8,1.1){$\theta^4$}
\put(7.8,0){\includegraphics[scale=0.09]{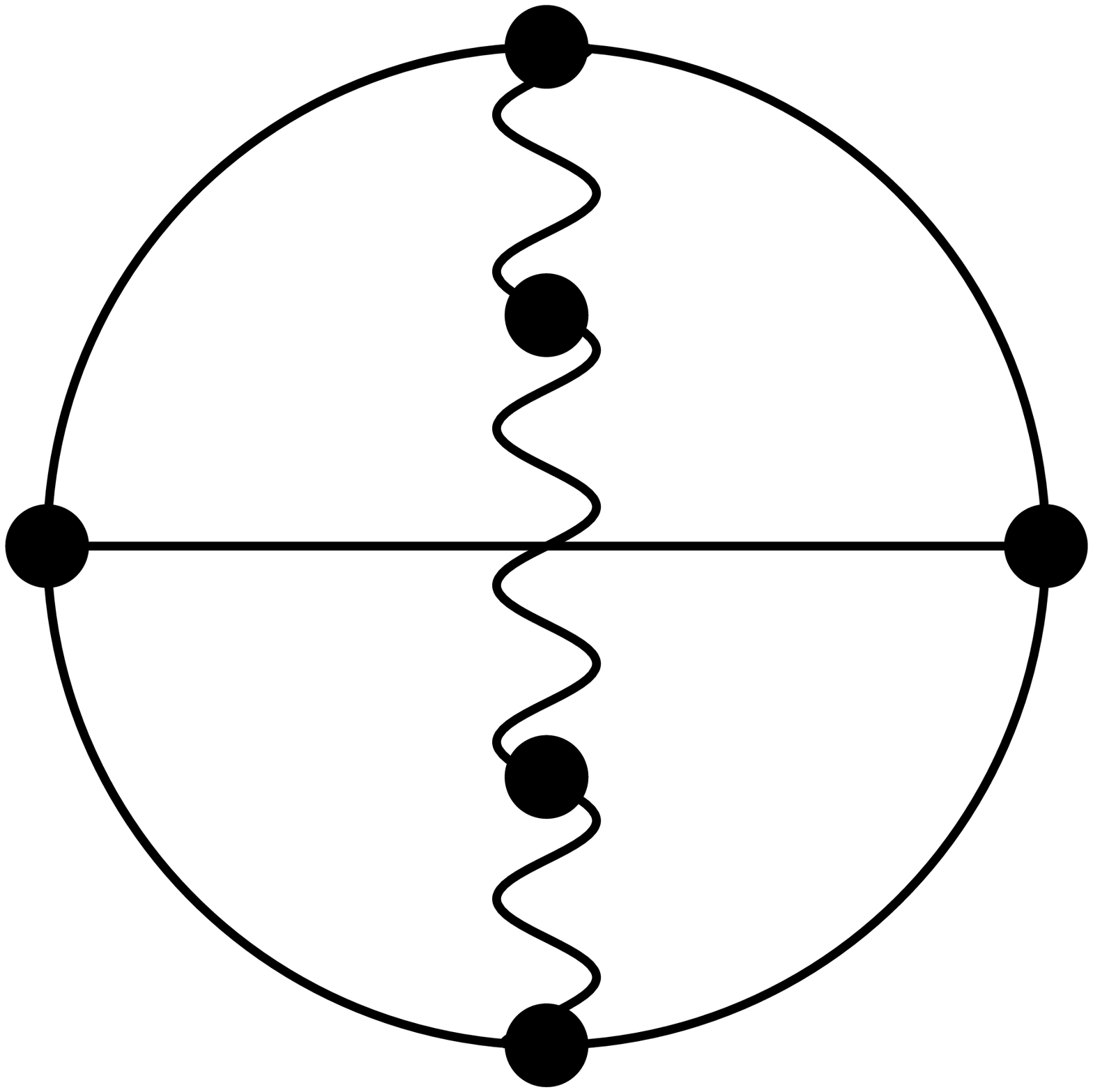}}
\put(8.2,0.4){$\theta^2$} \put(8.8,1.1){$\bar\theta^2$}
\put(12.6,1.8){\vector(1,0){0.7}}
\put(13.8,0.9){\includegraphics[scale=0.09]{graph2d1.eps}}
\put(14.6,0.5){$\theta^4$}
\end{picture}
\caption{Subgraphs of the supergraph (2) correspond to different positions of $\theta^2$ and $\bar\theta^2$. However, the sum of them is effectively reduced to a single supergraph in which $\theta^4$ can be placed in an arbitrary point and $g=g^*=1$.}\label{Figure_Subdiagrams_Of_Graph2}
\end{figure}

Calculating the supergraph (2) in Fig. \ref{Figure_Yukawa} we should take into account that $\theta^2$ and $\bar\theta^2$ can appear in different points. This produces a set of subgraphs presented in the curly brackets in Fig. \ref{Figure_Subdiagrams_Of_Graph2}. However, all these subgraphs differ only in the numeric coefficients. Really, they are quartic in $\theta$-s, so that these $\theta$-s can be shifted to an arbitrary point of the supergraph. (Terms with lower degrees of $\theta$, which can appear after such shifts, evidently vanish due to the integration over $d^4\theta$.) For example, it is possible to shift $\theta$-s as it is shown in the right hand side of Fig. \ref{Figure_Subdiagrams_Of_Graph2}.\footnote{If we consider an $L$ loop supergraph without external lines contributing to the effective action, then the terms which do not contain the derivatives of $\mbox{\sl g}$ and $\mbox{\sl g}^*$ are proportional to $(\bm{g} \bm{g}^*)^{L-1}$. Therefore, the corresponding contribution to the expression (\ref{Starting_Expression}) is obtained by inserting a factor $(L-1)^2 \theta^4$ to an arbitrary point of the supergraph containing the integration over the full superspace, see Fig. \ref{Figure_Subdiagrams_Of_Graph2} as an illustration. (The numerical coefficient should be calculated before the insertion of $\theta^4$.)}

The result for their sum (in the Euclidean space after the Wick rotation) can be written as

\begin{eqnarray}
&& \mbox{graph}(2) = 16 {\cal V}_4\, g g^*  \frac{d}{d\ln\Lambda} \int\frac{d^4Q}{(2\pi)^4}\frac{d^4L}{(2\pi)^4}\frac{d^4K}{(2\pi)^4}\, e_0^2 \lambda_0^{ijk} \lambda^*_{0imn} (T^B)_j{}^m (T^B)_k{}^n \frac{1}{K^2 R_{K} L^2 F_{L}}\qquad\nonumber\\
&& \times \frac{N(Q,K,L)}{Q^2 F_{Q} (Q+K)^2 F_{Q+K}(Q-L)^2 F_{Q-L} (Q+K-L)^2 F_{Q+K-L}},
\end{eqnarray}

\noindent
where, following Ref. \cite{Kazantsev:2018nbl}, we use the notation

\begin{eqnarray}\label{N_Function}
&& N(Q,K,L) \equiv L^2 F_{Q+K} F_{Q+K-L} -Q^2 \Big((Q+K)^2 -L^2\Big) F_{Q+K-L} \frac{F_{Q+K}-F_{Q}}{(Q+K)^2-Q^2}  \nonumber\\
&& -(Q-L)^2 \Big((Q+K-L)^2-L^2\Big) F_{Q+K} \frac{F_{Q+K-L}-F_{Q-L}}{(Q+K-L)^2-(Q-L)^2} +Q^2 (Q-L)^2 \\
&& \times \Big(L^2-(Q+K)^2-(Q+K-L)^2\Big) \left(\frac{F_{Q+K}-F_{Q}}{(Q+K)^2-Q^2}\right) \left(\frac{F_{Q+K-L}-F_{Q-L}}{(Q+K-L)^2-(Q-L)^2}\right).\qquad\nonumber
\end{eqnarray}

\noindent
As earlier, we should replace the factor $\lambda_0^{ijk} \lambda^*_{0imn} (T^B)_j{}^m (T^B)_k{}^n$ by a relevant differential operator. For constructing this differential operator we again mark the propagators with the independent momenta $Q_\mu$, $L_\mu$, and $K_\mu$, see Fig. \ref{Figure_Yukawa}. The beginnings of the lines which denote them correspond to the indices $m$, $i$, and $B$. They refer to the representations $R$ (in which the matter superfields lie), $\bar R$, and $Adj$, respectively. Then, the calculation of the first ``variation'' gives

\begin{eqnarray}
&& \lambda_0^{ijk} \lambda^*_{0imn} (T^B)_j{}^m (T^B)_k{}^n\ \to\  \lambda_0^{ijk} \lambda^*_{0ipn} (T^A)_m{}^p (T^B)_j{}^m (T^B)_k{}^n \frac{\partial}{\partial Q^\mu}\nonumber\\
&&\qquad - \lambda_0^{pjk} \lambda^*_{0imn} (T^A)_p{}^i (T^B)_j{}^m (T^B)_k{}^n \frac{\partial}{\partial L^\mu} -i \lambda_0^{ijk} \lambda^*_{0imn} (T^B)_j{}^m f^{ABC} (T^C)_k{}^n \frac{\partial}{\partial K^\mu},\qquad\quad
\end{eqnarray}

\noindent
where we take into account that $T_{\bar R}^A = -\big(T^A\big)^t$ (with $T^A$ being the generators of the representation $R$) and $\big(T^A_{Adj}\big)_{BC} = -if^{ABC}$. The second ``variation'' is calculated in a similar way. After some (rather non-trivial) transformations involving Eq. (\ref{Yukawa_Invariance}) we obtain that the differential operator for the considered graph has the form

\begin{eqnarray}\label{Graph2_Operator}
&& \hspace*{-9mm} \lambda_0^{ijk} \lambda^*_{0imn} (T^B)_j{}^m (T^B)_k{}^n\ \to\  \Big(\frac{1}{2} \lambda_0^{ijk} \lambda^*_{0ijl} \big(C(R)^2\big)_k{}^l - \lambda_0^{ipq} \lambda^*_{0imn} C(R)_p{}^m C(R)_q{}^n \Big)\frac{\partial}{\partial L^\mu} \Big( \frac{\partial}{\partial L^\mu} \nonumber\\
&& \hspace*{-9mm} + \frac{\partial}{\partial Q^\mu} \Big) - \frac{1}{2} \lambda_0^{ijk} \lambda^*_{0ijl} \big(C(R)^2\big)_k{}^l \frac{\partial}{\partial Q^\mu} \frac{\partial}{\partial Q_\mu}  -\frac{1}{2} C_2 \lambda_0^{ijk} \lambda^*_{0ijl} C(R)_k{}^l \frac{\partial}{\partial K^\mu} \Big(\frac{\partial}{\partial K_\mu} - \frac{\partial}{\partial Q_\mu}\Big).
\end{eqnarray}

\noindent
Then it is necessary to repeat the same algorithm as for the graph (1), namely,

1. replace $\lambda_0^{ijk} \lambda^*_{0imn} (T^B)_j{}^m (T^B)_k{}^n$ by the operator (\ref{Graph2_Operator});

2. multiply the result by $-2\pi/r {\cal V}_4$;

3. apply the operator (\ref{Integration_Operator}).

\noindent
The three-loop supergraphs are proportional to $g g^* = \rho$, so that in the considered case the integration gives\footnote{In general, an $L$-loop supergraph is proportional to $\rho^{L-2}$, and the integration gives the factor $(L-1)^{-2}$. This implies that in the general case to find a contribution to Eq. (\ref{Starting_Expression}), it is possible to start with a vacuum supergraph contributing to the effective action with $g=g^*=1$ and simply insert $\theta^4$ to an arbitrary point which contains integration over the full superspace. (Note that the integrations over $d^6x$ or $d^6\bar x$ in the Yukawa terms can always be converted to the integrals over the full superspace.)}

\begin{equation}\label{Integral_Of_Rho}
\int\limits_{+0}^1 \frac{d\rho}{\rho} \int\limits_{+0}^\rho d\rho\, \rho = \frac{1}{4}.
\end{equation}

\noindent
Thus, the contribution of the graph (2) to the function $\beta/\alpha_0^2$ takes the form

\begin{eqnarray}\label{Graph2_Result}
&&\hspace*{-9mm} \Delta_2\Big(\frac{\beta}{\alpha_0^2}\Big) = \frac{4\pi}{r} \frac{d}{d\ln\Lambda} \int\frac{d^4Q}{(2\pi)^4}\frac{d^4L}{(2\pi)^4}\frac{d^4K}{(2\pi)^4}\, e_0^2 \Biggl[ \lambda^{*}_{0lkj} \lambda_{0}^{lki} C_{2} C(R)_{i}{}^{j} \frac{\partial}{\partial K^\mu}\left(\frac{\partial}{\partial K_\mu} -\frac{\partial}{\partial Q_\mu}\right)\nonumber\\
&&\hspace*{-9mm}  -\Big( \lambda_{0jln}^{*} \lambda_{0}^{iln} \big(C(R)^2\big)_{i}{}^{j} - 2  \lambda^{*}_{0jln} \lambda_{0}^{imn} C(R)_{i}{}^{j} C(R)_{m}{}^{l} \Big) \frac{\partial}{\partial L^\mu}\left(\frac{\partial}{\partial L_\mu}+\frac{\partial}{\partial Q_\mu}\right) +  \lambda_{0jln}^{*} \lambda_{0}^{iln} \big(C(R)^2\big)_{i}{}^{j} \nonumber\\
&&\hspace*{-9mm} \times \frac{\partial}{\partial Q^\mu} \frac{\partial}{\partial Q_\mu}\Biggr] \frac{N(Q,K,L)}{K^2 R_{K} Q^2 F_{Q} (Q+K)^2 F_{Q+K} (Q+K-L)^2 F_{Q+K-L} (Q-L)^2 F_{Q-L} L^2 F_{L}}.
\end{eqnarray}

\noindent
We see that this result coincides with the one obtained in Ref. \cite{Kazantsev:2018nbl} by the straightforward calculation of superdiagrams with two external legs of the background gauge superfield.

The expression for the next graph (3) has the form

\begin{eqnarray}
&& \mbox{graph}(3) = 16 {\cal V}_4\, g g^* \frac{d}{d\ln\Lambda} \int \frac{d^4Q}{(2\pi)^4}\frac{d^4K}{(2\pi)^4} \frac{d^4L}{(2\pi)^4}\,  e_0^2 \lambda_0^{ijk} \lambda^*_{0ijl}\, \big(T^B\big)_k{}^m \big(T^B\big)_m{}^l\, \nonumber\\
&& \qquad\qquad\qquad\qquad\qquad\qquad\qquad \times\, \frac{L(Q,Q+K)}{K^2 R_K Q^2 F_Q^2 (Q+L)^2 F_{Q+L} (Q+K)^2 F_{Q+K} L^2 F_L},\qquad\qquad
\end{eqnarray}

\noindent
where

\begin{equation}\label{L_Function}
L(Q,P) \equiv F_Q F_P + \frac{F_P - F_Q}{P^2-Q^2}\Big(F_Q Q^2 + F_P P^2 \Big) + 2 Q^2 P^2 \left(\frac{F_P-F_Q}{P^2-Q^2}\right)^2.
\end{equation}

\noindent
Similar to the previous supergraphs, we replace the factor $\lambda_0^{ijk} \lambda^*_{0ijl}\, \big(T^B\big)_k{}^m \big(T^B\big)_m{}^l$ by a differential operator. To obtain this operator, we begin with calculating the first ``variation'' of the considered factor,

\begin{eqnarray}
&& \lambda_0^{ijk} \lambda^*_{0ijl}\, \big(T^B\big)_k{}^m \big(T^B\big)_m{}^l\ \to\  \lambda_0^{ijk} \lambda^*_{0ijp}\, \big(T^B\big)_k{}^m \big(T^B\big)_m{}^l (T^A)_l{}^p\frac{\partial}{\partial Q^\mu}\nonumber\\
&&\qquad\quad + \lambda_0^{ijk} \lambda^*_{0ipl}\, \big(T^B\big)_k{}^m \big(T^B\big)_m{}^l (T^A)_j{}^p\frac{\partial}{\partial L^\mu} - i \lambda_0^{ijk} \lambda^*_{0ijl}\, f^{ABC} \big(T^C\big)_k{}^m \big(T^B\big)_m{}^l \frac{\partial}{\partial K^\mu}.\qquad\qquad
\end{eqnarray}

\noindent
The second ``variation'' is constructed by a similar procedure. The result can be written in the form

\begin{eqnarray}\label{Second_Variation3}
&&\hspace*{-9mm} \lambda_0^{ijk} \lambda^*_{0ijl}\, \big(T^B\big)_k{}^m \big(T^B\big)_m{}^l\ \to\  \lambda_0^{ijk} \lambda^*_{0ijl}\, C_2 C(R)_k{}^l \frac{\partial}{\partial K^\mu}\Big(\frac{\partial}{\partial K_\mu} - \frac{\partial}{\partial Q_\mu} \Big) + \lambda_0^{ijk} \lambda^*_{0ijl}\, \big(C(R)^2\big)_k{}^l \nonumber\\
&&\hspace*{-9mm} \times\frac{\partial}{\partial Q^\mu}\Big(\frac{\partial}{\partial Q_\mu} - \frac{\partial}{\partial L_\mu} \Big) + \lambda_0^{ijk} \lambda^*_{0imn} C(R)_j{}^m C(R)_k{}^n \frac{\partial}{\partial L^\mu} \frac{\partial}{\partial L_\mu} + \frac{1}{2} \lambda_0^{ijk} \lambda^*_{0ijl} C_2 C(R)_k{}^l \frac{\partial}{\partial K^\mu} \frac{\partial}{\partial L_\mu}.\nonumber\\
\end{eqnarray}

\noindent
Proceeding according to the above described algorithm, we find the contribution of the supergraph (3) to the function $\beta/\alpha_0^2$,

\begin{eqnarray}\label{Graph3_Result}
&&\hspace*{-9mm} \Delta_3\Big(\frac{\beta}{\alpha_0^2}\Big) = -\frac{8\pi}{r} \frac{d}{d\ln\Lambda}\int\frac{d^4Q}{(2\pi)^4}\frac{d^4L}{(2\pi)^4}\frac{d^4K}{(2\pi)^4}\, e_0^2\Biggl[ \lambda^{*}_{0lkj} \lambda_{0}^{lki} C_{2}C(R)_{i}{}^{j} \frac{\partial}{\partial K^\mu}\biggl(\frac{\partial}{\partial K_\mu}-\frac{\partial}{\partial Q_\mu}\biggr)\nonumber\\
&&\hspace*{-9mm} + \lambda_{0jln}^{*} \lambda_{0}^{iln} \big(C(R)^2\big)_{i}{}^{j} \frac{\partial}{\partial Q^\mu}\biggl(\frac{\partial}{\partial Q_\mu}-\frac{\partial}{\partial L_\mu}\biggr) +  \lambda^{*}_{0jln} \lambda_{0}^{imn} C(R)_{i}{}^{j} C(R)_{m}{}^{l} \frac{\partial}{\partial L^\mu}\frac{\partial}{\partial L_\mu}\Biggr]\frac{1}{K^2 R_K}
\nonumber\\
&&\hspace*{-9mm} \times \frac{L(Q,Q+K)}{Q^2 F_Q^2 (Q+L)^2 F_{Q+L} (Q+K)^2 F_{Q+K} L^2 F_L}.
\end{eqnarray}

\noindent
Note that the last term in Eq. (\ref{Second_Variation3}) is not essential, because the corresponding contribution to $\beta/\alpha_0^2$ vanishes. (It changes the sign under the sequence of the variable changes $L^\mu \to L^\mu - Q^\mu$; $Q^\mu \to -Q^\mu$; $K^\mu \to - K^\mu$.) The result (\ref{Graph3_Result}) also coincides with the one obtained in Ref. \cite{Kazantsev:2018nbl}.

The expression for the supergraph (4) is

\begin{eqnarray}
&& \mbox{graph}(4) = -16 {\cal V}_4\, g g^* \frac{d}{d\ln\Lambda} \int \frac{d^4Q}{(2\pi)^4}\frac{d^4K}{(2\pi)^4} \frac{d^4L}{(2\pi)^4}\, e_0^2 \lambda_0^{ijk} \lambda^*_{0ijl}\, \nonumber\\
&& \qquad\qquad\qquad\qquad\qquad \times \big(T^B\big)_k{}^m \big(T^B\big)_m{}^l \frac{K(Q,K)}{K^2 R_K Q^2 F_Q^2 L^2 F_L (Q+L)^2 F_{Q+L}}.\qquad
\end{eqnarray}

\noindent
Here we use the same notation as in Ref. \cite{Kazantsev:2018nbl},

\begin{equation}\label{K_Function}
K(Q,K) \equiv \frac{F_{Q+K} - F_Q - 2Q^2 F_Q'/\Lambda^2}{(Q+K)^2 - Q^2} + \frac{2Q^2(F_{Q+K}-F_Q)}{\big((Q+K)^2-Q^2\big)^2},
\end{equation}

\noindent
where the prime and the subscript $Q$ denote the derivative with respect to $Q^2/\Lambda^2$. The corresponding operator is exactly the same as for the supergraph (3) and is given by Eq. (\ref{Second_Variation3}). Similarly to the case of the supergraph (3), the last term in this expression does not contribute to $\beta/\alpha_0^2$, so that

\begin{eqnarray}\label{Graph4_Result}
&&\hspace*{-9mm} \Delta_4\Big(\frac{\beta}{\alpha_0^2}\Big) = \frac{8\pi}{r} \frac{d}{d\ln\Lambda} \int \frac{d^4Q}{(2\pi)^4}\frac{d^4L}{(2\pi)^4} \frac{d^4K}{(2\pi)^4}\, e_0^2 \Biggl[ \lambda^{*}_{0lkj} \lambda_{0}^{lki} C_{2} C(R)_{i}{}^{j} \frac{\partial}{\partial K^\mu}\biggl(\frac{\partial}{\partial K_\mu}-\frac{\partial}{\partial Q_\mu}\biggr)\nonumber\\
&&\hspace*{-9mm} + \lambda_{0jln}^{*} \lambda_{0}^{iln} \big(C(R)^2\big)_{i}{}^{j} \frac{\partial}{\partial Q^\mu}\biggl(\frac{\partial}{\partial Q_\mu}-\frac{\partial}{\partial L_\mu}\biggr) + \lambda^{*}_{0jln} \lambda_{0}^{imn} C(R)_{i}{}^{j} C(R)_{m}{}^{l} \frac{\partial}{\partial L^\mu}\frac{\partial}{\partial L_\mu}\Biggr]\frac{1}{K^2 R_K}\nonumber\\
&&\hspace*{-9mm} \times \frac{K(Q,K)}{Q^2 F_Q^2 L^2 F_L (Q+L)^2 F_{Q+L}}.
\end{eqnarray}

\noindent
This result also agrees with the calculation of Ref. \cite{Kazantsev:2018nbl}.

The last supergraph (5) is given by the expression

\begin{eqnarray}\label{Graph5_Result}
&& \mbox{graph}(5) = -8 {\cal V}_4\, g g^* \frac{d}{d\ln\Lambda}\int \frac{d^4Q}{(2\pi)^4} \frac{d^4K}{(2\pi)^4}  \frac{d^4L}{(2\pi)^4} \lambda_0^{ijk} \lambda^*_{0ijl} \lambda_0^{mnl} \lambda^*_{0mnk}\nonumber\\
&&\qquad\qquad\qquad\qquad\qquad\qquad \times \frac{1}{Q^2 F_Q (K+Q)^2 F_{K+Q} L^2 F_L (K+L)^2 F_{K+L} K^2 F_K^2}.\qquad
\end{eqnarray}

\noindent
The first ``variation'' of the factor $\lambda_0^{ijk} \lambda^*_{0ijl} \lambda_0^{mnl} \lambda^*_{0mnk}$ is written as

\begin{eqnarray}
&& \lambda_0^{ijk} \lambda^*_{0ijl} \lambda_0^{mnl} \lambda^*_{0mnk}\ \to\ \lambda_0^{ijk} \lambda^*_{0pjl} \lambda_0^{mnl} \lambda^*_{0mnk} \big(T^A\big)_i{}^p \frac{\partial}{\partial Q^\mu}\qquad\nonumber\\
&&\qquad\qquad\qquad + \lambda_0^{ijk} \lambda^*_{0ijp} \lambda_0^{mnl} \lambda^*_{0mnk} \big(T^A\big)_l{}^p \frac{\partial}{\partial K^\mu} + \lambda_0^{ijk} \lambda^*_{0ijl} \lambda_0^{mnl} \lambda^*_{0pnk} \big(T^A\big)_m{}^p \frac{\partial}{\partial L^\mu}.\qquad
\end{eqnarray}

\noindent
The second ``variation'' can be found by a similar method, but, to simplify the resulting expression, it is necessary to involve the identities

\begin{eqnarray}
&&  \lambda_0^{ijk} \lambda^*_{0pjq} \lambda_0^{mnl} \lambda^*_{0mnk} \big(T^A\big)_i{}^p \big(T^A\big)_l{}^q = -\frac{1}{2} \lambda_0^{ijk} \lambda^*_{0ijp} \lambda_0^{mnl} \lambda^*_{0mnk} C(R)_l{}^p;\qquad\\
&& \lambda_0^{ijk} \lambda^*_{0pjl} \lambda_0^{mnl} \lambda^*_{0qnk} \big(T^A\big)_i{}^p \big(T^A\big)_m{}^q = \frac{1}{4} \lambda_0^{ijk} \lambda^*_{0ijp} \lambda_0^{mnl} \lambda^*_{0mnk} C(R)_l{}^p,
\end{eqnarray}

\noindent
which follow from Eq. (\ref{Yukawa_Invariance}). Using these identities and taking into account that the integrand of Eq. (\ref{Graph5_Result}) is symmetric in $Q$ and $L$, we find the required replacement

\begin{eqnarray}
&& \lambda_0^{ijk} \lambda^*_{0ijl} \lambda_0^{mnl} \lambda^*_{0mnk}\ \to\ 2\lambda_0^{ijk} \lambda^*_{0pjl} \lambda_0^{mnl} \lambda^*_{0mnk} C(R)_i{}^p \frac{\partial}{\partial Q^\mu} \frac{\partial}{\partial Q_\mu}\nonumber\\
&&\qquad\qquad\qquad + \lambda_0^{ijk} \lambda^*_{0ijp} \lambda_0^{mnl} \lambda^*_{0mnk} C(R)_l{}^p \Big(\frac{\partial}{\partial K^\mu} \frac{\partial}{\partial K_\mu} + \frac{1}{2} \frac{\partial}{\partial Q^\mu} \frac{\partial}{\partial L_\mu} - 2 \frac{\partial}{\partial K^\mu} \frac{\partial}{\partial Q_\mu}\Big).\qquad\quad
\end{eqnarray}

\noindent
Constructing the contribution of the graph (5) to the function $\beta/\alpha_0^2$ with the help of this operator and using the equations

\begin{eqnarray}
&& \frac{d}{d\ln\Lambda} \int \frac{d^4K}{(2\pi)^4} \frac{d^4L}{(2\pi)^4} \frac{d^4Q}{(2\pi)^4} \frac{\partial}{\partial Q^\mu} \frac{\partial}{\partial L_\mu}\nonumber\\
&&\qquad\qquad\qquad\qquad \times \frac{1}{K^2 F_K^2 Q^2 F_Q (Q+K)^2 F_{Q+K} L^2 F_L (L+K)^2 F_{L+K}} = 0;\\
&& \frac{d}{d\ln\Lambda} \int \frac{d^4K}{(2\pi)^4} \frac{d^4L}{(2\pi)^4} \frac{d^4Q}{(2\pi)^4} \frac{\partial}{\partial Q^\mu} \Big(2 \frac{\partial}{\partial K_\mu} - \frac{\partial}{\partial Q_\mu} \Big)\nonumber\\
&&\qquad\qquad\qquad\qquad \times \frac{1}{K^2 F_K^2 Q^2 F_Q (Q+K)^2 F_{Q+K} L^2 F_L (L+K)^2 F_{L+K}} = 0,\qquad\quad
\end{eqnarray}

\noindent
we obtain

\begin{eqnarray}
&&\hspace*{-9mm} \Delta_5\Big(\frac{\beta}{\alpha_0^2}\Big) = \frac{4\pi}{r} C(R)_i{}^j \frac{d}{d\ln\Lambda} \int \frac{d^4K}{(2\pi)^4} \frac{d^4L}{(2\pi)^4} \frac{d^4Q}{(2\pi)^4} \Biggl[\lambda_0^{iab} \lambda^*_{0kab} \lambda_0^{kcd} \lambda^*_{0jcd} \biggl(\frac{\partial}{\partial K^\mu} \frac{\partial}{\partial K_\mu} -\frac{\partial}{\partial Q^\mu} \frac{\partial}{\partial Q_\mu}\biggr) \nonumber\\
&&\hspace*{-9mm}  + 2\lambda_0^{iab} \lambda_{0jac}^* \lambda_0^{cde} \lambda^*_{0bde} \frac{\partial}{\partial Q^\mu}\frac{\partial}{\partial Q_\mu}\Biggr] \frac{1}{K^2 F_K^2 Q^2 F_Q (Q+K)^2 F_{Q+K} L^2 F_L (L+K)^2 F_{L+K}}.
\end{eqnarray}

\noindent
This expression also agrees with Refs. \cite{Shakhmanov:2017soc,Kazantsev:2018nbl}.

Thus, we see that the algorithm described in this paper allows reproducing all results obtained earlier by the direct summation of the superdiagrams with two external lines of the background gauge superfield. Certainly, this fact can be viewed as an evidence in favour of the correctness of the general consideration made in this paper.

\section{Conclusion}
\hspace*{\parindent}

We have proved that for ${\cal N}=1$ supersymmetric gauge theories the integrals giving the $\beta$-function defined in terms of the bare couplings are integrals of double total derivatives with respect to the loop momenta  in all orders in the case of using the regularization by higher covariant derivatives. This fact agrees with the results of numerous explicit calculations in the lowest orders and generalizes the similar statement for the Abelian case \cite{Stepanyantz:2011jy,Stepanyantz:2014ima}. The proof of the factorization into double total derivatives is a very important step towards the all-loop perturbative derivation of the exact NSVZ $\beta$-function. This derivation consists of the following main steps:

1. Using the finiteness of the triple ghost-gauge vertices (which has been demonstrated in Ref. \cite{Stepanyantz:2016gtk}) we rewrite the NSVZ equation in the equivalent form (\ref{NSVZ_Second_Form}).

2. The $\beta$-function defined in terms of the bare couplings is extracted from the difference between the effective action and the classical action by the formal substitution (\ref{V_Substitution}). Then, using the identity (\ref{Auxiliary_Theta_Identity}) and the background gauge invariance, the result is presented as an integral of a double total derivative in the momentum space. This integral is reduced to the sum of singular contributions which are given by integrals of the momentum $\delta$-functions. (This has been done in this paper.)

3. The remaining step is to sum the singular contributions and to prove that they produce the anomalous dimensions of the quantum superfields in Eq. (\ref{NSVZ_Second_Form}). Now this work is in progress.

As a result, we presumably obtain Eqs. (\ref{NSVZ_First_Form}) and (\ref{NSVZ_Second_Form}) for RGFs defined in terms of the bare couplings in the case of using the higher covariant derivative regularization (in agreement with the results of explicit multiloop calculations). Due to scheme independence of these RGFs (for a fixed regularization) this statement is valid for all renormalization prescriptions.

If the NSVZ relation is really valid for RGFs defined in terms of the bare couplings for theories regularized by higher covariant derivatives, then the all-order prescription for constructing the NSVZ scheme for RGFs defined in terms of the renormalized couplings is HD+MSL. This means using of the higher covariant derivative regularization supplemented by minimal subtractions of logarithms, when only powers of $\ln\Lambda/\mu$ are included into renormalization constants.

As a by-product of the proof presented in this paper we have obtained a simple method for constructing the loop integrals contributing to the $\beta$-function defined in terms of the bare couplings. Actually, it is necessary to calculate (a specially modified) supergraphs without external lines and replace the products of couplings and group factors by a certain differential operator specially constructed for each supergraph. The result is equal to the sum of a large number of superdiagrams which are obtained from the original supergraph by attaching two external lines of the background gauge superfield in all possible ways. Certainly, this drastically simplifies the calculations.

As an illustration of this method we considered all three-loop contributions containing the Yukawa couplings and compared the result with the one found by the standard calculation in Refs.  \cite{Shakhmanov:2017soc,Kazantsev:2018nbl}. The coincidence of the expressions obtained by both these methods confirms the correctness of the algorithm proposed in this paper.

\section*{Acknowledgments}
\hspace*{\parindent}

This work was supported by Foundation for Advancement of Theoretical Physics and Mathematics `BASIS', grant No. 19-1-1-45-1.

I would like to express my gratitude to S.S.Aleshin, A.E.Kazantsev, M.D.Kuzmichev, N.P.Meshcheriakov, S.V.Novgorodtsev, and I.E.Shirokov for valuable discussions and comments on the manuscript.

\appendix

\section{Proof of the identity (\ref{Auxiliary_Theta_Identity})}
\hspace*{\parindent}\label{Appendix_Identity}

For proving the identity (\ref{Auxiliary_Theta_Identity}) we commute $\theta$-s with the operators $A$ and $B$ using equations similar to Eq. (\ref{Shift_Identities}). It is important that $\theta_a \theta_b \theta_c = 0$ (where all $\theta$-s are taken in the same point of the superspace). Therefore,

\begin{eqnarray}\label{LHS}
&&\hspace*{-5mm} \theta^2 AB \theta^2 + 2(-1)^{P_A+P_B}\theta^a A \theta^2 B \theta_a - \theta^2 A \theta^2 B - A\theta^2 B \theta^2\vphantom{\Big(}\nonumber\\
&&\hspace*{-5mm} = \theta^2 [[AB,\theta^a\},\theta_a\}
+ 2(-1)^{P_A+P_B}[\theta^a,A\} \theta^2 [B, \theta_a\} - [\theta^a, [\theta_a, A\}\} \theta^2 B - A[\theta^a, [\theta_a, B\}\} \theta^2,\vphantom{\Big(}\qquad
\end{eqnarray}

\noindent
where

\begin{equation}
[X,Y\} \equiv XY - (-1)^{P_X P_Y} YX.
\end{equation}

\noindent
Anticommuting $\theta_a$ with supersymmetric covariant derivatives inside $A$ and $B$ we obtain expressions which do not explicitly depend on $\theta$. This implies that the right hand side of Eq. (\ref{LHS}) is proportional to the second degree of (explicitly written) $\theta$. After commuting the remaining $\theta^2$ to the left, the expression (\ref{LHS}) can be presented as

\begin{eqnarray}
&& \theta^2 \Big([[AB,\theta^a\},\theta_a\}
- 2(-1)^{P_B}[A,\theta^a\} [B, \theta_a\} - [\theta^a, [\theta_a, A\}\} B - A[\theta^a, [\theta_a, B\}\} \Big) + O(\theta)\nonumber\\
&& = \theta^2 \Big(A[[B,\theta^a\},\theta_a\} + [[A,\theta^a\},\theta_a\} B + 2 (-1)^{P_B} [A,\theta^a\}[B,\theta_a\}
- 2(-1)^{P_B}[A,\theta^a\} [B, \theta_a\}\qquad\nonumber\\
&& - [\theta^a, [\theta_a, A\}\} B - A[\theta^a, [\theta_a, B\}\} \Big) + O(\theta) = O(\theta).
\end{eqnarray}

\noindent
Thus, we have proved the identity (\ref{Auxiliary_Theta_Identity}).

\end{document}